\documentclass[galaxies,article,submit,moreauthors,pdftex,10pt,a4paper,color=usenames, dvipsnames]{mdpi}

\firstpage{1}
\pubvolume{xx}
\issuenum{1}
\articlenumber{1}
\pubyear{2018}
\copyrightyear{2018}
\usepackage{xspace}
\usepackage{amsmath}
\usepackage{graphicx}
\usepackage{subcaption}
\usepackage{threeparttable}
\usepackage[utf8]{inputenc}
\usepackage{aas_macros}

\history{Received: date; Accepted: date; Published: date}

\newcommand{\quotes}[1]{``#1''}

\def\gs{\mathrel{\raise0.35ex\hbox{$\scriptstyle >$}\kern-0.6em \lower0.40ex\hbox{{$\scriptstyle \sim$}}}}
\def\ls{\mathrel{\raise0.35ex\hbox{$\scriptstyle <$}\kern-0.6em \lower0.40ex\hbox{{$\scriptstyle \sim$}}}}

\def\oi{\mbox{\rm{[O\,\scalebox{.8}{I}]}}\xspace}

\def\h2{\mbox{{\sc H}$_2$}\xspace}

\def\cii{\mbox{\rm{[C\,\scalebox{.8}{II}]}}\xspace}

\def\ci{\mbox{\rm{C\,\scalebox{.8}{I}}}\xspace}

\def\fH2{\mbox{$f_{\rm{H}_2}$}\xspace}

\def\d0{\mbox{$D_0$}\xspace}

\def\rh2{\mbox{$R_{\rm{H}_2}$}\xspace}
\def\rrh2{\mbox{$r_{\rm{H}_2}$}\xspace}

\def\cmpc{\mbox{cm$^{-3}$}\xspace} 

\def\msun{\mbox{$\rm{M}_\odot$}\xspace}

\def\cloudy{\mbox{\rm{C\textsc{loudy}}}\xspace}
\def\arttwo{\mbox{\rm{ART$^2$}}\xspace}
\def\despotic{\mbox{\rm{DESPOTIC}}\xspace}
\def\maihem{\mbox{\rm{MAIHEM}}\xspace}
\def\lime{\mbox{\rm{LIME}}\xspace}
\def\radmcthreed{\mbox{\rm{RADMC-3D}}\xspace}

\def\ramsesrt{\texttt{RAMSES-RT}\xspace}

\def\starburst{\texttt{starburst99}\xspace}

\def\trident{\texttt{Trident}\xspace}

\newcommand{\nephthys}{\mbox{\sc nephthys}}
\newcommand{\mollie}{\mbox{MOLLIE}}

\newcommand{\gt}[1]{\mbox{>}}

\Title{Challenges and Techniques for Simulating Line Emission}

\Author{Karen P. Olsen $^{1}$\orcidA{},
 Andrea Pallottini $^{2,3}$\orcidC,
 Aida Wofford $^{4}$\orcidD,
 Marios Chatzikos $^{5}$\orcidE,
 Mitchell Revalski $^{6,7}$\orcidF,
 Francisco Guzm\'an $^{5}$\orcidI,
 Gerg\"o Popping$^{8}$\orcidJ,
 Enrique V\'azquez-Semadeni$^{9}$,
 Georgios E. Magdis$^{10}$\orcidL,
 Mark L. A. Richardson$^{11}$,
 Michaela Hirschmann$^{12}$,
 and William J. Gray$^{13}$
 }

\AuthorNames{Firstname Lastname, Firstname Lastname and Firstname Lastname}

\address{%
$^{1}$ \quad School of Earth and Space Exploration, Arizona State University, 781 South Terrace Road, Tempe, AZ 85287, USA; kpolsen@asu.edu\\
$^{2}$ \quad Centro Fermi, Museo Storico della Fisica e Centro Studi e Ricerche \quotes{Enrico Fermi}, Piazza del Viminale 1, Roma, 00184, Italy; andrea.palllottini@centrofermi.it\\
$^{3}$ \quad Scuola Normale Superiore, Piazza dei Cavalieri 7, 56126, Pisa, Italy; andrea.palllottini@sns.it\\
$^{4}$ \quad Instituto de Astronom\'ia, Universidad Nacional Aut\'onoma de M\'exico, Unidad Acad\'emica en Ensenada, Km 103 Carr. Tijuana$-$Ensenada, Ensenada 22860, M\'exico; awofford@astro.unam.mx\\
$^{5}$ \quad Department of Physics \& Astronomy, University of Kentucky, Lexington, KY 40506, United States; mchatzikos@gmail.com\\
$^{6}$ \quad Department of Physics \& Astronomy, Georgia State University, Atlanta, GA 30302, United States; revalski@astro.gsu.edu\\
$^{7}$ \quad National Science Foundation Graduate Research Fellow (DGE-1550139)\\
$^{8}$ \quad Max-Planck-Institut f\"ur Astronomie, K\"onigstuhl 17, D-69117 Heidelberg, Germany; popping@mpia.de\\
$^{9}$ \quad Instituto de Radioastronom\'ia y Astrof\'isica, UNAM, Morelia; e.vazquez@irya.unam.mx\\
$^{10}$ \quad Cosmic DAWN Centre, Niels Bohr Institute, University of Copenhagen, Juliane Maries Vej 30, 2100 Copenhagen, Denmark\\
$^{11}$ \quad Department of Astrophysics, American Museum of Natural History, 79th Street at Central Park West, New York, NY 10024, USA\\
$^{12}$ \quad Sorbonne Universit\'es, UPMC-CNRS, UMR7095, Institut d’Astrophysique de Paris, F-75014 Paris, France \\
$^{13}$ \quad CLASP, College of Engineering, University of Michigan, Ann Arbor, Michigan 48109
}

\corres{Correspondence: kpolsen@asu.edu}

\abstract{Modeling emission lines from the millimeter to the UV and producing synthetic spectra is crucial for a good understanding of observations, yet it is an art filled with hazards.
This is the proceedings of \quotes{Walking the Line}, a 3-day conference held in 2018 that brought together scientists working on different aspects of emission line simulations, in order to share knowledge and discuss the methodology.
Emission lines across the spectrum from the millimeter to the UV were discussed, with most of the focus on the interstellar medium, but also some topics on the circumgalactic medium.
The most important quality of a useful model is a good synergy with observations and experiments. 
Challenges in simulating line emission are identified, some of which are already being worked upon, and others that must be addressed in the future for models to agree with observations. 
Recent advances in several areas aiming at achieving that synergy are summarized here, from micro-physical to galactic and circum-galactic scale.
}
\keyword{simulation; line emission; galaxies; ISM; radiative transfer; hydrodynamic simulations; CGM; AGN}

\begin{document}



\section{Introduction}\label{intro}
Line emission from the Interstellar Medium (ISM) of galaxies carries information that is crucial in understanding galaxy evolution. 
Observing line emission across the electromagnetic spectrum allows us to characterize the mass, composition, and chemical state of the ISM, as well as to trace galaxy properties such
as star formation rate (SFR), metallicity and dynamics. 
For example, the emission from major cooling lines, such as H$\alpha$ or \cii, is sensitive to the physical conditions (densities, radiation field) and dynamics of the ISM. In addition, emission lines work on all physical scales, from galaxy dynamics and inflows to turbulent and collapse motions in star-forming clouds and cores. By systematically comparing spectral-line signatures of different physical models, one can correctly identify the physical processes occurring in these regions.
Furthermore, the emission from ionized interstellar gas contains particularly valuable information about the nature of the ionizing radiation sources in a galaxy. In fact, prominent optical emission lines are routinely used to estimate whether ionization is dominated by young massive stars (tracing SFR), an AGN or evolved, post-asymptotic giant branch (post-AGB) stars. Three of the most widely used line-ratio diagnostic ``BPT'' diagrams\footnote{``Baldwin, Phillips \& Terlevich'' (BPT) diagrams \cite{Baldwin1981}}, relate the [OIII]/H$\beta$ ratio to the [NII]/H$\alpha$, [SII]/H$\alpha$ and [OI]/H$\alpha$ ratios. These diagrams have proven useful in identifying the nature of the ionizing radiation in large samples of galaxies in the local Universe \citep{Kewley2001,Kauffmann2003}.
Complementary to line emission are the observations of absorption lines of the circumgalactic medium (CGM), which can give key information on the history of the feedback, in terms of chemical, ionization, and thermodynamical state of the outflowing/inflowing gas, that regulates the star formation process.
Gas kinematics, from both emission and absorption, give information about large scale gas flows. Thus galactic outflows, from active galactic nuclei (AGN) and starbursts, can be combined with CGM absorption line observations, to study the star formation history, AGN activity history, and feedback processes that regulate both the evolution of the galaxy and its environment.

\begin{figure}[H]
\centering
\includegraphics[width=0.6\linewidth]{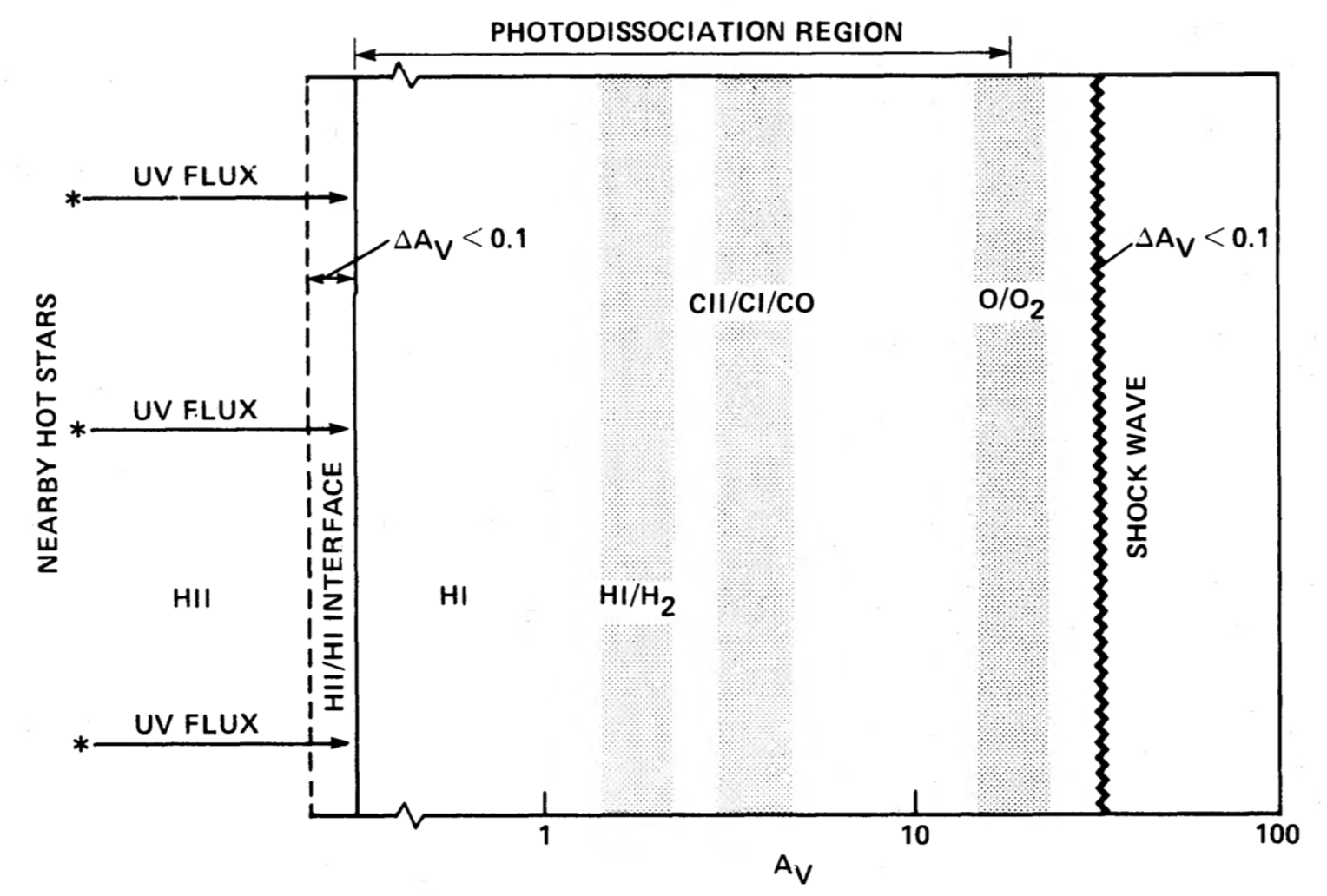}\\
\includegraphics[width=0.495\linewidth]{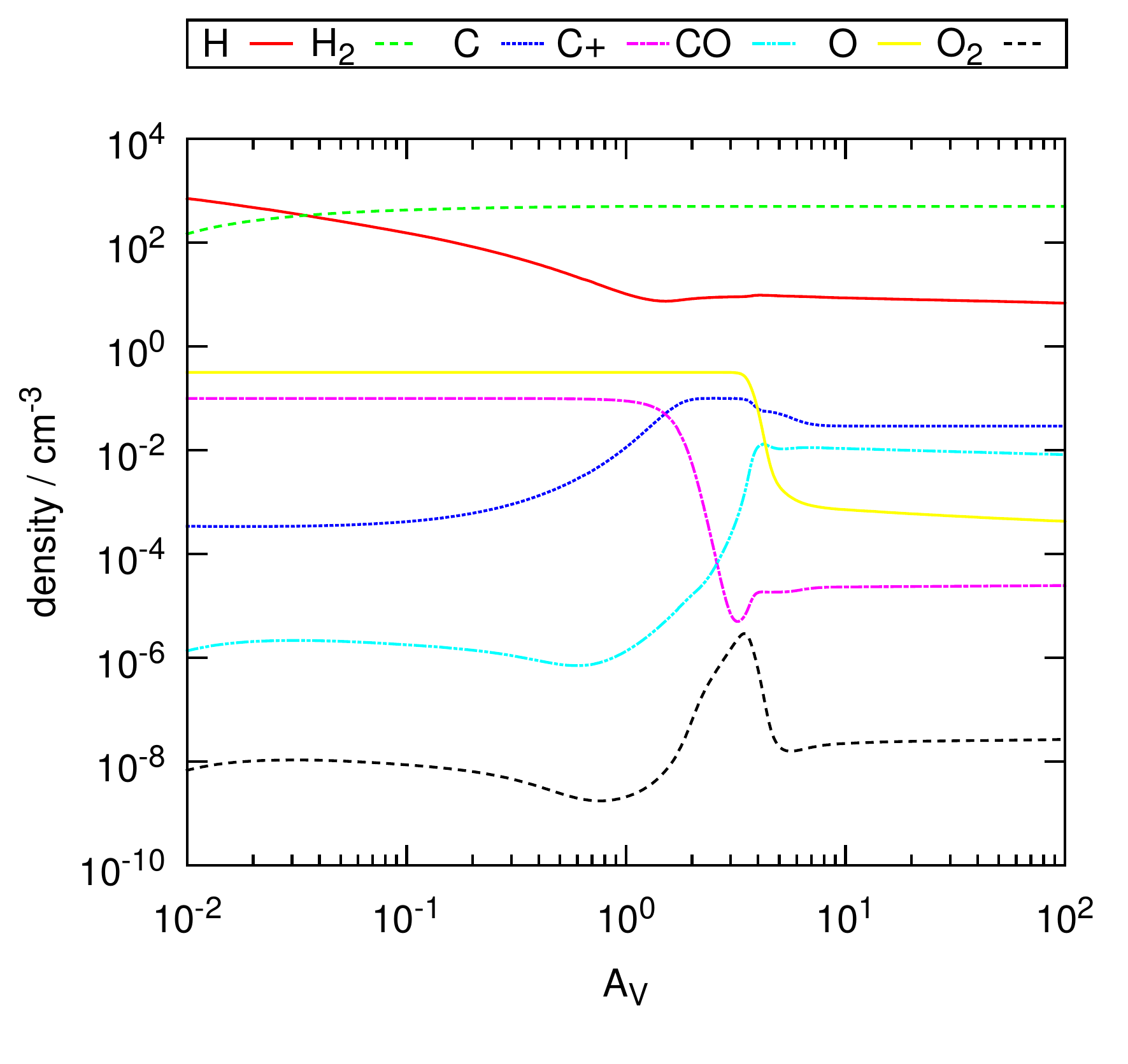}
\includegraphics[width=0.495\linewidth]{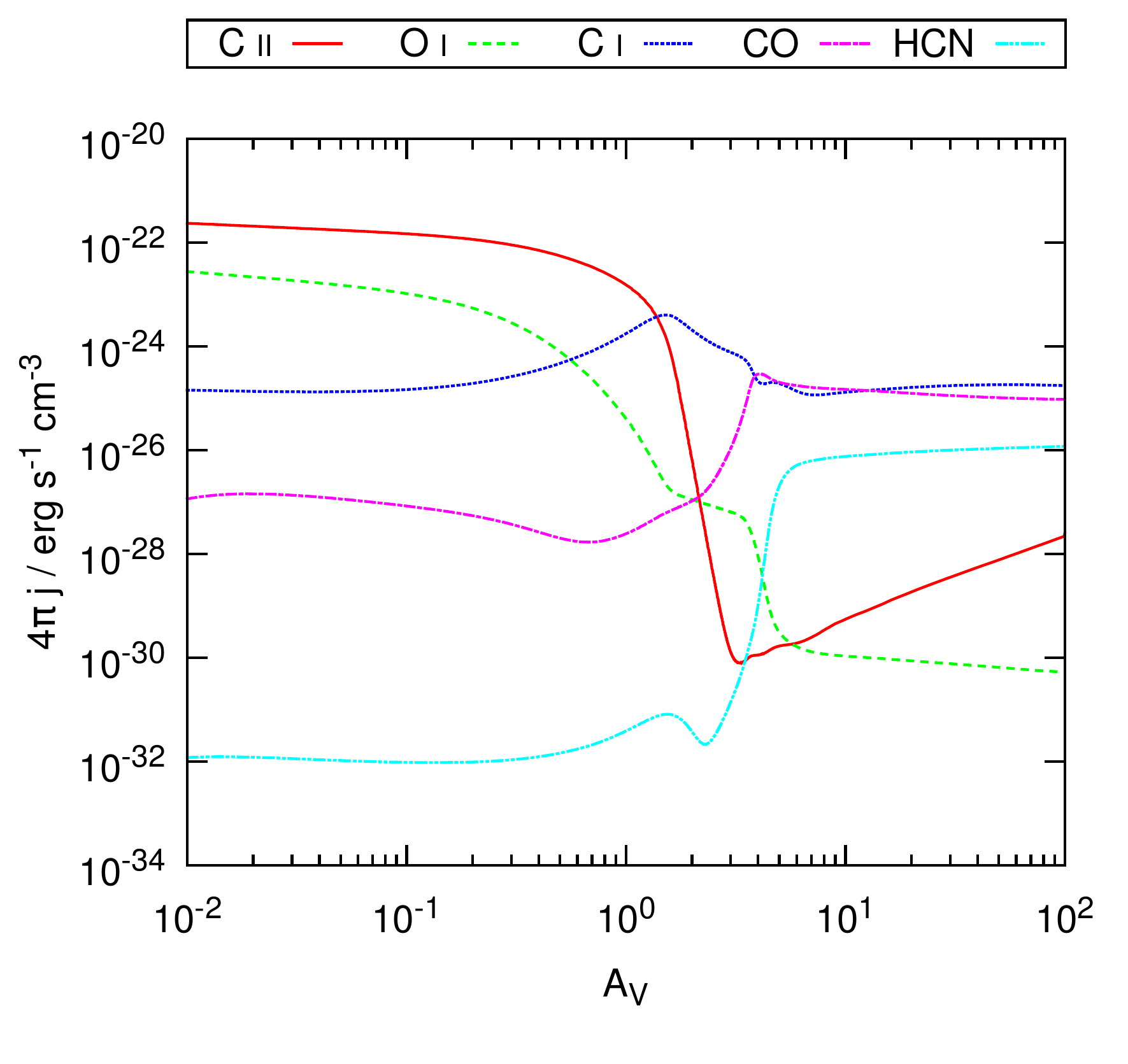}
\caption{{\it (Top)} Sketch of standard PDR structure, from innovative work in the 80's \citep[][\copyright AAS. Reproduced with permission]{Tielens1985} .
{\it (Bottom)} For comparison, an example of PDR structure computed this year with version C17 of \cloudy
\cite{Ferland2017} for model V1 of \cite{Rollig2007}, adapted to extend to $A_V = 100$.
{\it (Left panel)} Densities of important atoms and molecules in the PDR.
{\it (Right panel)} Local emissivities for some of the most important coolants.}
\label{fig:tielens85}
\end{figure}

\begin{figure}[H]
\centering
\includegraphics[width=0.55\linewidth]{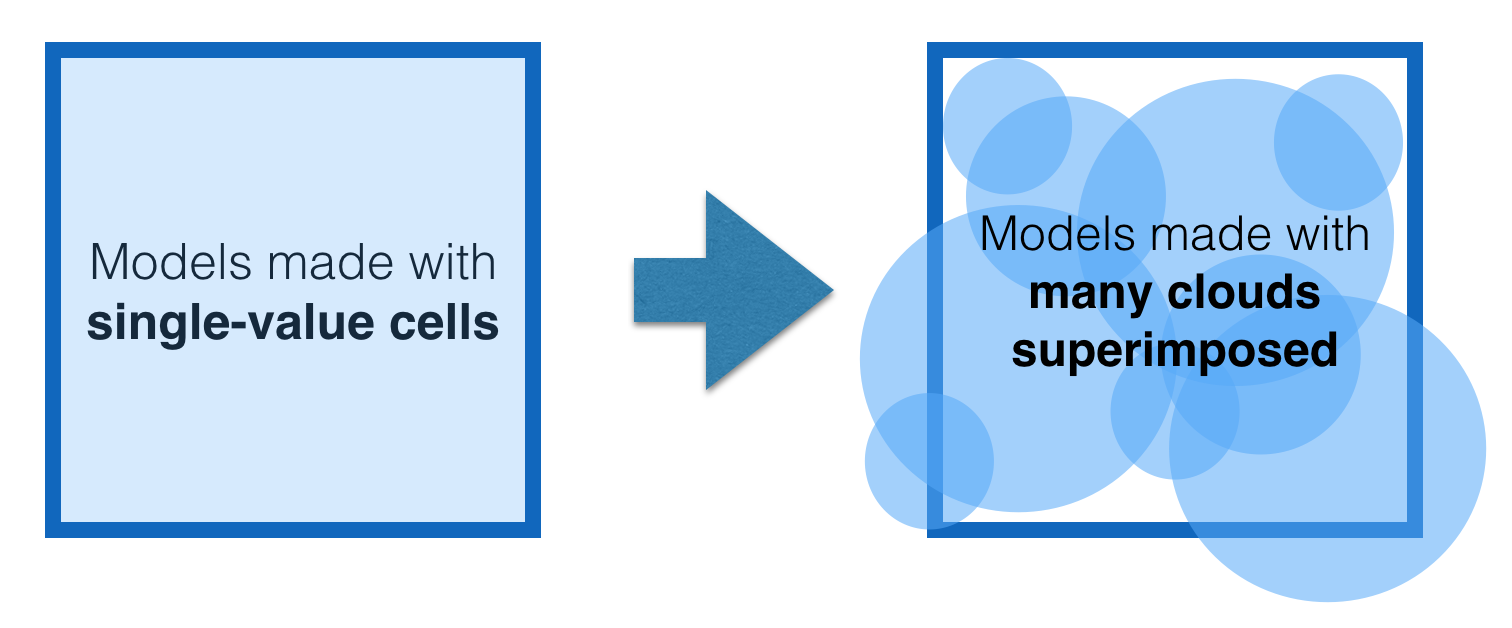}
\caption{Illustration of the difference between simulating line emission from a cell of constant density, temperature, metallicity etc., and using an ensemble of clouds superimposed to create a more realistic synthetic observation of a patch of the sky. }
\label{fig:superimposed_clouds}
\end{figure}

Looking back on the past three decades, the approach to creating synthetic observations of line emission has gone from simplified analytical modeling to complex simulations, increasing the reliability of the results. Since many important cooling lines emerge from the photodissociation regions (PDRs) of the ISM, modeling line emission from the PDRs has been an active field of research since the basic 1D PDR models came into place in the 80’s \citep[e.g.][]{Tielens1985,vanDishoeck1986,vanDishoeck1988a,vanDishoeck1988b,Sternberg1989}.
This early modeling work included the reprocessing of starlight in the UV to infrared continuum by dust grains and polycyclic aromatic hydrocarbons (PAHs), and the stratified layers of increasingly photo-dissociated species as one moves through the neutral gas of a cloud towards the HII region. This picture is illustrated in the top panel of Fig.\,\ref{fig:tielens85}.
In the late 90s, the models presented in the 80's \citep{Tielens1985,Hollenbach1991} were used as basis for a modeling effort to create line ratio diagnostic plots of the \oi, \cii, \ci and CO FIR emission lines \cite{Kaufman1999}. Later the modeling was improved with the online PDR Toolbox\footnote{\url{http://dustem.astro.umd.edu/pdrt/}} as a result \citep{Pound2008,kaufman2006}.
Still, one of the main problems when comparing these models to actual observations is that the models are based on patches of gas with constant density, metallicity, and radiation field strength whereas any given galaxy (or region in a galaxy) will be a superposition of different gas states and radiation conditions as illustrated in Fig.\,\ref{fig:superimposed_clouds} (Ideally, each cloud should also have a radial gradient, but this feature has been omitted from the Figure for the sake of clarity).
With improved simulations of galaxy (and cloud) formation, it became possible in the 21st century to use galaxy/cloud simulations as direct input to photoionization codes, that also calculate the line emission \citep[e.g.][]{Bolatto1999,Rollig2006,Narayanan2006,Popping2014,Vallini2015,Olsen2015,Popping2016,popping2018,vallini18}. This new approach gave way for a more realistic picture in which each region of the ISM is approximated by a combination of different gas conditions. Some of these simulations include radiative transfer and the non-equilibrium impact on the gas temperature of the local radiation field generated by nearby star formation \citep[e.g.][]{Rosdahl2018}, allowing for more accurate estimation of the nebular line emission. This method is being extended to simulations of the CGM, with increased resolution to capture the impact of outflows on accreting structure (see Sec.\,\ref{sec:44}).

Improved numerical techniques such as those mentioned above have also meant a slight division in the field of line modeling, between groups would who specialize in creating the photoionization codes and groups that apply these codes to hydrodynamical simulations. The photoiononization codes have evolved to also calculate line emission and are kept updated as experimental values for collision strengths, chemical rate coefficients and photoelectric heating rates are being revised and databases of atomic and molecular line data increase in size \citep[e.g.][]{Keto+04,Brinch2010,Keto_Rybicky10,Dullemond2012,Yajima2012,Ferland2013,Krumholz2014,Gray2015}.

Similar to the case of PDRs, significant progress has been achieved over the last few years in modelling nebular emission from ionized regions around young star clusters, AGN and post-AGN stellar populations in a full cosmological context. Yet, fully self-consistent models of this kind are currently limited by the performance of cosmological radiation-hydrodynamic simulations and insufficient spatial resolution on the scales of individual ionized regions around stars and active nuclei. To circumvent these limitations, some pioneer studies proposed the post-processing of cosmological hydrodynamic simulations and semi-analytic models with photoionization models to compute the cosmic evolution of nebular emission \citep{Kewley2013,Orsi2014,Shimizu2016}. Some further improvement has been achieved recently by \cite{hirschman2017}, not only accounting for the integrated nebular emission from young stars \citep[as in][]{Orsi2014,Shimizu2016}, but also from AGN and post-AGB stars, based on the star formation, AGN luminosity, gas density, and chemical enrichment histories of the simulated galaxies.

Across different aspects of simulating line emission, key questions remain to be answered:
\begin{itemize}
\item What are the best emission lines to trace various ISM properties and ionizing sources in galaxies?
\item How can we use emission lines to trace feedback and ISM evolution with redshift?
\item How should absorption features be correctly interpreted?
\item Where do we stand in deriving sub-grid physics and comparing codes?
\item How do we coordinate our efforts?
\end{itemize}

To address such questions, the conference \quotes{Walking the Line} was held in Phoenix, Arizona on March 14-16 2018. The main topic was Simulating Line Emission from Galaxies\footnote{\url{https://walk2018.weebly.com/}} and the program\footnote{\url{https://walk2018.weebly.com/program.html}} included 35\,min presentations from three invited speakers and 15\,min presentations from 24 of the remaining 27 participants. In total, nine participants were PhD students and 1/3 of all participants were women. Most talks are now publicly available with video recordings online\footnote{Slides from and video recordings of talks can be found at \url{https://zenodo.org/communities/walk2018/}}. In addition, the conference had daily discussion sessions stimulating in-depth exploration of new topics and sparking new connections. This way, the meeting was an opportunity to discuss the challenges that need to be solved in order to make more realistic simulations and a fairer comparison with observations. 

To take a specific case study where line emission has come to play an important role, one can consider the problem of measuring gas mass, which became a common thread for a large part of the workshop.
Measuring the gas mass of galaxies, the fuel for star formation, as a function of cosmic time is a crucial component in understanding the formation and evolution of galaxies. Different groups have observed the gas content of main-sequence galaxies through cosmic time (the bulk of the galaxy population that is responsible for forming new stars at any epoch in cosmic history) either through their $^{12}$CO (hereafter
CO) emission or the dust continuum \citep[e.g.,][]{Aravena2010,Daddi2010,Tacconi2010, Tacconi2013, Tacconi2018,Geach2011,Magdis2012,Santini2014,Bethermin2016,Decarli2016,Scoville2016}. Both approaches (CO and dust) rely on uncertain conversion factors between the observed luminosity and an estimated gas mass. 
ISM physical processes can further complicate the use of CO and dust continuum to reliably measure the gas mass of galaxies. For example, the contrast of the CO emission line and dust continuum against the cosmic microwave background becomes lower at higher redshifts \citep[e.g.,][]{daCunha2013,vallini18} and under the influence of cosmic rays the CO molecule can be destroyed \citep{Bisbas2015,Glover2016}. 
Reliable alternative measures of the gas mass in a well defined sample of galaxies are therefore essential to overcome the systematic uncertainties in the CO and dust conversion factors and to overcome the CMB contrast. Alternative options seen in the literature include for instance the emission from [CI] \citep{Bothwell2016,Popping2017}, [CII] (Zanella et al. in prep, 2018), from H$_{\rm 2}$O (e.g. Liu et al. 2017) and PAHs features (e.g. Cortzen et al. submitted 2018), and from optically thin isotopologues \citep[e.g.,][]{cormier2018}. 
The synergy between galaxy formation simulations, ISM chemistry simulations, and radiative transfer codes has the power to pave the way towards reliably measuring gas masses. This synergy allows for a controlled setting in which the conversion between sub-mm line/continuum strength and H$_{\rm 2}$ mass can be explored. This is important to 1) quantify under which physical conditions classical approaches to measure gas masses such as CO and dust continuum emission break down and 2) explore the robustness of other mid-IR and sub-mm emission lines such as PAHs, [CI], [CII] and H$_{\rm 2}$O (although one could think of other examples) as a tracer of molecular hydrogen mass.

In this paper we summarize the topics covered in the conference, with particular attention to the break-out sessions of this workshop. We report our conclusions on the state of the art of modelling emission lines, by going to progressively larger physical scales (Sec.s \ref{sec:2} -- \ref{sec:4}), concluding by discussing the possible ways to move forward as a community (Sec. \ref{discussion}).

\begin{table}[H]
\caption{Overview of the lines discussed at the workshop and referred to in the remainder of this paper.}
\label{table:2}
\centering
\begin{threeparttable}
\begin{tabular}{p{1cm}|p{2cm}|p{3cm}|p{4cm}|p{2cm}}
\toprule
\textbf{Name}	& Type & \textbf{Wavelength(s)\tnote{(1)}}	& \textbf{Tracer of} & \textbf{Reference for wavelength(s)}\\
Ly$\alpha$ 		& 	Recombination	& $1215.67\,$\AA 	& Ionized ISM & \cite{Sansonetti2004}	\\
C\,{\sc iv} 		& 	CE\tnote{(2)}		& $1548.19$, $1550.77\,$\AA 	& Stellar wind, ionized ISM	& \cite{Leitherer2011}	\\
O\,{\sc iii}] 		& 	CE\tnote{(2)}	& $1660.81$, $1666.15\,$\AA 	& Ionized ISM	& \cite{Leitherer2011}	\\
He\,{\sc ii} 		& 	Recombination & $1640.42\,$\AA 	& Stellar wind, ionized ISM	& \cite{Leitherer2011}	\\
$[$C\,{\sc iii}$]$ & 	CE\tnote{(2)}	& $1906.68$ \AA 	& ISM	& \cite{Leitherer2011}	\\
C\,{\,\sc iii}] 		& 	CE\tnote{(2)}	& $1908.73\,$\AA 	& Ionized ISM	& \cite{Leitherer2011}	\\
H$\beta$ 		& 	Recombination	& $4861.36\,$\AA 	& Ionized ISM & NIST\tnote{(3)}	\\
$[$O\,{\sc iii}$]$ & 	CE\tnote{(2)}	& $4958.91,5006.84$ \AA 	& Ionized ISM	& NIST\tnote{(3)}	\\
$[$O\,{\sc i}$]$ 		& 	Recombination	& $6300.30\,$\AA 	& Ionized ISM & NIST\tnote{(3)}	\\
H$\alpha$ 		& 	Recombination	& $6562.80\,$\AA 	& Ionized ISM & NIST\tnote{(3)}	\\
$[$N\,{\sc ii}$]$ & 	CE\tnote{(2)}	& $6548.05,\,6583.45$ \AA 	& Ionized ISM	& NIST\tnote{(3)}	\\
$[$S\,{\sc ii}$]$ & 	CE\tnote{(2)}	& $6716.44,\,6730.82$ \AA 	& Ionized ISM	& NIST\tnote{(3)}	\\
\ci 			& 	Fine-structure	& $609.14$, $370.42\,\mu$m 	& Atomic and molecular gas	& 	LAMDA\tnote{(4)}\\
\cii 			& 	Fine-structure	& $157.74\,\mu$m 	& All ISM	& 	LAMDA\tnote{(4)}\\
\oi 			& 	Fine-structure	& $63.18$, $145.53\,\mu$m 	& Atomic and molecular gas 	& 	LAMDA\tnote{(4)}\\
CO 				& 	Rotational		& $2.6$, $1.3$, $0.87$\,mm ... 	& Molecular gas	& 	LAMDA\tnote{(4)}\\
\midrule
\bottomrule
\end{tabular}
\begin{tablenotes}
\item[1] We give air wavelengths above 2000 \AA~ and below 20,000 \AA, and vacuum wavelengths otherwise.\vspace{3.5pt}
\item[2] Collisionally excited line\vspace{3.5pt}
\item[3] \url{https://physics.nist.gov/PhysRefData/ASD/lines_form.html}\vspace{3.5pt}
\item[4] \url{http://home.strw.leidenuniv.nl/~moldata/}\vspace{3.5pt}
\end{tablenotes}
\end{threeparttable}
\end{table}

Note that the number of emission (and absorption) lines observed from the ISM of galaxies is ever increasing, and we did not attempt to cover them all with this workshop. However, to give the reader a quick overview of the lines that were discussed at the workshop and are commonly used to diagnose the ISM and/or CGM, we present in Table \ref{table:2} a list of lines that will be mentioned in this paper.


\section{On the micro-physical level}\label{sec:2}

\subsection{Tools for solving level populations and line excitation}\label{sec:21}

Several software tools are available online for the calculation of line emission. While not covering all, the tools presented and discussed at this workshop are listed alphabetically in Table \ref{table:1} for quick reference and comparison. Before comparing the tools, it is worth noting that they are built to achieve different levels of precision at the cost of computation time. As such, codes like \lime, \mollie, and \radmcthreed are very accurate in 3D, but slow compared to \cloudy and \despotic which solve for chemistry and temperature in 1D and are intended for larger parameter studies that can be analyzed fast.

In terms of chemical networks, \cloudy is the most comprehensive tool with 625 species and a large range of permitted densities that allows for application to models on scales from clouds and HII regions to galaxies \citep[e.g.][]{Mittal2011,Pallottini2017,Xiao2018}. \despotic has been applied in post-processing to galaxy simulations mainly \citep[e.g.][]{Popping2016,Safranek-Shrader2017,popping2018} but also to smaller regions such as the Galactic central molecular zone \cite{Ginsburg2016}. 
As mentioned above, \cloudy and \despotic are restricted to problems in 1D whereas the remaining tools in Table \ref{table:1} work in 3D. 
\lime, \mollie, and \radmcthreed do exact radiative transfer in 3D and are typically used on smaller scales where non-symmetric features such as filaments and turbulence become important \citep[e.g][]{Penaloza2017,Penaloza2018}. 
\arttwo also does radiative transfer in 3D including the continuum from far-UV to radio wavelengths and incorporating the resonant line Ly$\alpha$ \citep[e.g][]{Li2008, Yajima2012,Yajima2014} (an updated version of the code including CO and some prominent far-infrared fine-structure lines such as CII, OI, OIII and NII will be out later this year). Finally, \maihem is unique in taking into account the  non-equilibrium effects of turbulence on line emission (c.f. Sec.\,\ref{sec:33}).

\begin{table}[H]
\caption{Overview of the computational tools represented at the workshop.}
\label{table:1}
\centering
\begin{threeparttable}
\begin{tabular}{p{2cm}|p{2cm}|p{2cm}|p{1cm}|p{4cm}|p{2cm}}
\toprule
\textbf{Name}	& \textbf{Reference}	& \textbf{Density regime} & \textbf{1D or 3D} & \textbf{Species in chemical network for calculating ionization states} & \textbf{Exact radiative transfer included?}\\
\midrule
\arttwo			& \cite{Yajima2012}, Li 2018 (in prep)		& $10^{-2}$-$10^7\,\cmpc$ 	& 	3D & Atomic database of \cloudy, molecular database of LAMDA	& yes\\
\cloudy\tnote{(1)} & \cite{Ferland2017}		& $10^{-6}$-$10^{15}\,\cmpc$ & 1D	& 625 species (including atomic ions); CHIANTI, Stout, LAMDA databases	& no\tnote{(9)}\\
\despotic\tnote{(2)}		& \cite{Krumholz2014}	& Cool atomic and molecular ISM 	& 1D	& C, O, H, and He, plus a super-species M that represents a composite of metallic elements	& no\tnote{(9)} \\
\lime\tnote{(3)}			& \cite{Brinch2010}		& $10^{2}$-$10^{12}\,\cmpc$ 	& 3D	& LAMDA database	& yes \\
\maihem\tnote{(4)}			& \cite{Gray2015}, \cite{Gray2016}, \cite{Gray2017} & Has been tested at $0.4$-$1200\,\cmpc$ 	& 3D	& 63 species (including atomic ions)\tnote{(5)}	& no\\
\mollie\tnote{(6)} & \cite{Keto+04, Keto_Rybicky10} & $10^{2}$-$10^{12}\,\cmpc$ & 3D & 18 molecular species\tnote{(7)} & yes\\
\radmcthreed\tnote{(8)}	& \cite{Dullemond2012} 	& No limits 	& 3D	& LAMDA database, but abundances can also be supplied by the user	& yes\\
\bottomrule
\end{tabular}
\begin{tablenotes}
	\item[1] \url{https://www.nublado.org/} \vspace{3.5pt}
	\item[2] \url{https://bitbucket.org/krumholz/despotic/} \vspace{3.5pt}
    \item[3] \url{http://www.nbi.dk/~brinch/index.php?page=lime} \vspace{3.5pt} 
	\item[4] \url{http://maihem.asu.edu/} \vspace{3.5pt}
    \item[5] H, H$^+$, He, He$^+$, He$^{2+}$, C-C$^{5+}$, N-N$^{6+}$, O-O$^{7+}$, Ne-Ne$^{9+}$, Na-Na$^{2+}$, Mg-Mg$^{3+}$, Si-Si$^{5+}$, S-S$^{4+}$, Ca-Ca$^{4+}$, Fe-Fe$^{4+}$, and electrons. \vspace{3.5pt}
    \item[6] \url{https://www.cfa.harvard.edu/~eketo/mollie} \vspace{3.5pt}
    \item[7]  CH$_3$CN, NH$_3$, N$_2$H$^+$, H$_2$O (para + ortho), CO, CS, $^{13}$CO, H$^{13}$CO$^+$, HCO$^+$, C$^{17}$O, C$^{18}$O, HCN, H$^{13}$CN, HC$^{15}$N, N$_2$D$^+$, SiO, H$_2$D$^+$.\vspace{3.5pt}
	\item[8] \url{http://www.ita.uni-heidelberg.de/~dullemond/software/radmc-3d/} \vspace{3.5pt}
    \item[9] Escape probability formalism.
\end{tablenotes}
\end{threeparttable}
\end{table}

\subsection{Correcting heavy-element energy levels}\label{sec:22}
Predictions on line emission modeling depends on the reliability of fundamental
atomic data. Software tools such as those discussed in the previous section use theoretical and experimental atomic and
molecular rates and energies to calculate emissivities and line wavelengths
respectively, mostly from external databases as CHIANTI \cite{dere97,delzanna15} or NIST.
Nowadays, theoretical energy levels of heavy many-electron ions
disagree with experimental values, leading to incorrectly predicted
wavelengths. This happens in all
the energy and wavelength ranges.  For instance, there are differences between
the energies of Fe II reported by the NIST and CHIANTI databases,
where several terms reported by NIST are missing
in CHIANTI. Similarly, efforts to fit the spectra of X-ray satellite
observations have accentuated the necessity of including transitions between
levels with high main quantum number $n$ in the theoretical calculations
\citep [e.g.][]{brickhouse2000, kaastra96, barshalom2000, delzanna04}. A
displacement of about 6\,\AA ~has been found in some X-ray lines due to
theoretical and experimental differences for energy levels \cite{delzanna04},
and observation-theory intensity ratios as large as 10 have been reported for
several lines in Fe ~XVI. It is believed that some of the
differences could be explained by the blending of the lines with emissions from
other ions \cite{keenan07}.  Moreover, experimental measurements using ion
traps (EBIT) have revealed significant differences with theoretical energies
\citep [see e.g.][]{kato09, beiersdorfer14, delzanna04}. This situation has
improved over the last decade. The high computational power reachable nowadays
allows calculating energies and collision strengths for groups of ions with the
same iso-sequence \citep[e.g.][]{liang09, liang10, wang14, delzanna16},
or the same type of transitions for all ions \cite{liedahl95}. As different
codes calculate new sets of data as well as fit spectra (e.g. FAC \citep{gu08},
GRASP \citep{dyall89}, AUTOSTRUCTURE \citep{badnell86}) and new data are stored
in databases such as CHIANTI \citep{dere97, delzanna15}) and HULLAC
\citep{barshalom2000}, there is better general agreement between theory and
experiments \citep[see for example][]{aggarwal14}. However, there are still
lines which remain unidentified, while others are poorly described because they
are blended or have wrongly assigned transitions \citep{delzannabadnell16}.

\begin{figure}[H]
\centering
\includegraphics[width=\textwidth]{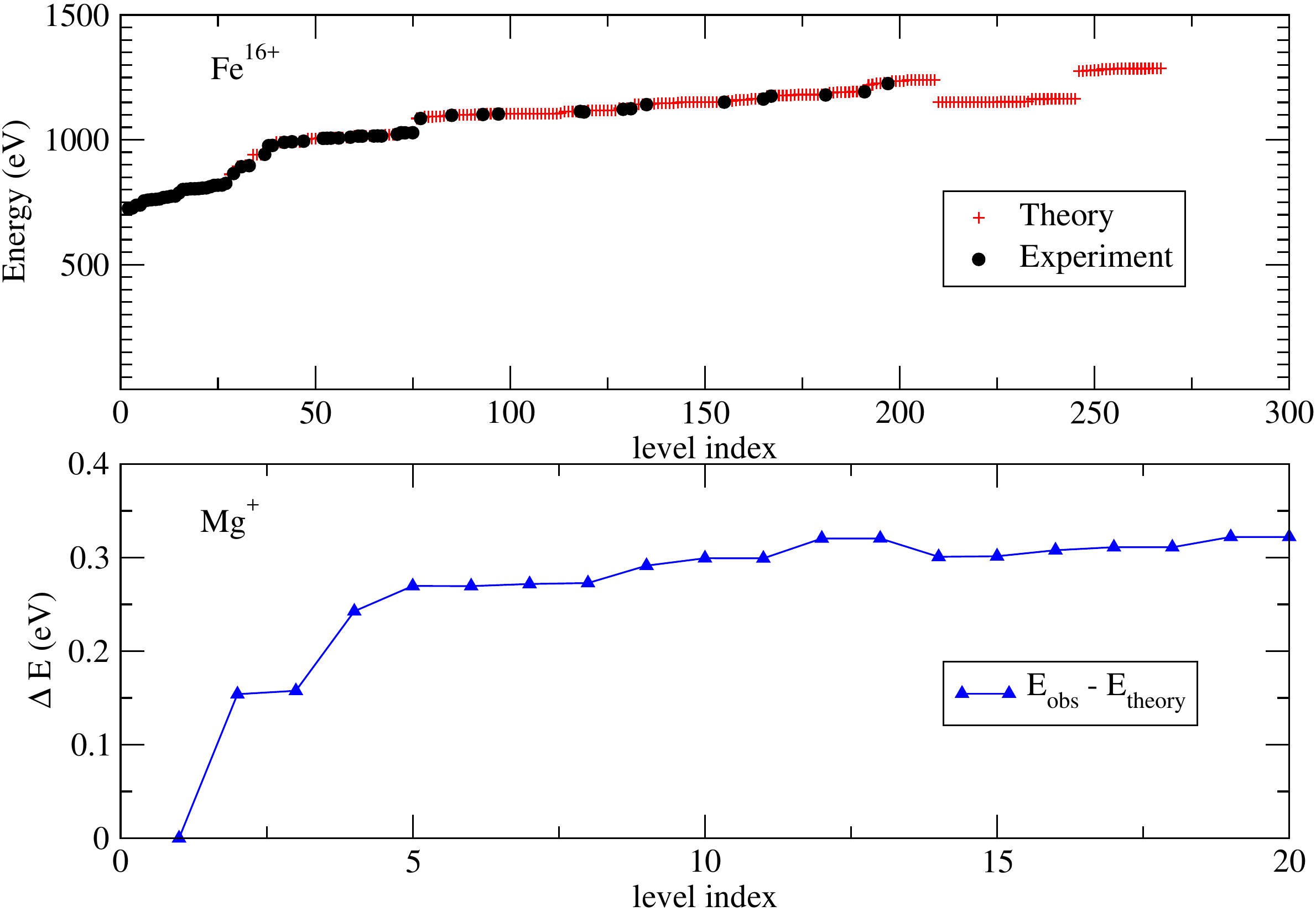}
\caption{Theoretical and experimental energies for levels obtained from the
  CHIANTI (v7, \cite{landi12}) database. ({\it Top}) Fe$^{16+}$: Observed energy
  levels (up to n=5): \cite{delzanna09}; observed energy levels (n=6,7) from the
  wavelength measurements of \cite{brown98}. ({\it Bottom}) Mg$^+$: theoretical
  energies:\cite{liang09}; observed energies from NIST: \cite{NIST_ASD}}
\label{f:energydiffs}
\end{figure}

To illustrate this, two possible cases of disagreement
between theory and experiment, both taken from CHIANTI (v7, \cite{landi12}), are
plotted in Fig. \ref{f:energydiffs}. In the upper panel, the theoretical energies
for Fe$^{16+}$ are, at most, roughly 1\% off the experimental values. However,
theory predicts many more levels that have not yet been observed. In the lower
panel, the few levels predicted for Mg$^+$ are systematically lower than
experimental values. Although the error for Mg$^+$ levels is around 3\%, in the
worst case, a line to the ground level can be shifted by as much as 25Å. 

\subsection{Radiative transfer of UV to infrared}\label{sec:23}

The reprocessing of UV light into infrared via dust grains is crucial for the energy balance in the ISM and thereby the line emission. In addition, it allows for a determination of the dust continuum emission which is as important an observable as line emission. However, radiative transport (RT) is a highly complex, non-local, and non-linear problem. It is considered one of the four grand challenges in computational astrophysics.\footnote{E.g., the Grand Challenge Problems in Computational Astrophysics conference series at
\href{http://www.ipam.ucla.edu/programs/long-programs/grand-challenge-problems-in-computational-astrophysics/}{https://www.ipam.ucla.edu/}.}
Considerable progress has been achieved over the last 20 years in terms of carrying out exact 3D radiative transfer simulations of dust, as outlined in \cite{dust3DRT}. Yet, modeling dust radiative transfer in large hydrodynamical simulations is very computationally expensive, and it is usually ignored. A few pioneering studies that account ``on the fly'' for radiative transfer and the effects of dust are \cite[e.g.][]{Gnedin2014,So2014,Wise2014,Norman2015,Pawlik2015,Ocvirk2016,Rosdahl2018}. \ramsesrt \cite{Rosdahl2013,RAMSES-RT-dust} implements radiative transfer in a cosmological code, albeit without a full treatment for dust emission. Examples of other codes that transfer radiation on the fly include EMMA \cite{EMMA} in a cosmological setting, as well as ORION \cite{ORION} and Athena \cite{ATHENA-RT} in an ISM setting. Generally speaking, the impact of dust on the energy balance of simulated galaxies is typically not accounted for when computed on the fly. Alternative approaches to modeling dust are therefore worth exploring.

Ideally, a full simulation of the hydrodynamical evolution of astrophysical clouds, including galaxies, needs to account for the propagation of radiation through the simulated volume, {\em and} for its interaction with matter. The latter requires detailed knowledge of the quantum structure of all ions, molecules, and dust present in the simulated gas, as well as keeping track of the chemical reactions that couple the various species. For a simulation with a few million cells or particles, the problem becomes extremely computationally expensive, and the non-local couplings induced by radiation render it practically intractable with the computational means available at present.

An alternate solution is to solve for the hydrodynamics and the microphysics separately, in post-processing. In this approach, a code such as those listed in Table\,\ref{table:1} is required to simulate the microphysics. \cloudy \cite{Ferland2017} performs 1D spectral synthesis simulations of astrophysical clouds, accounting for all ions of elements up to Zn, a large number of molecular species, and all important microphysical processes. The propagation of radiation is done with the escape probability formalism, which is an expedient, and generally accurate approximation \cite{Elizur2006-CEP}. However, a few pathological cases exist for which escape probabilities are known to not give correct results \cite{EPvsERT}. Exact radiative transfer methods are required if a code is to be used in conjunction with hydrodynamical codes. The \cloudy team will seek to implement exact radiative transfer, eventually with full treatment of the interactions of dust and radiation. Even with these improvements, however, due to their 1D nature these simulations will not capture the time-varying three-dimensional radiative fields that exist in nature. An exact 3D treatment of radiative transfer will be needed to connect the microphysics and 3D hydrodynamical models.


\section{Cloud-scale simulations}\label{sec:3}

\subsection{The internal density and velocity profile of molecular clouds}\label{sec:31}
In calculating emission line strengths from molecular clouds, the adopted radial profiles of densities and radial velocity profile of the gas can drastically change the result. For example, at the scale of molecular clouds and their substructure, the so-called ``infall profiles'' are routinely used in identifying collapsing dense cores \citep[see, e.g.,] [] {Evans99}. These profiles consist of self-absorbed lines with a blue excess in moderately optically thick lines, in conjunction with single-peaked profiles of similar total width in optically thin lines. These profiles have been interpreted in terms of a dense molecular cloud core undergoing ``inside-out collapse'' \cite{Shu77}, where the inner regions of the core are undergoing collapse, while the external envelope is in hydrostatic equilibrium, the two regions mediated an expanding rarefaction front. The assumed underlying radial velocity profile is essential for the generation of the lines. Recent studies \citep[e.g.,] [Loughnane et al. 2018, submitted] {Keto+15} have focused on the effect different assumed radial density and velocity profiles have on the emitted lines, and favored the collapse of marginally-unstable structures, characterized by ``outside-in'' velocity profiles rather than the canonical inside-out one. Thus, traditional line modeling has to {\it assume} a velocity profile, but there may be a degeneracy such that different combinations of density and velocity profiles may produce similar spectral lines. Synthetic observations of numerical simulations are thus essential in order to break the degeneracy, since they produce dynamically self-consistent density and velocity profiles, restricting the range of physically acceptable choices and thus facilitating the identification of the actual physical conditions that produce the signature spectral lines.

\subsection{Simulating the ionizing UV field that clouds are embedded in}\label{sec:32}

\begin{figure}[H]
\centering
\includegraphics[width=\textwidth]{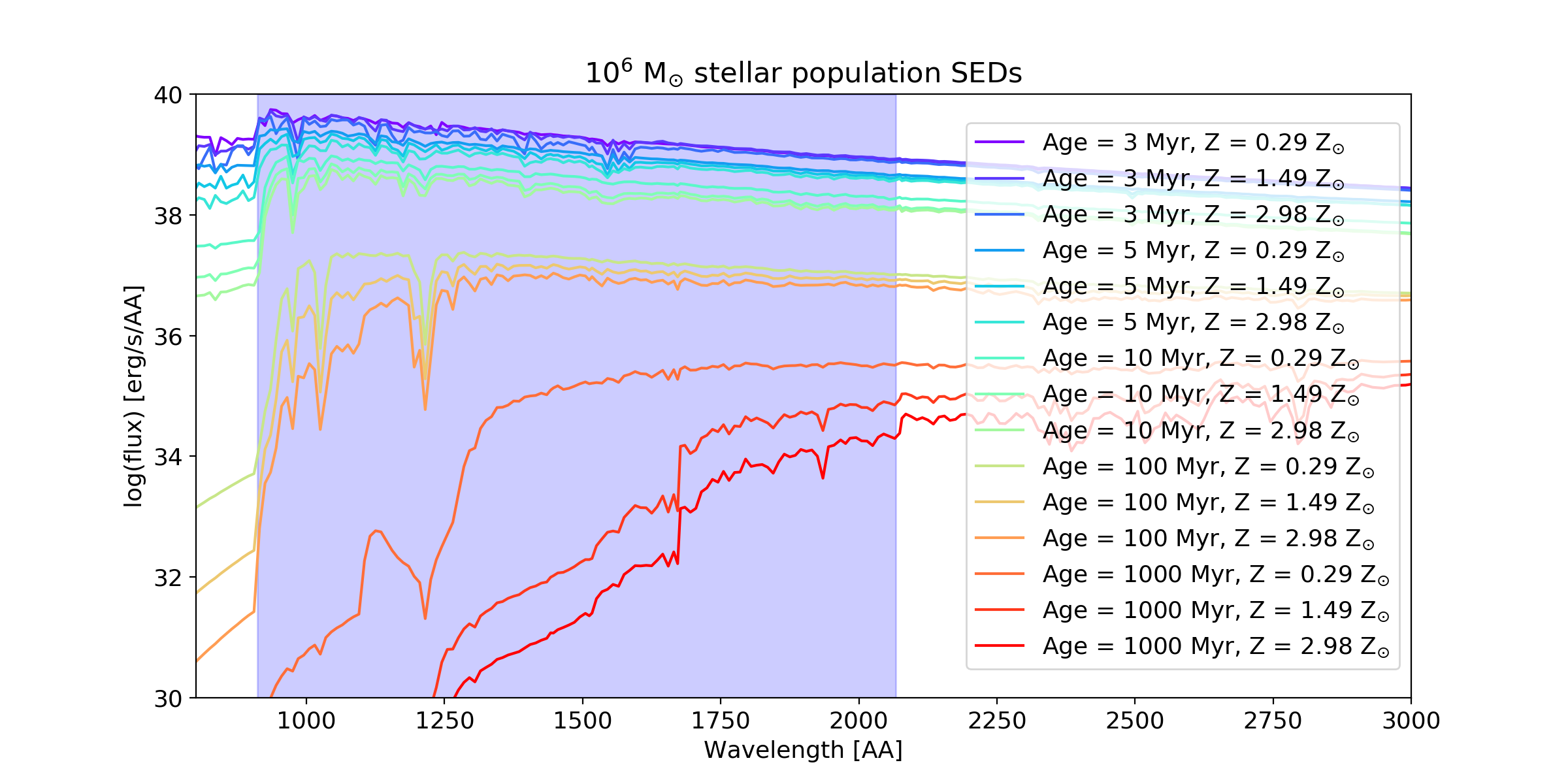}
\caption{Example of SEDs for stellar populations of increasing age and metallicity, calculated with the \starburst tool (see the text). A Kroupa IMF \cite{Kroupa2001} at masses $0.1$-$100$\,\msun was used together with an instantaneous burst of star formation which happened at age = 0 Myr. The Geneva high-mass loss tracks \cite{Meynet1994} were used. The shaded area indicates the region in wavelength used to integrate the SED for obtaining the FUV flux in e.g. Habing units \cite{Habing1968}, setting the chemical state of the gas. We note that comparing these models of different metallicities at the same age can be misleading, because stars of different metallicities have different lifetimes, i.e., a star of 3 Myr with solar metallicity will not be in the same evolutionary stage as a star of 3 Myr with 0.1 times solar metallicity.}\label{sb99}
\end{figure}

The standard way to model the ionizing UV radiation, which  ultimately determines the chemistry of the ISM, is to use stellar population synthesis (SPS) codes. In the SPS methodology, the integrated spectrum from populations of stars with given characteristics (IMF, star formation history, metallicity, etc.) is derived from stellar evolutionary tracks, which give the H-R diagram positions of stars of given masses with time, and from individual stellar spectra, which can be empirical or theoretical. If theoretical, they are computed with stellar atmosphere codes. Fig.\,\ref{sb99} shows examples of model predictions for populations of stars resulting from an instantaneous burst of star formation, which were computed with \starburst\footnote{\url{http://www.stsci.edu/science/starburst99/docs/default.htm}}. 

However, using SPS codes comes with a few cautions worth knowing. For example, SPS codes from independent research groups account for different stellar astrophysics (rotation, close-binarity, etc., see e.g. \cite{Eldridge2017}) and predict different ionizing spectra (e.g., Fig. 2 of \cite{Wofford2016}). 
In addition, the ionizing continuum from individual ionizing stars cannot be observed directly, because it is strongly absorbed in the ISM. This makes constraining model predictions of the ionizing spectrum of populations of stars difficult. The escape fractions of ionizing photons from a few distant galaxies have been estimated from direct detections of ionizing photons (e.g., \cite{Izotov2016}). However, these are insufficient to constrain the ionizing spectrum from massive star populations, because they are just escape fractions. Thus, we must rely on theory. However, there remain large uncertainties in massive star evolution (trajectories in H-R diagram and stellar atmospheres), particularly for post-MS phases and stars more massive than 20 $M_\odot$.

\subsection{Direct observations of low-metallicity massive stars}\label{sec:33}
Massive ($M>8\,M_\odot$) extremely metal-poor stars ($Z/Z_\odot\le 0.1$) are expected to be present in extremely metal-poor galaxies, i.e., galaxies with ionized-gas oxygen abundances of $\le$solar/10. Since such galaxies are farther than $\sim$10 Mpc away, they are too distant to enable observations of their individual massive stars. This is a problem for understanding a variety of astrophysical objects which range from the first stars and galaxies which re-ionized the universe to the metal-poor stars that give rise to the formation of heavy binary black holes. Direct observations of individual, massive, extremely metal-poor stars will only be possible if telescopes such as LUVOIR\footnote{The Large UV/Optical/IR Surveyor: \url{https://asd.gsfc.nasa.gov/luvoir/}} and HabEx\footnote{Habitable Exoplanet Imaging Mission: \url{https://www.jpl.nasa.gov/habex/}}, which are being proposed to NASA, ever see the light.

\subsection{Implementing turbulence and shocks in simulations of the ISM}\label{sec:34}

\begin{figure}[H]
\centering
\includegraphics[width=15 cm]{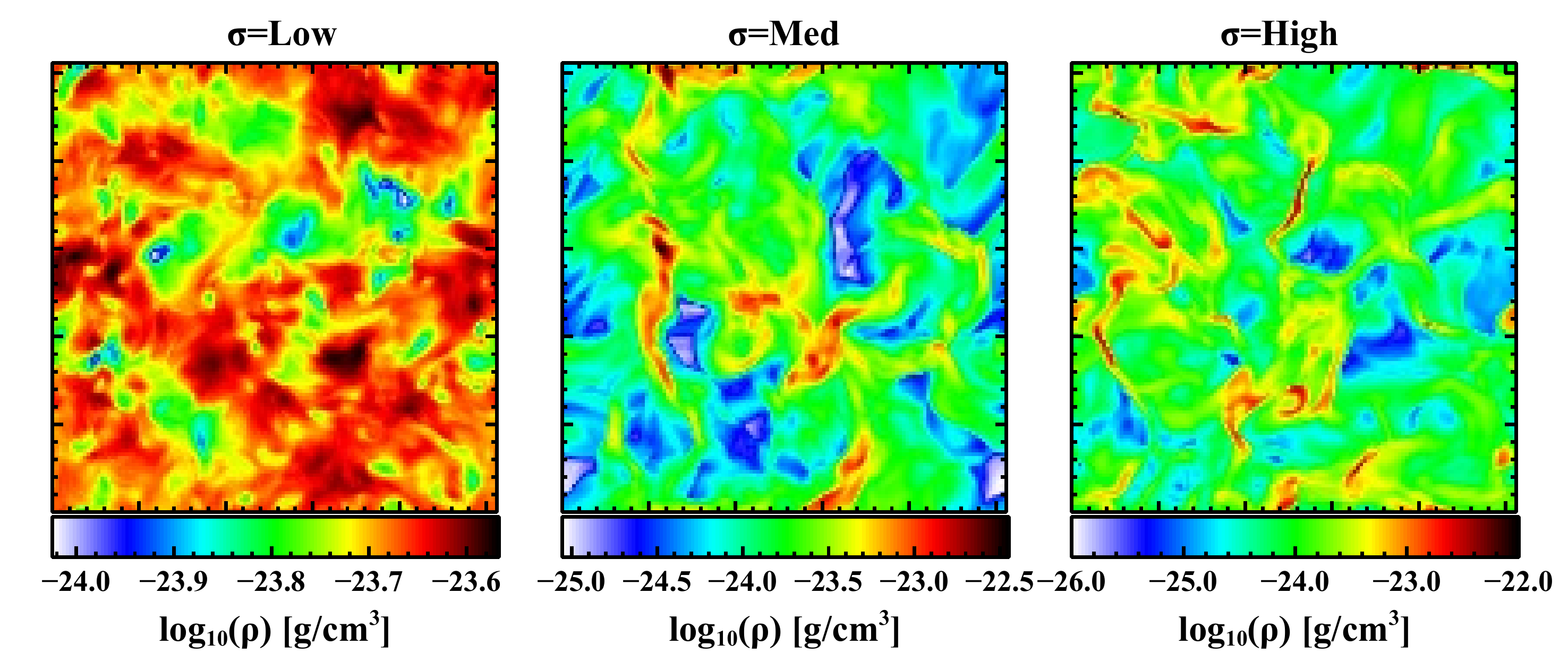}
\caption{An illustration of how the gas density structure changes with increasing levels of turbulence, $\sigma$. At low $\sigma$ the turbulence is subsonic and no shocks are formed creating a nearly uniform density distribution. At intermediate and high $\sigma$, shocks form and create regions of very high and low density.}\label{fig:turb}
\end{figure}

In addition to the importance of modeling stars as described in the previous two subsections, another important source of heating and ionization is turbulence and shocks. The introduction of density gradients and additional sources of ionization through these processes can have a significant impact on the observed spectrum, and hence the emission line diagnostics. Examples of how different levels of turbulence affect the density structure of ISM gas is shown in Fig.\,\ref{fig:turb}.

One solution to understanding the effects of turbulence is through the publicly available MAIHEM\footnote{\url{http://maihem.asu.edu/}} code. This FLASH (version 4.4) based code tracks ionizations, recombinations, and species by species radiative cooling for hydrogen, helium, carbon, nitrogen, oxygen, neon, sulfur, calcium, and iron \citep{Gray2015,Gray2016,Gray2017}. By including a solenoidal stirring mechanism, turbulence is introduced into the gas that then propagates through all spatial scales.

At the meeting, there was an interest in applying MAIHEM to Baldwin-Phillips-Terlevich (BPT) diagrams \citep{Baldwin1981,Veilleux1987} and other UV diagnostics. Specifically, turbulence in the gas might help explain the observed dispersion in some spatially-resolved BPT diagrams of nearby AGN \cite{Revalski2018}, including a larger scatter seen in the neutral optical [O~I] emission line as compared to [N~II] and [S~II].

\begin{figure}[H]
\includegraphics[width=7.5 cm]{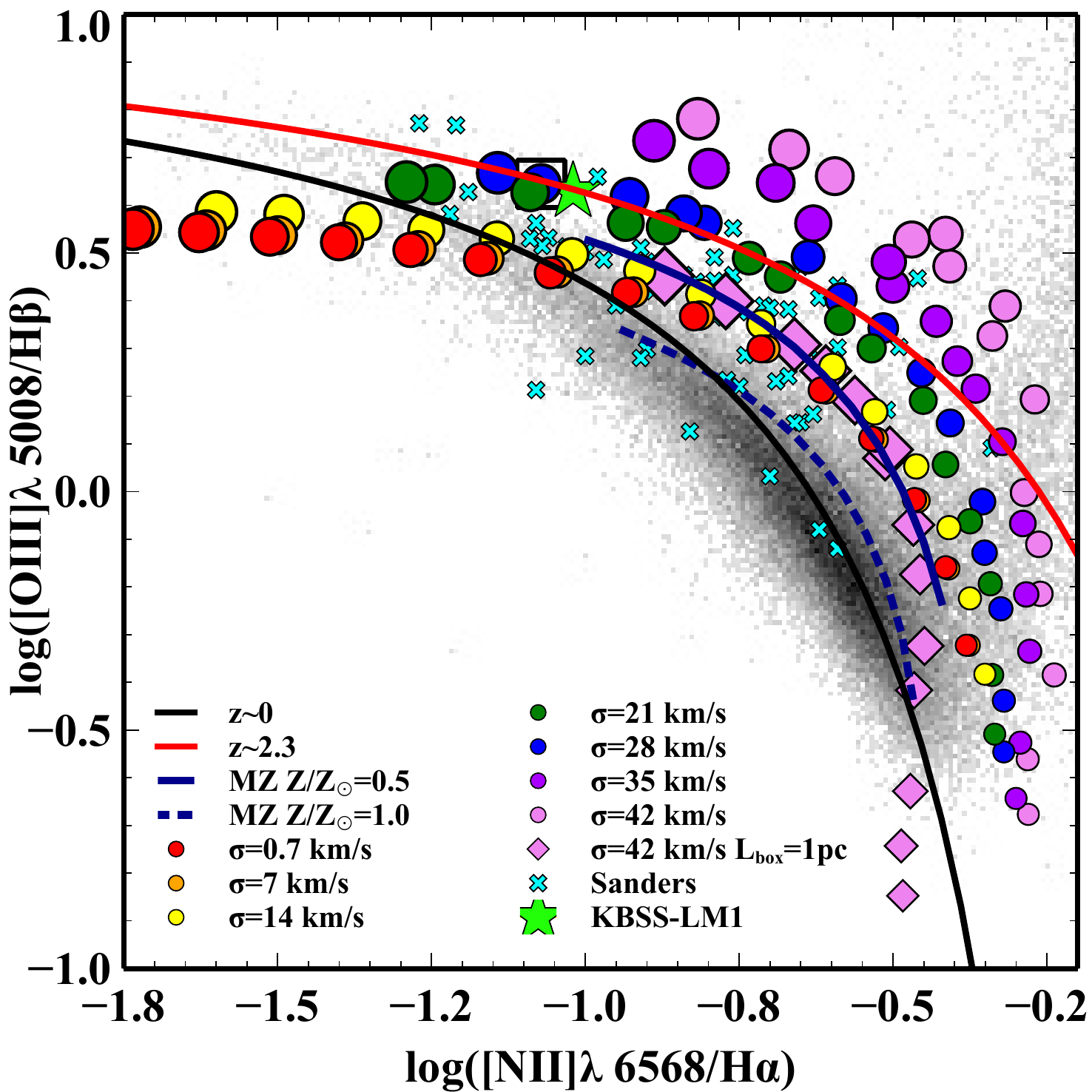}
\includegraphics[width=7.5 cm]{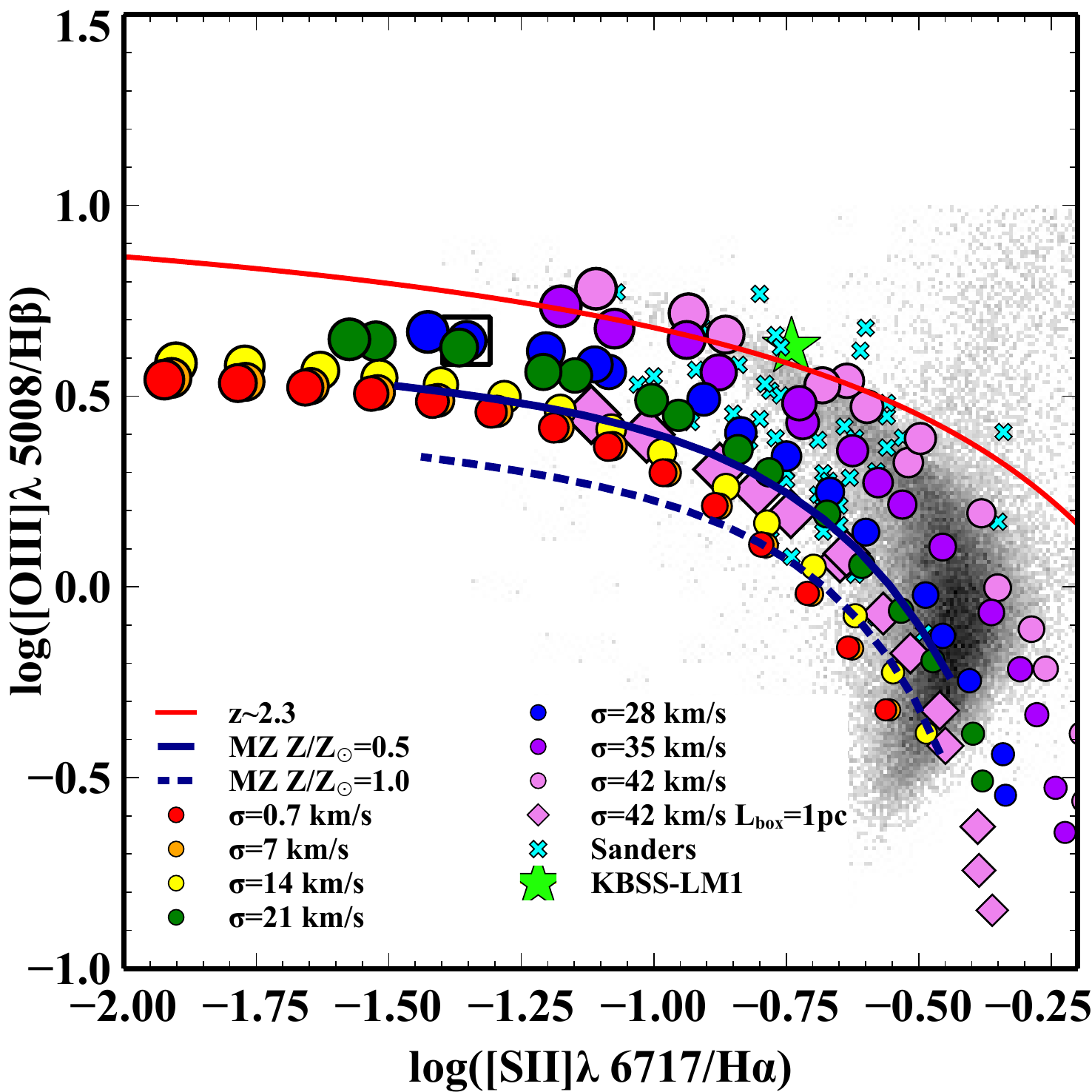}
\caption{The effects of turbulence on the classic [N~II] and [S~II] BPT diagrams. These figures are as presented at the workshop, see \cite{Gray2017} for more details. }\label{bpts}
\end{figure}

In addition, the treatment of shocks is particularly complex as it requires understanding the physical conditions in the emission line gas before, during, and after the shock event. These events can cause collisional ionization and the resulting recombinations may release additional ionizing photons that interact with surrounding material, depending on the shock velocity. A new database of shock models calculated using the MAPPINGS V\footnote{\url{https://miocene.anu.edu.au/mappings/}} code \cite{Sutherland2017} was presented at the workshop, and details can be found by consulting the documentation available on the Mexican Million Models database (3MdB) website\footnote{\url{https://sites.google.com/site/mexicanmillionmodels/}}.


\section{Galaxy-scale simulations}\label{sec:4}

\subsection{From cloud to galaxy scale simulations}\label{sec:41}

Cloud scale simulations starting at masses of $10^4$-$10^6$\,\msun have shown us how gas structures emerge and evolve for different inflow rates from converging streams of diffuse gas, corresponding to large-scale velocity dispersions in galaxy simulations \citep[e.g.] [] {BP+99, HP99, KI02, AH05, Heitsch+05, VS+07, Banerjee+09}. 
However, these simulations are mostly made at CNM initial conditions, i.e. in a limited temperature range. 
Projects dealing with line emission on galaxy scales would benefit greatly from seeing the impact on small-scale simulations from different temperatures reflecting different radiation backgrounds and metallicities. 
The community could also benefit from more ISM simulation work on scales in between cloud and galaxy sizes \citep[e.g.,] [] {Dobbs+12} and also observations of individual molecular clouds in external galaxies (e.g., PHANGS survey\footnote{\url{https://sites.google.com/view/phangs/home}} and the LMT effort\footnote{\url{http://www.lmtgtm.org/}}).

When simulating line emission from objects extracted from analytical/cosmological simulations, it is often a challenge to identify the minimum amount of information you need in order to get a decent estimate of the emission. For example, some simulations calculate abundances of specific elements on the fly, whereas others only contain one number for the metallicity and an abundance table must be adopted. Including several elements separately instead of a fixed abundance table, has been shown to make a significant difference for modeling of the fine-structure \cii line \cite{olsen17}.

Additionally, other possible uncertainties are present when considering galaxy simulations that include in  their modeling both chemical networks \citep[e.g.][]{pallottini:2017chem,capelo:2018} and radiative transfer \citep[e.g.][]{rosdahl:2015MNRAS,katz2017}; while such simulations can capture dynamical and thermodynamical effects due to photochemistry, they are not refined enough to calculate line emission, which is to be done in post-processing. However, in post-processing line calculations typically consider photoionization equilibrium and it is difficult to account for the past thermal and ionizational history of the simulated gas. 

One of the more valuable conclusions from the discussions on galaxy-scales simulations, was the importance of simulating more than one emission line simultaneously. By simulating different lines, arising in different ISM phases, and comparing with observations, one ensures that the post-process recipes not only satisfy what is seen in {\it one} ISM phase, but is consistent across the entire galaxy.

\subsection{Mapping simulated galaxy samples to observed samples}\label{sec:42}
An important step in simulating line emission from galaxies and from interstellar clouds is to compare with observed analogues of the model systems. Here, a number of issues continue to stand out, of which we discussed the following:

\begin{itemize}
\item[1] Observed SFRs come from tracers such as H$\alpha$, UV, IR and radio sensitive to a time averaged SFR of 10 to a few hundred Myr. Model SFRs are often the instantaneous SFRs. The question here is whether models should use time averaged SFRs or produce continuum and line emission to measure SFRs like observers do? One of the better solutions is to do both and identify any possible biases that are introduced by using observational methods.

\item[2] Model metallicities are often $M_{\rm metal}/M_{\rm gas|star}$, whereas observed metallicities are determined using a number of ionized emission lines and presented as e.g. $12+\log(\rm{O/H})$. As with SFR, the best option at the moment might be to directly calculate $12+\log(\rm{O/H})$ if the oxygen abundance is tracked by the simulation  used, although even this approach has problems as the observed $12+\log(\rm{O/H})$ will ultimately be luminosity-weighted which is hard to replicate in a simulation.

\item[3] A realistic comparison between modeled and observed UV/optical emission lines (and continuum) requires the correct treatment of dust absorption within simulations. The amount of intervening dust in galaxy scale simulations between the photon source and the observer is ideally calculated self-consistently using dust chemical networks  \citep[e.g.,][]{Grassi2017,Popping2017dust,McKinnon2017,Aoyama2018}. However, often the amount of intervening dust is scaled as a function of the gas-phase metallicity by assuming a fixed grain size distribution. Finally, the geometry of the galaxy and the relative star-to-dust location is critical in correctly determining the observed properties \citep[e.g.,][]{Behrens2018,Narayanan2018}, including sub-grid modeling of the dust properties of the birth-clouds of young stars.

\item[4] Atomic hydrogen emission at 21cm is out of reach of current instrumentation beyond at $z>0.3$ (SKA and SKA pathfinders such as APERTIF, MEERKAT, and ASKAP will be a remedy for this). This hampers the validation of the models that predict an evolution of atomic to molecular gas fraction with molecular gas dominating the gas budget within the optical scale of the galaxies at higher redshifts.

\item[5] Analyzing a large sample of strongly star-forming SDSS galaxies with nebular He\,{\sc ii} $\lambda$4686 emission has shown that the luminosity of this line can only be reproduced with single bursts of star formation of 20 percent solar metallicity or higher, and ages of 4-5 Myr, when the extreme UV continuum is dominated by Wolf-Rayet stars \cite{Shirazi2012}. He\,{\sc ii} $\lambda$1640 in the UV is 10 times stronger than the optical He\,{\sc ii} $\lambda$4686 line, and has been observed in broad (FWHM 1000 km s$^{-1}$) and narrow emission for tens of dwarf star-forming galaxies also selected from SDSS. Reproducing the luminosity of the narrow nebular He\,{\sc ii} $\lambda$1640 emission from these galaxies has been challenging, even with state-of-the-art spectral synthesis models which combine the newest Charlot \& Bruzual population synthesis models, which include very massive (300 $M_\odot$) stars, with photoionization models, as described in \cite{Gutkin2016}. This failure is reported for instance in \cite{Senchyna2017}. In this workshop, Aida Wofford presented the case of one of the most metal poor nearby galaxies known, SBS 0335-052E, which has a metallicity of solar/20. None of the models which they tried were able to reproduce the luminosity of the He\,{\sc ii} $\lambda$1640 line. This is a problem because this line will be one of the only diagnostics of massive stars in future observations with large telescopes such as JWST, TMT, and e-ELT, which will obtain rest-frame UV spectra for thousands of galaxies, at redshifts between 10 and 15, in the era of re-ionization, when the first stars and galaxies formed.

\item[6] How do AGN affect the comparison between observed and simulated galaxies? Comparing galaxies of similar SFR can become problematic, as the ionizing radiation and presence of a radio jet will enhance H$\alpha$ and radio emission that is typically used as a SFR diagnostic. 
In addition, the radiation will heat dust that is often used in mass estimations. See the following section for more on the effect of AGN and mass outflows. 
Although not directly discussed at this workshop, X-ray dominated regions (XDRs) are another important result of having an AGN present, and they must be modeled in order to reproduce certain emission lines. 
For example, this is extremely relevant for high-J CO lines, and it is in fact still not clear if high-J CO lines are influenced more by the presence of X-ray Photons or shocks \cite{Gallerani2014}. Important theoretical work on modeling XDRs was published in 2005 \cite{Meijerink2005} and \cloudy has also been used to model XDRs \cite[e.g.]{Mingozzi2018}.

\end{itemize}

\subsection{The effects of AGN and mass outflows on line emission}\label{sec_agn_effect}\label{sec:43}

The presence of an AGN can profoundly affect every aspect of observed emission line profiles including their velocities, widths, symmetry, and relative flux ratios. The AGN emits radiation anisotropically, and under the proper physical conditions can effectively couple with the ISM. This gives rise to outflows \citep{crenshaw2003} of molecular and ionized gas that are likely radiatively driven. Understanding these outflows is critical as they may deliver feedback to the galaxy by clearing the inner regions of star forming gas, and assist in establishing the observed scaling relationships between galaxies and their supermassive black holes (SMBHs) \citep{hopkins2005, kormendy2013, batiste2017, fiore2017}.

\begin{figure}[!t]
  \includegraphics[width=\textwidth]{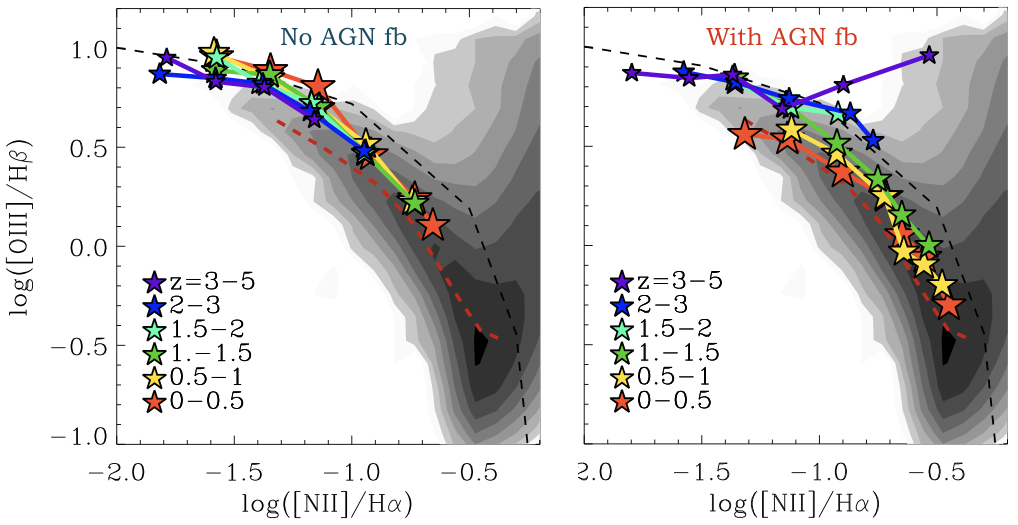}
  \caption{Average [OIII]/H$\beta$ emission-line ratio in bins of [NII]/H$\alpha$ for the star-forming subset (i.e. with log (BHAR/SFR) $< -4$) of the massive galaxies simulated with (right panel) and without AGN feedback (left panel) in different redshift ranges (connected stars of different colors). To guide the eye, SDSS data are shown in grey (adapted from \citep{hirschman2017}).}\label{f:agn_hirschmann}
\end{figure}

At the workshop a subset of topics were covered, highlighting the physical mechanisms that couple the ISM to the AGN radiation field and affect the observed emission line properties.
As an example, recent results from post-processing high-resolution, cosmological zoom-in simulations of massive galaxies with nebular emission models of galaxies and AGN were presented at the workshop \cite{hirschman2017,Choi2016}. One of the main results is that at least for massive galaxies, AGN-driven outflows seem to be a necessary component in order to reproduce the observed cosmic evolution of [OIII]/H$\beta$  at fixed [NII]/H$\alpha$ via strongly regulating the star formation history, which controls nebular emission from young stars via the ionization parameter (see Fig. \ref{f:agn_hirschmann}).

There was also a particular focus on what can be gleaned from spatially resolved observations and models. In the local universe, spatially resolved BPT diagrams have proven their value in understanding how the physical conditions of the emitting gas change with distance from the SMBH, and can be used to constrain \cloudy photoionization models of ionized mass outflows \citep{Revalski2018}. A first theoretical modeling of such spatially resolved BPT diagrams, using cosmological zoom-in simulations of massive galaxies shows that central low ionization emission in massive galaxies is not necessarily caused by an AGN, but can be also due to the ionizing radiation coming from post-AGB stars (Hirschmann et al., in prep.). For some galaxies, direct signatures for AGN-driven outflows are also visible in nebular emission line maps. These studies have primarily focused on the {\it optical} nebular emission line gas; however, observations reveal that powerful outflows of molecular gas may equal or even dominate the feedback in some galaxies. A conundrum in this area has been understanding how molecular outflows remain stable over time, as the physical conditions in the gas suggest many molecules should be destroyed. New modeling with advanced chemistry networks has found that molecules can form in-situ within the outflow, providing a significant opacity source that helps to drive the molecular outflows to the observed velocities \citep{richings2018}. Additional work in this area is needed to fully understand how the presence of an AGN modifies the standard emission line diagnostics for determining properties such as dust temperature, metallicity, and the inferred star formation rates.


\begin{figure}[!t]
  \includegraphics[width=\textwidth]{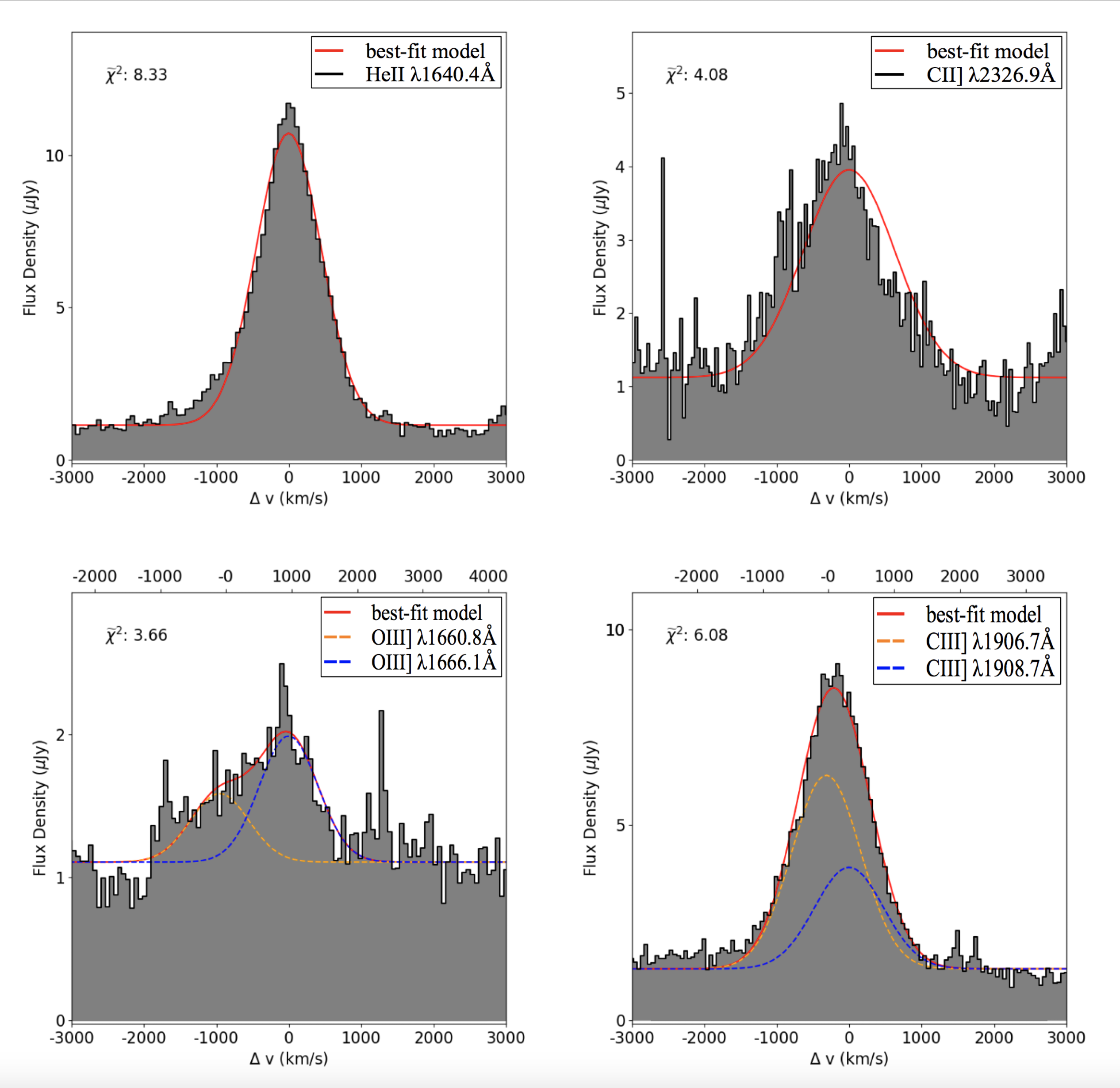}
  \caption{From Kolwa et al. in prep. Line emission from gas in the central 0.6 arcsec (4.75 kpc) of MRC 0943-242. This gas is ionized by the local AGN radiation field. Line widths larger than 1000 km s$^{-1}$ suggest a turbulent medium. For the top two figures the red profiles are best-fit single Gaussians. For the bottom two figures the doublets are fit with identical redshift. For the bottom two figures, the velocity scale of the bottom (top) horizontal axis is set to zero at the blue (orange) Gaussian centroid.}\label{f:agn_kolwa}
\end{figure}

\begin{figure}[!t]
  \includegraphics[width=0.49\textwidth]{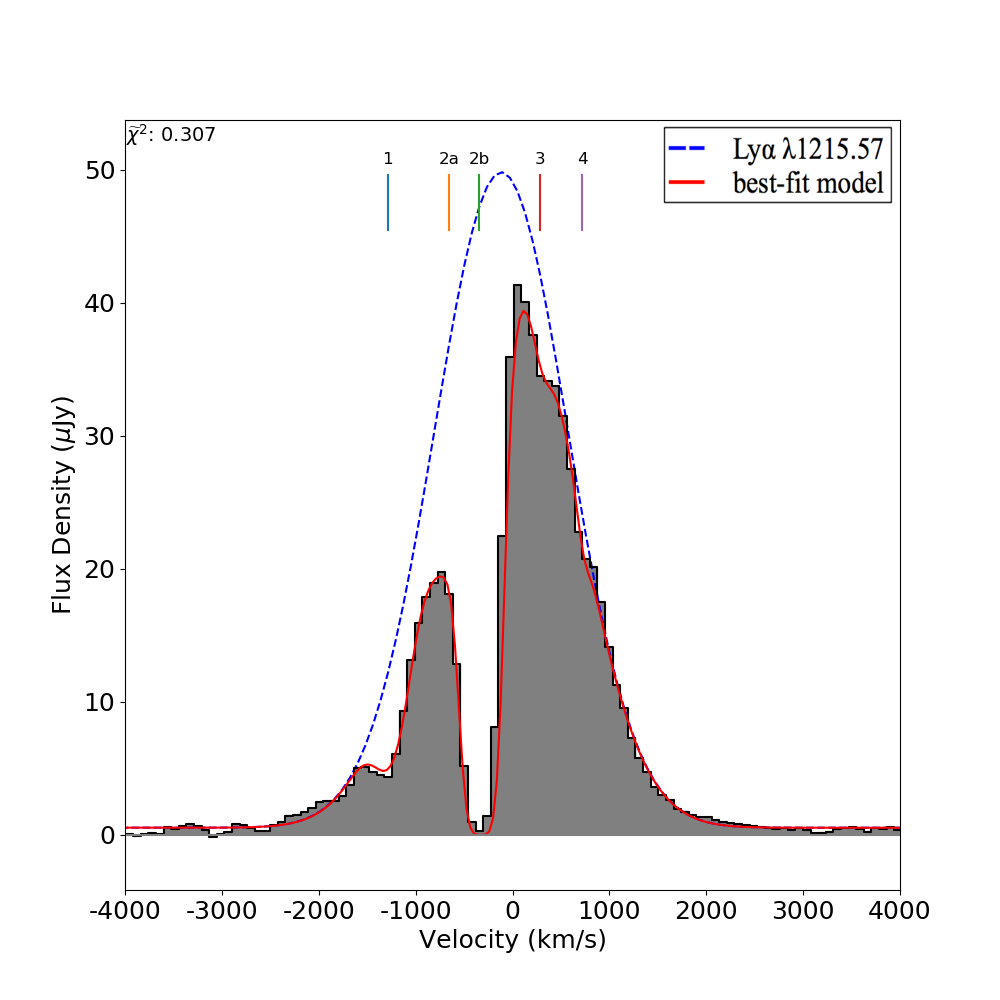}
  \includegraphics[width=0.49\textwidth]{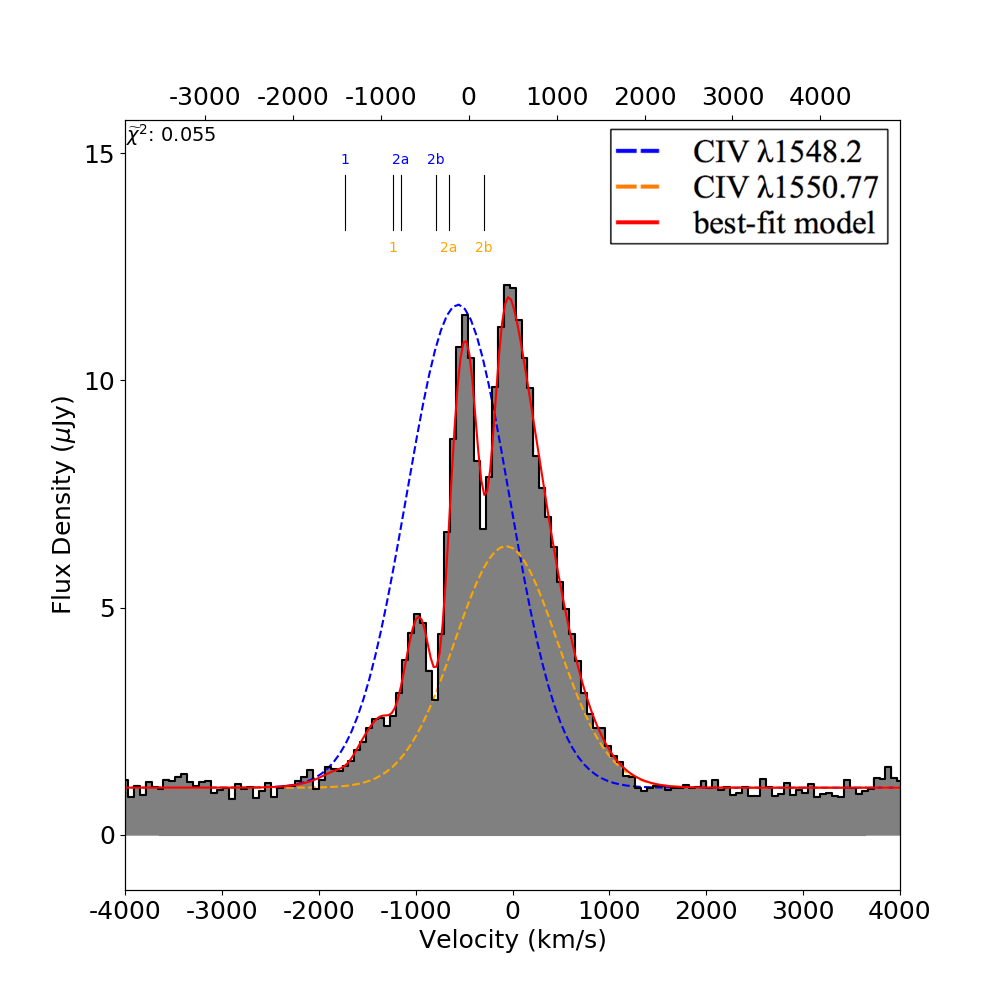}
  \caption{From Kolwa et al. in prep. Line emission in the same spectra as Fig.\, \ref{f:agn_kolwa}, however with clear components of absorption as well. The absorbing components' centroids are marked with vertical lines. Absorbing components 1, 2a, and 2b are seen in both the Ly$\alpha$ (left) and CIV (right) emission. The best fit models (red line) result from the Gaussian emission with multiple Gaussian absorption profiles. For the right figure, the velocity scale of the bottom (top) horizontal axis is set to zero at the blue (orange) Gaussian centroid, and each centroid has its absorption component centroids marked with corresponding font color.}\label{f:agn_kolwa2}
\end{figure}

A special case combining the effect of AGN, going from the host galaxy to the surrounding CGM, is the high-redshift radio galaxy MRC 0943-242. S. Kolwa presented nuclear emission and CGM absorption observations of MRC 0943-242 (see Fig.\,\ref{f:agn_kolwa}), whose emission is spatially integrated over a 0.6 arcsec (4.75 kpc) radius aperture centered at what is assumed to be an AGN. Hence, the lines  reveal gas in the ISM in close proximity to the nucleus of the galaxy and ionized by the AGN radiation. The emitting gas is turbulent with a velocity width larger than 1000 km s$^{-1}$. Models and simulations are important in understanding the physical mechanisms behind the turbulence in the ionized gas caused by processes surrounding the AGN. Further, Fig.\,\ref{f:agn_kolwa2} shows absorption from the gas of various column densities in the CGM. The kinematics of this gas will be studied soon with IFU observations. What impact the AGN has on heating or removing the galaxy gas may be uncovered by combining observations of the ISM emission excited by the AGN with emission from CGM absorbers possibly resulting from AGN outflows. Further, these observations may discriminate between different models of the origin and fate of CGM gas.

\subsection{Simulating line absorption from the CGM}\label{sec:44}

While the workshop focus was not specifically on absorption line profiles, or the CGM, several talks and break-out discussions emphasized the connection between modeling emission lines and absorption lines, and the value in collaborations between these two fields. In particular, the CGM can give further constraints on the nature of the galaxy that could be important for converting emission line observations to other characteristics of the galaxy, such as gas mass. 

\begin{figure}[H]
\centering
\includegraphics[width=0.5\columnwidth]{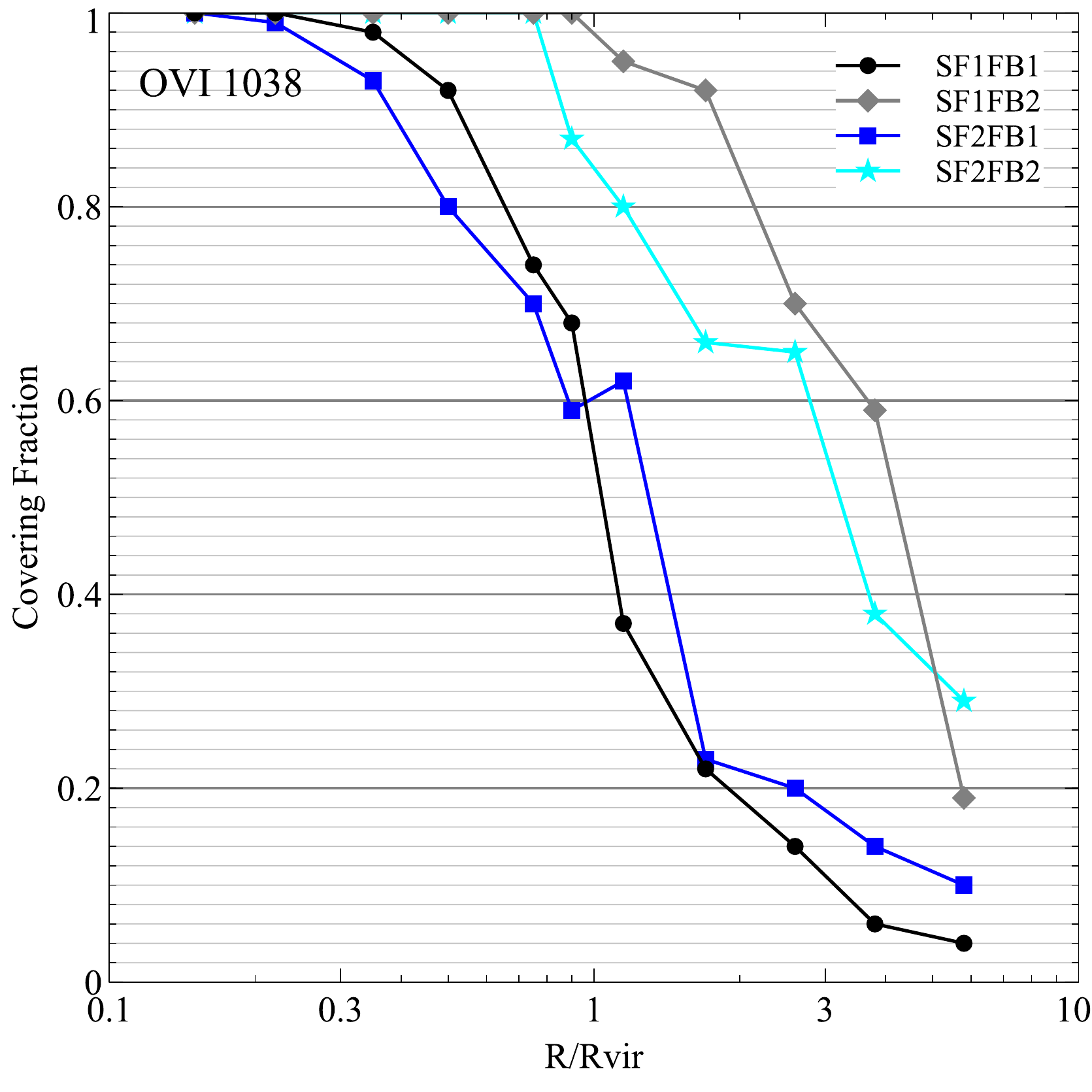}
\caption{CGM OVI covering fractions in a sub-L$*$ simulated galaxy at z=3 for different subgrid models. {\tt SF\#FB\#} corresponds to whether the simulations uses star formation model 1 or 2 ({\tt SF1} and {\tt SF2}), and stellar feedback model 1 or 2 ({\tt FB1} and {\tt FB2}). {\tt SF1} uses a density threshold for triggering star formation. {\tt FB1} corresponds to feedback from a stellar population with a Salpeter IMF and the massive stars exploding with 10$^{51}$ erg per 10 M$_{\odot}$ and the energy injected in the simulation matching the Sedov solution \citep{Dubois2008}. Both of these subgrid models depend on the resolution of the simulations (here 10 pc). {\tt SF2} uses a variable star formation efficiency for gas that is not supported against collapse by turbulence, following \citep{Padoan2011, Federrath2012}. {\tt FB2} is consistent with a Chabrier IMF, and either the energy or momentum solution is injected depending on which is resolved by the simulations \citep{Kimm2014}.}\label{f:cgm_ovi}
\end{figure}

Simulations of the CGM and its corresponding absorption line profiles are progressing, even given the difficulty of resolving this medium. Observations of distant quasars with absorption from the CGM of foreground galaxies provide useful metrics to help improve simulations. The characteristics of these observations are described well in \cite{tumlinson2017} and references therein. In particular, observations find an abundance of OVI in the CGM of galaxies of a range of masses, only well matched in simulations at high galaxy masses above 10$^{10.5}$ M$_{\odot}$ \citep{Suresh2017}. Other simulations post-process their outputs with \trident \citep{hummels2017} to produce mock absorption line profiles and compare them with quasar absorption line observations. The predicted CGM column densities and covering fractions of ions such as OVI depend on the models of star formation and stellar feedback (see Fig.\,\ref{f:cgm_ovi}; Chuniaud et al. in prep). Star formation models influence the amount of metals produced in the simulations, while feedback models impact their ability to leave the galaxy and persist in the CGM. 

Observations also find that these OVI lines typically have suprathermal line widths \citep{Werk2016}. However, this could be due to blending of multiple components along the line of sight. Simulations could reveal the nature of these broad components, but this may depend on resolution. The dependence of the simulated CGM on the numerical resolution is a debated topic. It is unclear what are the scales of the cool gas in the CGM, which could range from tens of pc down to sub-pc scale \citep[e.g.][]{McCourt2018}. Increasing the resolution in the CGM leads to better resolving the central density peaks of outflowing and inflowing material in the CGM. The initially slightly higher density has faster cooling rates, which is amplified over time. Thus, simulations with better resolution in the CGM, such as in NEPHTHYS presented by M. L. A. Richardson, form more gas clumps, mostly in inflowing gas. Inflows are enriched from outflows, leading to enhanced cooling. However even the high resolution CGM simulations shown by Richardson et al. (\nephthys) which have 100 physical pc resolution, may by numerically supported. Observed clumps in their simulations are roughly 10 - 20 cells across. 

Many models of the CGM assume the ionizing radiation source to be the background UV reionization field, but the central galaxy can dominate the background over periods of time, including during AGN flaring or starburst episodes \citep[e.g.][]{Suresh2017,Oppenheimer2018}. Non-equilibrium treatment of the ionization state could make this effect be longer-lasting. Future work, including the production of look-up tables to convert simulation data to predicted ionization states, will need to consider this.

Finally, the CGM will be better constrained in the future with the ELT and other next-generation observatories that will be able to use Lyman-break galaxies, far more numerous than quasars, to have multiple line-of-sight absorption profiles per foreground galaxy. This will allow for spatially resolved study of the CGM, and give more instruction for combining different foreground galaxy column density to make general CGM column density profiles.


\section{Discussion: How can we as a community move forward?}\label{discussion}

This workshop was the first of its kind (to our knowledge) to attempt to create a community of astronomers simulating line emission and working on the comparison with observations. In order to continue the sharing of knowledge within the community, we discussed ways in which to make the material of the workshop available to a larger group and encourage collaborations in the future. 
For this workshop, it was decided to use zenodo.org where all speakers had the option to upload their slides together with a recording of their talk (where available) to a community space\footnote{\url{https://zenodo.org/communities/walk2018/}}. This approach automatically provides a DOI for each upload for easy reference and we hope that future workshops of this kind will likewise attempt to make the presentations publicly available.

It should also be mentioned that the specific problems of simulating line emission from plasma are not only faced by the astronomical community. Within the field of plasma physics, collisional-radiative models are used to analyze plasma kinetics and spectra as well as to design plasma experiments in the laboratory. Yet, the two fields, i.e. that of astronomy and that of plasma physics, are currently far from fully benefiting from each other. We therefore wish to highlight corresponding established workshop series for non-LTE modeling, namely the "NLTE Code Comparison" workshops\footnote{\url{https://nlte.nist.gov/}} and the "Spectral Line Shapes in Plasmas" workshops\footnote{\url{http://plasma-gate.weizmann.ac.il/slsp/}}.

\subsection{Getting involved with the community of developers}
As Table\,\ref{table:1} shows, there are several software tools available for simulating line emission, each with their own group of developers behind. However, in order to achieve the best usage of these tools and adapting them to astrophysically relevant scenarios, it is crucial that users are able to connect with the developers to request support and updates, or even to adjust the underlying code themselves. Here we detail how this is done for two examples listed in Table\,\ref{table:1}.

\subsubsection{The \cloudy community}
Small tutorials on using \cloudy have been uploaded to youtube by individuals not part of the \cloudy developer team and such video uploads are much appreciated by many and strongly encouraged by the \cloudy community. It was suggested to establish a \cloudy-branded youtube channel or youtube playlist for these videos.
Also, many people are not aware that you can edit the databases used by \cloudy to for example include new species.
The \cloudy quickstart guide has been noted to be an excellent tool for introducing new researchers and students to making predictions with the software. Maintaining it and keeping it up-to-date will be a priority.
The \cloudy team is always looking for more participation on the Yahoo Groups forum. A goal is that questions and discussions will take place there among users, rather than solely between a user and cloudy developers. An interactive wiki has also been suggested as a place for users to post about specific cases or issues in their \cloudy research.

\subsubsection{LIME}
\lime resides on the code hosting platform GitHub (\url{https://github.com/lime-rt/lime}) from where it can be downloaded. A small team of active developers maintain and develop the package and minor releases appear regularly, approximately once per year. Users that experience problems when using LIME or have ideas to improve on the code, are welcome to create issues on GitHub where they will be answered by one or more developers. It is also possible to fork the code and develop it for specific needs. Documentation exists in the form of a written PDF manual as well as an issue tracker. Both are available on GitHub. A bi-annually summer school on Monte Carlo techniques in radiative transfer problems, aimed at PhD-students, is held at University of St Andrews School of Physics \& Astronomy and this school offers both a lecture and tutorials on how to use \lime.  


\section{Conclusions}\label{con}
This paper reviews recent progress and identifies current problems in the field of simulating line emission from the ISM and CGM. 
Conclusions presented here are a result of discussion sessions at the workshop “Walking the Line” held in Phoenix Arizona on March 14-16 2018. 
This international meeting was the first of its kind to gather astronomers specializing in the modeling of line emission and brought together 30 scientists from grad student to professor level. 
Most talks with follow-up questioning sessions have been made available online, but as a more digestible release of knowledge, we have condensed the outcome on the most relevant topics in this paper. 
Returning to the questions posed in the introduction, below are the main conclusions, focusing on the methods used to give the most reliable answers to each question. 

{\it What are the best emission lines to trace various ISM properties and ionized sources in galaxies?} The best method to answer this question of course depends on the type and scale of the emitting source in the ISM considered. In order to provide the reader with an overview, though not complete, of tools used to calculate line emission either on-the-fly in a simulation or in post process, we listed the software tools that were discussed at the workshop. One outstanding issue that can affect these tools, mostly when considering line emission in the X-ray but also at longer wavelengths \cite{delzanna18}, is to correct theoretical heavy-element energy levels that do not agree with experimental values. 

Key to deriving the line emission from any cloud is the UV field irradiating that cloud. However, our knowledge on the ionizing continuum of stars is severely hampered by the observational difficulties and thus currently restricted to theoretical predictions. Direct observations of individual, massive, low-metallicity stars will only be possible with telescopes such as LUVOIR and HabEx.
Once the UV field is known, radiative transfer of UV into infrared light is a crucial part of any method used, especially when studying neutral or molecular regions. While still too computationally expensive to run alongside galaxy-scale simulations, work is ongoing to include it more regularly in post-process as part of any line emission simulation. 

In order to compare simulated galaxies with observed ones, we need to put more thought into the choice of physical properties being compared. For example, a SFR that is modeled as instantaneous can hardly be compared to an observed SFR that is essentially derived as an average over 10 to a few-hundred Myr. Many other parameters such as the effects of an AGN or the inclination of a galaxy can distort an attempt to compare model galaxies with observed ones. Finally, it was concluded that the aim should always be to simulate more than one emission line simultaneously, in order to assure consistency across an the entire galaxy.

{\it How can we use emission lines to trace feedback and ISM evolution with redshift?} Feedback from AGN in the form of radiation and outflows has a profound effect on the nebular line emission. As an example, simulations of massive galaxies with and without AGN were presented in which the presence of an AGN could clearly be identified from the [OIII]/H$\beta$ and [NII]/H$\alpha$ emission-line ratios. AGN-driven outflows thus strongly regulates the star formation history, at least for massive galaxies, which controls nebular emission from young stars via the ionization parameter. As an observational example, the high-redshift galaxy MRC 0943-242 is believed to contain an AGN, with line emission (and absorption) revealing ionized ISM in close proximity to the nucleus as well as outflows and high levels of turbulence.

In close relation to stellar and AGN feedback, turbulence and shocks are another important source of heating and ionization. Recently developed tools for the treatment of turbulence and shocks were presented to aid the line emission simulations in cases where necessitated.

On cloud-scale level, simulating the line emission from collapsing dense cores can reveal whether the molecular cloud is undergoing an ``inside-out’’ or ``outside-in’’ collapse. Work is undergoing to break this degeneracy by making synthetic observations of signature spectral lines for comparison with observations.

{\it How should absorption features be correctly interpreted?} Simulations of line absorption through the CGM are faced by its own set of problem, where for example the treatment of the ionizing field from the galaxy source can play a crucial role but has yet to be fully treated in a non-equilibrium manner. Another choice to be made when simulating the CGM is the level of spatial resolution needed, as better resolution leads generally to more gas clumps and a point of convergence must be found.

{\it Where do we stand in deriving sub-grid physics and comparing codes?} When studying ISM evolution with redshift, line emission must be derived from cosmological simulations which capture the evolution but comes at low spatial resolution compared to semi-analytical models. In addition, full chemistry and radiative transfer must typically be applied in post process. The uncertainties in the approach could be alleviated by more work on scales between galaxies and clouds, because simulations of the latter come at higher resolution and could hence be used as a benchmark on larger scales. 

{\it How do we coordinate our efforts?} We hope to have set an example for more theoretical workshops of this kind to come, and look forward to seeing the results of the ongoing projects presented here. 
The developers of the software tools used to derive line emission, increasingly encourage their users to report problems or suggest improvements to their code via online platforms. 
With the list of caveats identified in this workshop, albeit long, we continue to believe that the future of simulating line emission is bright, especially when research groups in the field come together to find and solve common problems.


\acknowledgments{The authors thank the two anonymous referees for their thoughtful comments and suggestions. The authors thank all participants at the Walking the Line 2018 conference for presenting their work and stimulating discussions. The authors also thank Mark Krumholz, Livia Vallini, Christian Brinch, Robert Loughnane, Peter van Hoof, Moupiya Maji, William Gray, and Sthabile Kolwa for providing additional info and comments to this paper. The workshop was sponsored in part by the School of Earth and Space Exploration and prof. Rogier Windhorst at Arizona State University.}

\authorcontributions{The conference was led by K.O. who planned it with nine other members on the science organizing committee, including E.V.-S.; K.O. contributed to sections \ref{intro}, \ref{sec:21}, \ref{discussion} and \ref{con}; A.P. contributed to the sections \ref{intro}, \ref{sec:41} and \ref{sec:42}; A.W. contributed to sections \ref{sec:32} and \ref{sec:33}; M.C. contributed to sections \ref{sec:23} and \ref{discussion}; M.R. contributed to sections \ref{sec:34}, \ref{sec:42} and \ref{sec:43}; F.G. contributed to sections \ref{sec:22} and \ref{discussion}; G.P. contributed to the introduction and section \ref{sec:42}; E.V.-S. contributed to section \ref{sec:31}; G.E.M. contributed to the introduction; M.L.A.R. contributed to section \ref{sec:4}; M.H. contributed to section \ref{sec:43}, and W.G. contributed to section \ref{sec:44}.}

\conflictsofinterest{The authors declare no conflict of interest.}

\reftitle{References}

 \externalbibliography{yes}
 \bibliography{bibs}

\begin{thebibliography}{-------}
\providecommand{\natexlab}[1]{#1}

\bibitem[{Baldwin} \em{et~al.}(1981){Baldwin}, {Phillips}, and
  {Terlevich}]{Baldwin1981}
{Baldwin}, J.A.; {Phillips}, M.M.; {Terlevich}, R.
\newblock {Classification parameters for the emission-line spectra of
  extragalactic objects}.
\newblock {\em \pasp} {\bf 1981}, {\em 93},~5--19.
\newblock
  doi:{\changeurlcolor{black}\href{https://doi.org/10.1086/130766}{\detokenize{10.1086/130766}}}.

\bibitem[{Kewley} \em{et~al.}(2001){Kewley}, {Dopita}, {Sutherland}, {Heisler},
  and {Trevena}]{Kewley2001}
{Kewley}, L.J.; {Dopita}, M.A.; {Sutherland}, R.S.; {Heisler}, C.A.; {Trevena},
  J.
\newblock {Theoretical Modeling of Starburst Galaxies}.
\newblock {\em \apj} {\bf 2001}, {\em 556},~121--140,
  \href{http://xxx.lanl.gov/abs/astro-ph/0106324}{{\normalfont
  [astro-ph/0106324]}}.
\newblock
  doi:{\changeurlcolor{black}\href{https://doi.org/10.1086/321545}{\detokenize{10.1086/321545}}}.

\bibitem[{Kauffmann} \em{et~al.}(2003){Kauffmann}, {Heckman}, {Tremonti},
  {Brinchmann}, {Charlot}, {White}, {Ridgway}, {Brinkmann}, {Fukugita}, {Hall},
  {Ivezi{\'c}}, {Richards}, and {Schneider}]{Kauffmann2003}
{Kauffmann}, G.; {Heckman}, T.M.; {Tremonti}, C.; {Brinchmann}, J.; {Charlot},
  S.; {White}, S.D.M.; {Ridgway}, S.E.; {Brinkmann}, J.; {Fukugita}, M.;
  {Hall}, P.B.; {Ivezi{\'c}}, {\v Z}.; {Richards}, G.T.; {Schneider}, D.P.
\newblock {The host galaxies of active galactic nuclei}.
\newblock {\em \mnras} {\bf 2003}, {\em 346},~1055--1077,
  \href{http://xxx.lanl.gov/abs/astro-ph/0304239}{{\normalfont
  [astro-ph/0304239]}}.
\newblock
  doi:{\changeurlcolor{black}\href{https://doi.org/10.1111/j.1365-2966.2003.07154.x}{\detokenize{10.1111/j.1365-2966.2003.07154.x}}}.

\bibitem[{Tielens} and {Hollenbach}(1985)]{Tielens1985}
{Tielens}, A.G.G.M.; {Hollenbach}, D.
\newblock {Photodissociation regions. I - Basic model. II - A model for the
  Orion photodissociation region}.
\newblock {\em \apj} {\bf 1985}, {\em 291},~722--754.
\newblock
  doi:{\changeurlcolor{black}\href{https://doi.org/10.1086/163111}{\detokenize{10.1086/163111}}}.

\bibitem[{van Dishoeck} and {Black}(1986)]{vanDishoeck1986}
{van Dishoeck}, E.F.; {Black}, J.H.
\newblock {Comprehensive models of diffuse interstellar clouds - Physical
  conditions and molecular abundances}.
\newblock {\em \apjs} {\bf 1986}, {\em 62},~109--145.
\newblock
  doi:{\changeurlcolor{black}\href{https://doi.org/10.1086/191135}{\detokenize{10.1086/191135}}}.

\bibitem[{van Dishoeck} and {Black}(1988{\natexlab{a}})]{vanDishoeck1988a}
{van Dishoeck}, E.F.; {Black}, J.H.
\newblock {The Photodissociation of Interstellar Co/}.
\newblock  Molecular Clouds, Milky-Way and External Galaxies; {Dickman}, R.L.;
  {Snell}, R.L.; {Young}, J.S., Eds.,  1988, Vol. 315, {\em Lecture Notes in
  Physics, Berlin Springer Verlag}, p. 168.
\newblock
  doi:{\changeurlcolor{black}\href{https://doi.org/10.1007/3-540-50438-9_255}{\detokenize{10.1007/3-540-50438-9_255}}}.

\bibitem[{van Dishoeck} and {Black}(1988{\natexlab{b}})]{vanDishoeck1988b}
{van Dishoeck}, E.F.; {Black}, J.H.
\newblock {The photodissociation and chemistry of interstellar CO}.
\newblock {\em \apj} {\bf 1988}, {\em 334},~771--802.
\newblock
  doi:{\changeurlcolor{black}\href{https://doi.org/10.1086/166877}{\detokenize{10.1086/166877}}}.

\bibitem[{Sternberg} and {Dalgarno}(1989)]{Sternberg1989}
{Sternberg}, A.; {Dalgarno}, A.
\newblock {The infrared response of molecular hydrogen gas to ultraviolet
  radiation - High-density regions}.
\newblock {\em \apj} {\bf 1989}, {\em 338},~197--233.
\newblock
  doi:{\changeurlcolor{black}\href{https://doi.org/10.1086/167193}{\detokenize{10.1086/167193}}}.

\bibitem[{Hollenbach} \em{et~al.}(1991){Hollenbach}, {Takahashi}, and
  {Tielens}]{Hollenbach1991}
{Hollenbach}, D.J.; {Takahashi}, T.; {Tielens}, A.G.G.M.
\newblock {Low-density photodissociation regions}.
\newblock {\em \apj} {\bf 1991}, {\em 377},~192--209.
\newblock
  doi:{\changeurlcolor{black}\href{https://doi.org/10.1086/170347}{\detokenize{10.1086/170347}}}.

\bibitem[{Kaufman} \em{et~al.}(1999){Kaufman}, {Wolfire}, {Hollenbach}, and
  {Luhman}]{Kaufman1999}
{Kaufman}, M.J.; {Wolfire}, M.G.; {Hollenbach}, D.J.; {Luhman}, M.L.
\newblock {Far-Infrared and Submillimeter Emission from Galactic and
  Extragalactic Photodissociation Regions}.
\newblock {\em \apj} {\bf 1999}, {\em 527},~795--813,
  \href{http://xxx.lanl.gov/abs/astro-ph/9907255}{{\normalfont
  [astro-ph/9907255]}}.
\newblock
  doi:{\changeurlcolor{black}\href{https://doi.org/10.1086/308102}{\detokenize{10.1086/308102}}}.

\bibitem[{Pound} and {Wolfire}(2008)]{Pound2008}
{Pound}, M.W.; {Wolfire}, M.G.
\newblock {The Photo Dissociation Region Toolbox}.
\newblock  Astronomical Data Analysis Software and Systems XVII; {Argyle},
  R.W.; {Bunclark}, P.S.; {Lewis}, J.R., Eds.,  2008, Vol. 394, {\em
  Astronomical Society of the Pacific Conference Series}, p. 654.

\bibitem[{Kaufman} \em{et~al.}(2006){Kaufman}, {Wolfire}, and
  {Hollenbach}]{kaufman2006}
{Kaufman}, M.J.; {Wolfire}, M.G.; {Hollenbach}, D.J.
\newblock {[Si II], [Fe II], [C II], and H$_{2}$ Emission from Massive
  Star-forming Regions}.
\newblock {\em \apj} {\bf 2006}, {\em 644},~283--299.
\newblock
  doi:{\changeurlcolor{black}\href{https://doi.org/10.1086/503596}{\detokenize{10.1086/503596}}}.

\bibitem[{Ferland} \em{et~al.}(2017){Ferland}, {Chatzikos}, {Guzm{\'a}n},
  {Lykins}, {van Hoof}, {Williams}, {Abel}, {Badnell}, {Keenan}, {Porter}, and
  {Stancil}]{Ferland2017}
{Ferland}, G.J.; {Chatzikos}, M.; {Guzm{\'a}n}, F.; {Lykins}, M.L.; {van Hoof},
  P.A.M.; {Williams}, R.J.R.; {Abel}, N.P.; {Badnell}, N.R.; {Keenan}, F.P.;
  {Porter}, R.L.; {Stancil}, P.C.
\newblock {The 2017 Release Cloudy}.
\newblock {\em \rmxaa} {\bf 2017}, {\em 53},~385--438,
  \href{http://xxx.lanl.gov/abs/1705.10877}{{\normalfont [1705.10877]}}.

\bibitem[{R{\"o}llig} \em{et~al.}(2007){R{\"o}llig}, {Abel}, {Bell}, {Bensch},
  {Black}, {Ferland}, {Jonkheid}, {Kamp}, {Kaufman}, {Le Bourlot}, {Le Petit},
  {Meijerink}, {Morata}, {Ossenkopf}, {Roueff}, {Shaw}, {Spaans}, {Sternberg},
  {Stutzki}, {Thi}, {van Dishoeck}, {van Hoof}, {Viti}, and
  {Wolfire}]{Rollig2007}
{R{\"o}llig}, M.; {Abel}, N.P.; {Bell}, T.; {Bensch}, F.; {Black}, J.;
  {Ferland}, G.J.; {Jonkheid}, B.; {Kamp}, I.; {Kaufman}, M.J.; {Le Bourlot},
  J.; {Le Petit}, F.; {Meijerink}, R.; {Morata}, O.; {Ossenkopf}, V.; {Roueff},
  E.; {Shaw}, G.; {Spaans}, M.; {Sternberg}, A.; {Stutzki}, J.; {Thi}, W.F.;
  {van Dishoeck}, E.F.; {van Hoof}, P.A.M.; {Viti}, S.; {Wolfire}, M.G.
\newblock {A photon dominated region code comparison study}.
\newblock {\em \aap} {\bf 2007}, {\em 467},~187--206,
  \href{http://xxx.lanl.gov/abs/astro-ph/0702231}{{\normalfont
  [astro-ph/0702231]}}.
\newblock
  doi:{\changeurlcolor{black}\href{https://doi.org/10.1051/0004-6361:20065918}{\detokenize{10.1051/0004-6361:20065918}}}.

\bibitem[{Bolatto} \em{et~al.}(1999){Bolatto}, {Jackson}, and
  {Ingalls}]{Bolatto1999}
{Bolatto}, A.D.; {Jackson}, J.M.; {Ingalls}, J.G.
\newblock {A Semianalytical Model for the Observational Properties of the
  Dominant Carbon Species at Different Metallicities}.
\newblock {\em \apj} {\bf 1999}, {\em 513},~275--286,
  \href{http://xxx.lanl.gov/abs/astro-ph/9812181}{{\normalfont
  [astro-ph/9812181]}}.
\newblock
  doi:{\changeurlcolor{black}\href{https://doi.org/10.1086/306849}{\detokenize{10.1086/306849}}}.

\bibitem[{R{\"o}llig} \em{et~al.}(2006){R{\"o}llig}, {Ossenkopf}, {Jeyakumar},
  {Stutzki}, and {Sternberg}]{Rollig2006}
{R{\"o}llig}, M.; {Ossenkopf}, V.; {Jeyakumar}, S.; {Stutzki}, J.; {Sternberg},
  A.
\newblock {[CII] 158 {$\mu$}m emission and metallicity in photon dominated
  regions}.
\newblock {\em \aap} {\bf 2006}, {\em 451},~917--924,
  \href{http://xxx.lanl.gov/abs/astro-ph/0601682}{{\normalfont
  [astro-ph/0601682]}}.
\newblock
  doi:{\changeurlcolor{black}\href{https://doi.org/10.1051/0004-6361:20053845}{\detokenize{10.1051/0004-6361:20053845}}}.

\bibitem[{Narayanan} \em{et~al.}(2006){Narayanan}, {Kulesa}, {Boss}, and
  {Walker}]{Narayanan2006}
{Narayanan}, D.; {Kulesa}, C.A.; {Boss}, A.; {Walker}, C.K.
\newblock {Molecular Line Emission from Gravitationally Unstable Protoplanetary
  Disks}.
\newblock {\em \apj} {\bf 2006}, {\em 647},~1426--1436,
  \href{http://xxx.lanl.gov/abs/astro-ph/0605329}{{\normalfont
  [astro-ph/0605329]}}.
\newblock
  doi:{\changeurlcolor{black}\href{https://doi.org/10.1086/505619}{\detokenize{10.1086/505619}}}.

\bibitem[{Popping} \em{et~al.}(2014){Popping}, {P{\'e}rez-Beaupuits}, {Spaans},
  {Trager}, and {Somerville}]{Popping2014}
{Popping}, G.; {P{\'e}rez-Beaupuits}, J.P.; {Spaans}, M.; {Trager}, S.C.;
  {Somerville}, R.S.
\newblock {The nature of the ISM in galaxies during the star-formation activity
  peak of the Universe}.
\newblock {\em \mnras} {\bf 2014}, {\em 444},~1301--1317,
  \href{http://xxx.lanl.gov/abs/1310.1476}{{\normalfont [1310.1476]}}.
\newblock
  doi:{\changeurlcolor{black}\href{https://doi.org/10.1093/mnras/stu1506}{\detokenize{10.1093/mnras/stu1506}}}.

\bibitem[{Vallini} \em{et~al.}(2015){Vallini}, {Gallerani}, {Ferrara},
  {Pallottini}, and {Yue}]{Vallini2015}
{Vallini}, L.; {Gallerani}, S.; {Ferrara}, A.; {Pallottini}, A.; {Yue}, B.
\newblock {On the [CII]-SFR Relation in High Redshift Galaxies}.
\newblock {\em \apj} {\bf 2015}, {\em 813},~36,
  \href{http://xxx.lanl.gov/abs/1507.00340}{{\normalfont [1507.00340]}}.
\newblock
  doi:{\changeurlcolor{black}\href{https://doi.org/10.1088/0004-637X/813/1/36}{\detokenize{10.1088/0004-637X/813/1/36}}}.

\bibitem[{Olsen} \em{et~al.}(2015){Olsen}, {Greve}, {Narayanan}, {Thompson},
  {Toft}, and {Brinch}]{Olsen2015}
{Olsen}, K.P.; {Greve}, T.R.; {Narayanan}, D.; {Thompson}, R.; {Toft}, S.;
  {Brinch}, C.
\newblock {Simulator of Galaxy Millimeter/Submillimeter Emission
  (S{\'{I}}GAME): The [C ii]-SFR Relationship of Massive z = 2 Main Sequence
  Galaxies}.
\newblock {\em \apj} {\bf 2015}, {\em 814},~76,
  \href{http://xxx.lanl.gov/abs/1507.00362}{{\normalfont [1507.00362]}}.
\newblock
  doi:{\changeurlcolor{black}\href{https://doi.org/10.1088/0004-637X/814/1/76}{\detokenize{10.1088/0004-637X/814/1/76}}}.

\bibitem[{Popping} \em{et~al.}(2016){Popping}, {van Kampen}, {Decarli},
  {Spaans}, {Somerville}, and {Trager}]{Popping2016}
{Popping}, G.; {van Kampen}, E.; {Decarli}, R.; {Spaans}, M.; {Somerville},
  R.S.; {Trager}, S.C.
\newblock {Sub-mm emission line deep fields: CO and [C II] luminosity functions
  out to z = 6}.
\newblock {\em \mnras} {\bf 2016}, {\em 461},~93--110,
  \href{http://xxx.lanl.gov/abs/1602.02761}{{\normalfont [1602.02761]}}.
\newblock
  doi:{\changeurlcolor{black}\href{https://doi.org/10.1093/mnras/stw1323}{\detokenize{10.1093/mnras/stw1323}}}.

\bibitem[{Popping} \em{et~al.}(2018){Popping}, {Narayanan}, {Somerville},
  {Faisst}, and {Krumholz}]{popping2018}
{Popping}, G.; {Narayanan}, D.; {Somerville}, R.S.; {Faisst}, A.L.; {Krumholz},
  M.R.
\newblock {The art of modeling CO, [CI], and [CII] in cosmological galaxy
  formation models}.
\newblock {\em ArXiv e-prints} {\bf 2018},
  \href{http://xxx.lanl.gov/abs/1805.11093}{{\normalfont [1805.11093]}}.

\bibitem[{Vallini} \em{et~al.}(2018){Vallini}, {Pallottini}, {Ferrara},
  {Gallerani}, {Sobacchi}, and {Behrens}]{vallini18}
{Vallini}, L.; {Pallottini}, A.; {Ferrara}, A.; {Gallerani}, S.; {Sobacchi},
  E.; {Behrens}, C.
\newblock {CO line emission from galaxies in the Epoch of Reionization}.
\newblock {\em \mnras} {\bf 2018}, {\em 473},~271--285,
  \href{http://xxx.lanl.gov/abs/1709.03993}{{\normalfont [1709.03993]}}.
\newblock
  doi:{\changeurlcolor{black}\href{https://doi.org/10.1093/mnras/stx2376}{\detokenize{10.1093/mnras/stx2376}}}.

\bibitem[{Rosdahl} \em{et~al.}(2018){Rosdahl}, {Katz}, {Blaizot}, {Kimm},
  {Michel-Dansac}, {Garel}, {Haehnelt}, {Ocvirk}, and {Teyssier}]{Rosdahl2018}
{Rosdahl}, J.; {Katz}, H.; {Blaizot}, J.; {Kimm}, T.; {Michel-Dansac}, L.;
  {Garel}, T.; {Haehnelt}, M.; {Ocvirk}, P.; {Teyssier}, R.
\newblock {The SPHINX Cosmological Simulations of the First Billion Years: the
  Impact of Binary Stars on Reionization}.
\newblock {\em ArXiv e-prints} {\bf 2018},
  \href{http://xxx.lanl.gov/abs/1801.07259}{{\normalfont [1801.07259]}}.

\bibitem[{Keto} \em{et~al.}(2004){Keto}, {Rybicki}, {Bergin}, and
  {Plume}]{Keto+04}
{Keto}, E.; {Rybicki}, G.B.; {Bergin}, E.A.; {Plume}, R.
\newblock {Radiative Transfer and Starless Cores}.
\newblock {\em \apj} {\bf 2004}, {\em 613},~355--373,
  \href{http://xxx.lanl.gov/abs/astro-ph/0407433}{{\normalfont
  [astro-ph/0407433]}}.
\newblock
  doi:{\changeurlcolor{black}\href{https://doi.org/10.1086/422987}{\detokenize{10.1086/422987}}}.

\bibitem[{Brinch} and {Hogerheijde}(2010)]{Brinch2010}
{Brinch}, C.; {Hogerheijde}, M.R.
\newblock {LIME - a flexible, non-LTE line excitation and radiation transfer
  method for millimeter and far-infrared wavelengths}.
\newblock {\em \aap} {\bf 2010}, {\em 523},~A25,
  \href{http://xxx.lanl.gov/abs/1008.1492}{{\normalfont
  [arXiv:astro-ph.SR/1008.1492]}}.
\newblock
  doi:{\changeurlcolor{black}\href{https://doi.org/10.1051/0004-6361/201015333}{\detokenize{10.1051/0004-6361/201015333}}}.

\bibitem[{Keto} and {Rybicki}(2010)]{Keto_Rybicky10}
{Keto}, E.; {Rybicki}, G.
\newblock {Modeling Molecular Hyperfine Line Emission}.
\newblock {\em \apj} {\bf 2010}, {\em 716},~1315--1322,
  \href{http://xxx.lanl.gov/abs/1004.1617}{{\normalfont [1004.1617]}}.
\newblock
  doi:{\changeurlcolor{black}\href{https://doi.org/10.1088/0004-637X/716/2/1315}{\detokenize{10.1088/0004-637X/716/2/1315}}}.

\bibitem[{Dullemond} \em{et~al.}(2012){Dullemond}, {Juhasz}, {Pohl},
  {Sereshti}, {Shetty}, {Peters}, {Commercon}, and {Flock}]{Dullemond2012}
{Dullemond}, C.P.; {Juhasz}, A.; {Pohl}, A.; {Sereshti}, F.; {Shetty}, R.;
  {Peters}, T.; {Commercon}, B.; {Flock}, M.
\newblock {RADMC-3D: A multi-purpose radiative transfer tool}.
\newblock Astrophysics Source Code Library,  2012,
  \href{http://xxx.lanl.gov/abs/1202.015}{{\normalfont [1202.015]}}.

\bibitem[{Yajima} \em{et~al.}(2012){Yajima}, {Li}, {Zhu}, and
  {Abel}]{Yajima2012}
{Yajima}, H.; {Li}, Y.; {Zhu}, Q.; {Abel}, T.
\newblock {ART$^{2}$: coupling Ly{$\alpha$} line and multi-wavelength continuum
  radiative transfer}.
\newblock {\em \mnras} {\bf 2012}, {\em 424},~884--901,
  \href{http://xxx.lanl.gov/abs/1109.4891}{{\normalfont [1109.4891]}}.
\newblock
  doi:{\changeurlcolor{black}\href{https://doi.org/10.1111/j.1365-2966.2012.21228.x}{\detokenize{10.1111/j.1365-2966.2012.21228.x}}}.

\bibitem[{Ferland} \em{et~al.}(2013){Ferland}, {Porter}, {van Hoof},
  {Williams}, {Abel}, {Lykins}, {Shaw}, {Henney}, and {Stancil}]{Ferland2013}
{Ferland}, G.J.; {Porter}, R.L.; {van Hoof}, P.A.M.; {Williams}, R.J.R.;
  {Abel}, N.P.; {Lykins}, M.L.; {Shaw}, G.; {Henney}, W.J.; {Stancil}, P.C.
\newblock {The 2013 Release of Cloudy}.
\newblock {\em \rmxaa} {\bf 2013}, {\em 49},~137--163,
  \href{http://xxx.lanl.gov/abs/1302.4485}{{\normalfont
  [arXiv:astro-ph.GA/1302.4485]}}.

\bibitem[{Krumholz}(2014)]{Krumholz2014}
{Krumholz}, M.R.
\newblock {DESPOTIC - a new software library to Derive the Energetics and
  SPectra of Optically Thick Interstellar Clouds}.
\newblock {\em \mnras} {\bf 2014}, {\em 437},~1662--1680,
  \href{http://xxx.lanl.gov/abs/1304.2404}{{\normalfont
  [arXiv:astro-ph.IM/1304.2404]}}.
\newblock
  doi:{\changeurlcolor{black}\href{https://doi.org/10.1093/mnras/stt2000}{\detokenize{10.1093/mnras/stt2000}}}.

\bibitem[{Gray} \em{et~al.}(2015){Gray}, {Scannapieco}, and {Kasen}]{Gray2015}
{Gray}, W.J.; {Scannapieco}, E.; {Kasen}, D.
\newblock {Atomic Chemistry in Turbulent Astrophysical Media. I. Effect of
  Atomic Cooling}.
\newblock {\em \apj} {\bf 2015}, {\em 801},~107,
  \href{http://xxx.lanl.gov/abs/1502.01019}{{\normalfont [1502.01019]}}.
\newblock
  doi:{\changeurlcolor{black}\href{https://doi.org/10.1088/0004-637X/801/2/107}{\detokenize{10.1088/0004-637X/801/2/107}}}.

\bibitem[{Kewley} \em{et~al.}(2013){Kewley}, {Dopita}, {Leitherer}, {Dav{\'e}},
  {Yuan}, {Allen}, {Groves}, and {Sutherland}]{Kewley2013}
{Kewley}, L.J.; {Dopita}, M.A.; {Leitherer}, C.; {Dav{\'e}}, R.; {Yuan}, T.;
  {Allen}, M.; {Groves}, B.; {Sutherland}, R.
\newblock {Theoretical Evolution of Optical Strong Lines across Cosmic Time}.
\newblock {\em \apj} {\bf 2013}, {\em 774},~100,
  \href{http://xxx.lanl.gov/abs/1307.0508}{{\normalfont [1307.0508]}}.
\newblock
  doi:{\changeurlcolor{black}\href{https://doi.org/10.1088/0004-637X/774/2/100}{\detokenize{10.1088/0004-637X/774/2/100}}}.

\bibitem[{Orsi} \em{et~al.}(2014){Orsi}, {Padilla}, {Groves}, {Cora}, {Tecce},
  {Gargiulo}, and {Ruiz}]{Orsi2014}
{Orsi}, {\'A}.; {Padilla}, N.; {Groves}, B.; {Cora}, S.; {Tecce}, T.;
  {Gargiulo}, I.; {Ruiz}, A.
\newblock {The nebular emission of star-forming galaxies in a hierarchical
  universe}.
\newblock {\em \mnras} {\bf 2014}, {\em 443},~799--814,
  \href{http://xxx.lanl.gov/abs/1402.5145}{{\normalfont [1402.5145]}}.
\newblock
  doi:{\changeurlcolor{black}\href{https://doi.org/10.1093/mnras/stu1203}{\detokenize{10.1093/mnras/stu1203}}}.

\bibitem[{Shimizu} \em{et~al.}(2016){Shimizu}, {Inoue}, {Okamoto}, and
  {Yoshida}]{Shimizu2016}
{Shimizu}, I.; {Inoue}, A.K.; {Okamoto}, T.; {Yoshida}, N.
\newblock {Nebular line emission from z=7 galaxies in a cosmological
  simulation: rest-frame UV to optical lines}.
\newblock {\em \mnras} {\bf 2016}, {\em 461},~3563--3575,
  \href{http://xxx.lanl.gov/abs/1509.00800}{{\normalfont [1509.00800]}}.
\newblock
  doi:{\changeurlcolor{black}\href{https://doi.org/10.1093/mnras/stw1423}{\detokenize{10.1093/mnras/stw1423}}}.

\bibitem[{Hirschmann} \em{et~al.}(2017){Hirschmann}, {Charlot}, {Feltre},
  {Naab}, {Choi}, {Ostriker}, and {Somerville}]{hirschman2017}
{Hirschmann}, M.; {Charlot}, S.; {Feltre}, A.; {Naab}, T.; {Choi}, E.;
  {Ostriker}, J.P.; {Somerville}, R.S.
\newblock {Synthetic nebular emission from massive galaxies - I: origin of the
  cosmic evolution of optical emission-line ratios}.
\newblock {\em \mnras} {\bf 2017}, {\em 472},~2468--2495,
  \href{http://xxx.lanl.gov/abs/1706.00010}{{\normalfont [1706.00010]}}.
\newblock
  doi:{\changeurlcolor{black}\href{https://doi.org/10.1093/mnras/stx2180}{\detokenize{10.1093/mnras/stx2180}}}.

\bibitem[{Aravena} \em{et~al.}(2010){Aravena}, {Carilli}, {Daddi}, {Wagg},
  {Walter}, {Riechers}, {Dannerbauer}, {Morrison}, {Stern}, and
  {Krips}]{Aravena2010}
{Aravena}, M.; {Carilli}, C.; {Daddi}, E.; {Wagg}, J.; {Walter}, F.;
  {Riechers}, D.; {Dannerbauer}, H.; {Morrison}, G.E.; {Stern}, D.; {Krips}, M.
\newblock {Cold Molecular Gas in Massive, Star-forming Disk Galaxies at z =
  1.5}.
\newblock {\em \apj} {\bf 2010}, {\em 718},~177--183,
  \href{http://xxx.lanl.gov/abs/1005.4965}{{\normalfont [1005.4965]}}.
\newblock
  doi:{\changeurlcolor{black}\href{https://doi.org/10.1088/0004-637X/718/1/177}{\detokenize{10.1088/0004-637X/718/1/177}}}.

\bibitem[{Daddi} \em{et~al.}(2010){Daddi}, {Bournaud}, {Walter}, {Dannerbauer},
  {Carilli}, {Dickinson}, {Elbaz}, {Morrison}, {Riechers}, {Onodera}, {Salmi},
  {Krips}, and {Stern}]{Daddi2010}
{Daddi}, E.; {Bournaud}, F.; {Walter}, F.; {Dannerbauer}, H.; {Carilli}, C.L.;
  {Dickinson}, M.; {Elbaz}, D.; {Morrison}, G.E.; {Riechers}, D.; {Onodera},
  M.; {Salmi}, F.; {Krips}, M.; {Stern}, D.
\newblock {Very High Gas Fractions and Extended Gas Reservoirs in z = 1.5 Disk
  Galaxies}.
\newblock {\em \apj} {\bf 2010}, {\em 713},~686--707,
  \href{http://xxx.lanl.gov/abs/0911.2776}{{\normalfont
  [arXiv:astro-ph.CO/0911.2776]}}.
\newblock
  doi:{\changeurlcolor{black}\href{https://doi.org/10.1088/0004-637X/713/1/686}{\detokenize{10.1088/0004-637X/713/1/686}}}.

\bibitem[{Tacconi} \em{et~al.}(2010){Tacconi}, {Genzel}, {Neri}, {Cox},
  {Cooper}, {Shapiro}, {Bolatto}, {Bouch{\'e}}, {Bournaud}, {Burkert},
  {Combes}, {Comerford}, {Davis}, {Schreiber}, {Garcia-Burillo},
  {Gracia-Carpio}, {Lutz}, {Naab}, {Omont}, {Shapley}, {Sternberg}, and
  {Weiner}]{Tacconi2010}
{Tacconi}, L.J.; {Genzel}, R.; {Neri}, R.; {Cox}, P.; {Cooper}, M.C.;
  {Shapiro}, K.; {Bolatto}, A.; {Bouch{\'e}}, N.; {Bournaud}, F.; {Burkert},
  A.; {Combes}, F.; {Comerford}, J.; {Davis}, M.; {Schreiber}, N.M.F.;
  {Garcia-Burillo}, S.; {Gracia-Carpio}, J.; {Lutz}, D.; {Naab}, T.; {Omont},
  A.; {Shapley}, A.; {Sternberg}, A.; {Weiner}, B.
\newblock {High molecular gas fractions in normal massive star-forming galaxies
  in the young Universe}.
\newblock {\em \nat} {\bf 2010}, {\em 463},~781--784,
  \href{http://xxx.lanl.gov/abs/1002.2149}{{\normalfont
  [arXiv:astro-ph.CO/1002.2149]}}.
\newblock
  doi:{\changeurlcolor{black}\href{https://doi.org/10.1038/nature08773}{\detokenize{10.1038/nature08773}}}.

\bibitem[{Tacconi} \em{et~al.}(2013){Tacconi}, {Neri}, {Genzel}, {Combes},
  {Bolatto}, {Cooper}, {Wuyts}, {Bournaud}, {Burkert}, {Comerford}, {Cox},
  {Davis}, {F{\"o}rster Schreiber}, {Garc{\'{\i}}a-Burillo}, {Gracia-Carpio},
  {Lutz}, {Naab}, {Newman}, {Omont}, {Saintonge}, {Shapiro Griffin}, {Shapley},
  {Sternberg}, and {Weiner}]{Tacconi2013}
{Tacconi}, L.J.; {Neri}, R.; {Genzel}, R.; {Combes}, F.; {Bolatto}, A.;
  {Cooper}, M.C.; {Wuyts}, S.; {Bournaud}, F.; {Burkert}, A.; {Comerford}, J.;
  {Cox}, P.; {Davis}, M.; {F{\"o}rster Schreiber}, N.M.;
  {Garc{\'{\i}}a-Burillo}, S.; {Gracia-Carpio}, J.; {Lutz}, D.; {Naab}, T.;
  {Newman}, S.; {Omont}, A.; {Saintonge}, A.; {Shapiro Griffin}, K.; {Shapley},
  A.; {Sternberg}, A.; {Weiner}, B.
\newblock {Phibss: Molecular Gas Content and Scaling Relations in z \~{} 1-3
  Massive, Main-sequence Star-forming Galaxies}.
\newblock {\em \apj} {\bf 2013}, {\em 768},~74,
  \href{http://xxx.lanl.gov/abs/1211.5743}{{\normalfont
  [arXiv:astro-ph.CO/1211.5743]}}.
\newblock
  doi:{\changeurlcolor{black}\href{https://doi.org/10.1088/0004-637X/768/1/74}{\detokenize{10.1088/0004-637X/768/1/74}}}.

\bibitem[{Tacconi} \em{et~al.}(2018){Tacconi}, {Genzel}, {Saintonge}, {Combes},
  {Garc{\'{\i}}a-Burillo}, {Neri}, {Bolatto}, {Contini}, {F{\"o}rster
  Schreiber}, {Lilly}, {Lutz}, {Wuyts}, {Accurso}, {Boissier}, {Boone},
  {Bouch{\'e}}, {Bournaud}, {Burkert}, {Carollo}, {Cooper}, {Cox}, {Feruglio},
  {Freundlich}, {Herrera-Camus}, {Juneau}, {Lippa}, {Naab}, {Renzini},
  {Salome}, {Sternberg}, {Tadaki}, {{\"U}bler}, {Walter}, {Weiner}, and
  {Weiss}]{Tacconi2018}
{Tacconi}, L.J.; {Genzel}, R.; {Saintonge}, A.; {Combes}, F.;
  {Garc{\'{\i}}a-Burillo}, S.; {Neri}, R.; {Bolatto}, A.; {Contini}, T.;
  {F{\"o}rster Schreiber}, N.M.; {Lilly}, S.; {Lutz}, D.; {Wuyts}, S.;
  {Accurso}, G.; {Boissier}, J.; {Boone}, F.; {Bouch{\'e}}, N.; {Bournaud}, F.;
  {Burkert}, A.; {Carollo}, M.; {Cooper}, M.; {Cox}, P.; {Feruglio}, C.;
  {Freundlich}, J.; {Herrera-Camus}, R.; {Juneau}, S.; {Lippa}, M.; {Naab}, T.;
  {Renzini}, A.; {Salome}, P.; {Sternberg}, A.; {Tadaki}, K.; {{\"U}bler}, H.;
  {Walter}, F.; {Weiner}, B.; {Weiss}, A.
\newblock {PHIBSS: Unified Scaling Relations of Gas Depletion Time and
  Molecular Gas Fractions}.
\newblock {\em \apj} {\bf 2018}, {\em 853},~179,
  \href{http://xxx.lanl.gov/abs/1702.01140}{{\normalfont [1702.01140]}}.
\newblock
  doi:{\changeurlcolor{black}\href{https://doi.org/10.3847/1538-4357/aaa4b4}{\detokenize{10.3847/1538-4357/aaa4b4}}}.

\bibitem[{Geach} \em{et~al.}(2011){Geach}, {Smail}, {Moran}, {MacArthur},
  {Lagos}, and {Edge}]{Geach2011}
{Geach}, J.E.; {Smail}, I.; {Moran}, S.M.; {MacArthur}, L.A.; {Lagos}, C.d.P.;
  {Edge}, A.C.
\newblock {On the Evolution of the Molecular Gas Fraction of Star-Forming
  Galaxies}.
\newblock {\em \apjl} {\bf 2011}, {\em 730},~L19,
  \href{http://xxx.lanl.gov/abs/1102.3694}{{\normalfont
  [arXiv:astro-ph.CO/1102.3694]}}.
\newblock
  doi:{\changeurlcolor{black}\href{https://doi.org/10.1088/2041-8205/730/2/L19}{\detokenize{10.1088/2041-8205/730/2/L19}}}.

\bibitem[{Magdis} \em{et~al.}(2012){Magdis}, {Daddi}, {Sargent}, {Elbaz},
  {Gobat}, {Dannerbauer}, {Feruglio}, {Tan}, {Rigopoulou}, {Charmandaris},
  {Dickinson}, {Reddy}, and {Aussel}]{Magdis2012}
{Magdis}, G.E.; {Daddi}, E.; {Sargent}, M.; {Elbaz}, D.; {Gobat}, R.;
  {Dannerbauer}, H.; {Feruglio}, C.; {Tan}, Q.; {Rigopoulou}, D.;
  {Charmandaris}, V.; {Dickinson}, M.; {Reddy}, N.; {Aussel}, H.
\newblock {The Molecular Gas Content of z = 3 Lyman Break Galaxies: Evidence of
  a Non-evolving Gas Fraction in Main-sequence Galaxies at z \gt 2}.
\newblock {\em \apjl} {\bf 2012}, {\em 758},~L9,
  \href{http://xxx.lanl.gov/abs/1209.1484}{{\normalfont [1209.1484]}}.
\newblock
  doi:{\changeurlcolor{black}\href{https://doi.org/10.1088/2041-8205/758/1/L9}{\detokenize{10.1088/2041-8205/758/1/L9}}}.

\bibitem[{Santini} \em{et~al.}(2014){Santini}, {Maiolino}, {Magnelli}, {Lutz},
  {Lamastra}, {Li Causi}, {Eales}, {Andreani}, {Berta}, {Buat}, {Cooray},
  {Cresci}, {Daddi}, {Farrah}, {Fontana}, {Franceschini}, {Genzel}, {Granato},
  {Grazian}, {Le Floc'h}, {Magdis}, {Magliocchetti}, {Mannucci}, {Menci},
  {Nordon}, {Oliver}, {Popesso}, {Pozzi}, {Riguccini}, {Rodighiero}, {Rosario},
  {Salvato}, {Scott}, {Silva}, {Tacconi}, {Viero}, {Wang}, {Wuyts}, and
  {Xu}]{Santini2014}
{Santini}, P.; {Maiolino}, R.; {Magnelli}, B.; {Lutz}, D.; {Lamastra}, A.; {Li
  Causi}, G.; {Eales}, S.; {Andreani}, P.; {Berta}, S.; {Buat}, V.; {Cooray},
  A.; {Cresci}, G.; {Daddi}, E.; {Farrah}, D.; {Fontana}, A.; {Franceschini},
  A.; {Genzel}, R.; {Granato}, G.; {Grazian}, A.; {Le Floc'h}, E.; {Magdis},
  G.; {Magliocchetti}, M.; {Mannucci}, F.; {Menci}, N.; {Nordon}, R.; {Oliver},
  S.; {Popesso}, P.; {Pozzi}, F.; {Riguccini}, L.; {Rodighiero}, G.; {Rosario},
  D.J.; {Salvato}, M.; {Scott}, D.; {Silva}, L.; {Tacconi}, L.; {Viero}, M.;
  {Wang}, L.; {Wuyts}, S.; {Xu}, K.
\newblock {The evolution of the dust and gas content in galaxies}.
\newblock {\em \aap} {\bf 2014}, {\em 562},~A30,
  \href{http://xxx.lanl.gov/abs/1311.3670}{{\normalfont [1311.3670]}}.
\newblock
  doi:{\changeurlcolor{black}\href{https://doi.org/10.1051/0004-6361/201322835}{\detokenize{10.1051/0004-6361/201322835}}}.

\bibitem[{B{\'e}thermin} \em{et~al.}(2016){B{\'e}thermin}, {De Breuck},
  {Gullberg}, {Aravena}, {Bothwell}, {Chapman}, {Gonzalez}, {Greve}, {Litke},
  {Ma}, {Malkan}, {Marrone}, {Murphy}, {Spilker}, {Stark}, {Strandet},
  {Vieira}, {Wei{\ss}}, and {Welikala}]{Bethermin2016}
{B{\'e}thermin}, M.; {De Breuck}, C.; {Gullberg}, B.; {Aravena}, M.;
  {Bothwell}, M.S.; {Chapman}, S.C.; {Gonzalez}, A.H.; {Greve}, T.R.; {Litke},
  K.; {Ma}, J.; {Malkan}, M.; {Marrone}, D.P.; {Murphy}, E.J.; {Spilker}, J.S.;
  {Stark}, A.A.; {Strandet}, M.; {Vieira}, J.D.; {Wei{\ss}}, A.; {Welikala}, N.
\newblock {An ALMA view of the interstellar medium of the z = 4.77 lensed
  starburst SPT-S J213242-5802.9}.
\newblock {\em \aap} {\bf 2016}, {\em 586},~L7,
  \href{http://xxx.lanl.gov/abs/1601.01682}{{\normalfont [1601.01682]}}.
\newblock
  doi:{\changeurlcolor{black}\href{https://doi.org/10.1051/0004-6361/201527739}{\detokenize{10.1051/0004-6361/201527739}}}.

\bibitem[{Decarli} \em{et~al.}(2016){Decarli}, {Walter}, {Aravena}, {Carilli},
  {Bouwens}, {da Cunha}, {Daddi}, {Elbaz}, {Riechers}, {Smail}, {Swinbank},
  {Weiss}, {Bacon}, {Bauer}, {Bell}, {Bertoldi}, {Chapman}, {Colina}, {Cortes},
  {Cox}, {G{\'o}nzalez-L{\'o}pez}, {Inami}, {Ivison}, {Hodge}, {Karim},
  {Magnelli}, {Ota}, {Popping}, {Rix}, {Sargent}, {van der Wel}, and {van der
  Werf}]{Decarli2016}
{Decarli}, R.; {Walter}, F.; {Aravena}, M.; {Carilli}, C.; {Bouwens}, R.; {da
  Cunha}, E.; {Daddi}, E.; {Elbaz}, D.; {Riechers}, D.; {Smail}, I.;
  {Swinbank}, M.; {Weiss}, A.; {Bacon}, R.; {Bauer}, F.; {Bell}, E.F.;
  {Bertoldi}, F.; {Chapman}, S.; {Colina}, L.; {Cortes}, P.C.; {Cox}, P.;
  {G{\'o}nzalez-L{\'o}pez}, J.; {Inami}, H.; {Ivison}, R.; {Hodge}, J.;
  {Karim}, A.; {Magnelli}, B.; {Ota}, K.; {Popping}, G.; {Rix}, H.W.;
  {Sargent}, M.; {van der Wel}, A.; {van der Werf}, P.
\newblock {ALMA Spectroscopic Survey in the Hubble Ultra Deep Field: Molecular
  gas reservoirs in high-redshift galaxies}.
\newblock {\em ArXiv e-prints} {\bf 2016},
  \href{http://xxx.lanl.gov/abs/1607.06771}{{\normalfont [1607.06771]}}.

\bibitem[{Scoville} \em{et~al.}(2016){Scoville}, {Sheth}, {Aussel}, {Vanden
  Bout}, {Capak}, {Bongiorno}, {Casey}, {Murchikova}, {Koda},
  {{\'A}lvarez-M{\'a}rquez}, {Lee}, {Laigle}, {McCracken}, {Ilbert}, {Pope},
  {Sanders}, {Chu}, {Toft}, {Ivison}, and {Manohar}]{Scoville2016}
{Scoville}, N.; {Sheth}, K.; {Aussel}, H.; {Vanden Bout}, P.; {Capak}, P.;
  {Bongiorno}, A.; {Casey}, C.M.; {Murchikova}, L.; {Koda}, J.;
  {{\'A}lvarez-M{\'a}rquez}, J.; {Lee}, N.; {Laigle}, C.; {McCracken}, H.J.;
  {Ilbert}, O.; {Pope}, A.; {Sanders}, D.; {Chu}, J.; {Toft}, S.; {Ivison},
  R.J.; {Manohar}, S.
\newblock {ISM Masses and the Star formation Law at Z~=~1 to 6: ALMA
  Observations of Dust Continuum in 145 Galaxies in the COSMOS Survey Field}.
\newblock {\em \apj} {\bf 2016}, {\em 820},~83,
  \href{http://xxx.lanl.gov/abs/1511.05149}{{\normalfont [1511.05149]}}.
\newblock
  doi:{\changeurlcolor{black}\href{https://doi.org/10.3847/0004-637X/820/2/83}{\detokenize{10.3847/0004-637X/820/2/83}}}.

\bibitem[{da Cunha} \em{et~al.}(2013){da Cunha}, {Groves}, {Walter}, {Decarli},
  {Weiss}, {Bertoldi}, {Carilli}, {Daddi}, {Elbaz}, {Ivison}, {Maiolino},
  {Riechers}, {Rix}, {Sargent}, and {Smail}]{daCunha2013}
{da Cunha}, E.; {Groves}, B.; {Walter}, F.; {Decarli}, R.; {Weiss}, A.;
  {Bertoldi}, F.; {Carilli}, C.; {Daddi}, E.; {Elbaz}, D.; {Ivison}, R.;
  {Maiolino}, R.; {Riechers}, D.; {Rix}, H.W.; {Sargent}, M.; {Smail}, I.
\newblock {On the Effect of the Cosmic Microwave Background in High-redshift
  (Sub-)millimeter Observations}.
\newblock {\em \apj} {\bf 2013}, {\em 766},~13,
  \href{http://xxx.lanl.gov/abs/1302.0844}{{\normalfont [1302.0844]}}.
\newblock
  doi:{\changeurlcolor{black}\href{https://doi.org/10.1088/0004-637X/766/1/13}{\detokenize{10.1088/0004-637X/766/1/13}}}.

\bibitem[{Bisbas} \em{et~al.}(2015){Bisbas}, {Papadopoulos}, and
  {Viti}]{Bisbas2015}
{Bisbas}, T.G.; {Papadopoulos}, P.P.; {Viti}, S.
\newblock {Effective Destruction of CO by Cosmic Rays: Implications for Tracing
  H$_{2}$ Gas in the Universe}.
\newblock {\em \apj} {\bf 2015}, {\em 803},~37,
  \href{http://xxx.lanl.gov/abs/1502.04198}{{\normalfont [1502.04198]}}.
\newblock
  doi:{\changeurlcolor{black}\href{https://doi.org/10.1088/0004-637X/803/1/37}{\detokenize{10.1088/0004-637X/803/1/37}}}.

\bibitem[{Glover} and {Clark}(2016)]{Glover2016}
{Glover}, S.C.O.; {Clark}, P.C.
\newblock {Is atomic carbon a good tracer of molecular gas in metal-poor
  galaxies?}
\newblock {\em \mnras} {\bf 2016}, {\em 456},~3596--3609,
  \href{http://xxx.lanl.gov/abs/1509.01939}{{\normalfont [1509.01939]}}.
\newblock
  doi:{\changeurlcolor{black}\href{https://doi.org/10.1093/mnras/stv2863}{\detokenize{10.1093/mnras/stv2863}}}.

\bibitem[{Bothwell} \em{et~al.}(2016){Bothwell}, {Aguirre}, {Aravena},
  {Bethermin}, {Bisbas}, {Chapman}, {De Breuck}, {Gonzalez}, {Greve},
  {Hezaveh}, {Ma}, {Malkan}, {Marrone}, {Murphy}, {Spilker}, {Strandet},
  {Vieira}, and {Weiss}]{Bothwell2016}
{Bothwell}, M.S.; {Aguirre}, J.E.; {Aravena}, M.; {Bethermin}, M.; {Bisbas},
  T.G.; {Chapman}, S.C.; {De Breuck}, C.; {Gonzalez}, A.H.; {Greve}, T.R.;
  {Hezaveh}, Y.; {Ma}, J.; {Malkan}, M.; {Marrone}, D.P.; {Murphy}, E.J.;
  {Spilker}, J.S.; {Strandet}, M.; {Vieira}, J.D.; {Weiss}, A.
\newblock {ALMA observations of atomic carbon in z\~{}4 dusty star-forming
  galaxies}.
\newblock {\em ArXiv e-prints} {\bf 2016},
  \href{http://xxx.lanl.gov/abs/1612.04380}{{\normalfont [1612.04380]}}.

\bibitem[{Popping} \em{et~al.}(2017){Popping}, {Decarli}, {Man}, {Nelson},
  {B{\'e}thermin}, {De Breuck}, {Mainieri}, {van Dokkum}, {Gullberg}, {van
  Kampen}, {Spaans}, and {Trager}]{Popping2017}
{Popping}, G.; {Decarli}, R.; {Man}, A.W.S.; {Nelson}, E.J.; {B{\'e}thermin},
  M.; {De Breuck}, C.; {Mainieri}, V.; {van Dokkum}, P.G.; {Gullberg}, B.; {van
  Kampen}, E.; {Spaans}, M.; {Trager}, S.C.
\newblock {ALMA reveals starburst-like interstellar medium conditions in a
  compact star-forming galaxy at z 2 using [CI] and CO}.
\newblock {\em \aap} {\bf 2017}, {\em 602},~A11,
  \href{http://xxx.lanl.gov/abs/1703.05764}{{\normalfont [1703.05764]}}.
\newblock
  doi:{\changeurlcolor{black}\href{https://doi.org/10.1051/0004-6361/201730391}{\detokenize{10.1051/0004-6361/201730391}}}.

\bibitem[{Cormier} \em{et~al.}(2018){Cormier}, {Bigiel}, {Jim{\'e}nez-Donaire},
  {Leroy}, {Gallagher}, {Usero}, {Sandstrom}, {Bolatto}, {Hughes}, {Kramer},
  {Krumholz}, {Meier}, {Murphy}, {Pety}, {Rosolowsky}, {Schinnerer}, {Schruba},
  {Sliwa}, and {Walter}]{cormier2018}
{Cormier}, D.; {Bigiel}, F.; {Jim{\'e}nez-Donaire}, M.J.; {Leroy}, A.K.;
  {Gallagher}, M.; {Usero}, A.; {Sandstrom}, K.; {Bolatto}, A.; {Hughes}, A.;
  {Kramer}, C.; {Krumholz}, M.R.; {Meier}, D.S.; {Murphy}, E.J.; {Pety}, J.;
  {Rosolowsky}, E.; {Schinnerer}, E.; {Schruba}, A.; {Sliwa}, K.; {Walter}, F.
\newblock {Full-disc $^{13}$CO(1-0) mapping across nearby galaxies of the
  EMPIRE survey and the CO-to-H$_{2}$ conversion factor}.
\newblock {\em \mnras} {\bf 2018}, {\em 475},~3909--3933,
  \href{http://xxx.lanl.gov/abs/1801.03105}{{\normalfont [1801.03105]}}.
\newblock
  doi:{\changeurlcolor{black}\href{https://doi.org/10.1093/mnras/sty059}{\detokenize{10.1093/mnras/sty059}}}.

\bibitem[{Sansonetti} \em{et~al.}(2004){Sansonetti}, {Kerber}, {Reader}, and
  {Rosa}]{Sansonetti2004}
{Sansonetti}, C.J.; {Kerber}, F.; {Reader}, J.; {Rosa}, M.R.
\newblock {Characterization of the Far-ultraviolet Spectrum of Pt/Cr-Ne Hollow
  Cathode Lamps as Used on the Space Telescope Imaging Spectrograph on Board
  the Hubble Space Telescope}.
\newblock {\em \apjs} {\bf 2004}, {\em 153},~555--579.
\newblock
  doi:{\changeurlcolor{black}\href{https://doi.org/10.1086/421874}{\detokenize{10.1086/421874}}}.

\bibitem[{Leitherer} \em{et~al.}(2011){Leitherer}, {Tremonti}, {Heckman}, and
  {Calzetti}]{Leitherer2011}
{Leitherer}, C.; {Tremonti}, C.A.; {Heckman}, T.M.; {Calzetti}, D.
\newblock {An Ultraviolet Spectroscopic Atlas of Local Starbursts and
  Star-forming Galaxies: The Legacy of FOS and GHRS}.
\newblock {\em \aj} {\bf 2011}, {\em 141},~37,
  \href{http://xxx.lanl.gov/abs/1011.0385}{{\normalfont [1011.0385]}}.
\newblock
  doi:{\changeurlcolor{black}\href{https://doi.org/10.1088/0004-6256/141/2/37}{\detokenize{10.1088/0004-6256/141/2/37}}}.

\bibitem[{Mittal} \em{et~al.}(2011){Mittal}, {O'Dea}, {Ferland}, {Oonk},
  {Edge}, {Canning}, {Russell}, {Baum}, {B{\"o}hringer}, {Combes}, {Donahue},
  {Fabian}, {Hatch}, {Hoffer}, {Johnstone}, {McNamara}, {Salom{\'e}}, and
  {Tremblay}]{Mittal2011}
{Mittal}, R.; {O'Dea}, C.P.; {Ferland}, G.; {Oonk}, J.B.R.; {Edge}, A.C.;
  {Canning}, R.E.A.; {Russell}, H.; {Baum}, S.A.; {B{\"o}hringer}, H.;
  {Combes}, F.; {Donahue}, M.; {Fabian}, A.C.; {Hatch}, N.A.; {Hoffer}, A.;
  {Johnstone}, R.; {McNamara}, B.R.; {Salom{\'e}}, P.; {Tremblay}, G.
\newblock {Herschel observations of the Centaurus cluster - the dynamics of
  cold gas in a cool core}.
\newblock {\em \mnras} {\bf 2011}, {\em 418},~2386--2402,
  \href{http://xxx.lanl.gov/abs/1108.2757}{{\normalfont [1108.2757]}}.
\newblock
  doi:{\changeurlcolor{black}\href{https://doi.org/10.1111/j.1365-2966.2011.19634.x}{\detokenize{10.1111/j.1365-2966.2011.19634.x}}}.

\bibitem[{Pallottini} \em{et~al.}(2017){Pallottini}, {Ferrara}, {Gallerani},
  {Vallini}, {Maiolino}, and {Salvadori}]{Pallottini2017}
{Pallottini}, A.; {Ferrara}, A.; {Gallerani}, S.; {Vallini}, L.; {Maiolino},
  R.; {Salvadori}, S.
\newblock {Zooming on the internal structure of $z\simeq6$ galaxies}.
\newblock {\em \mnras} {\bf 2017}, {\em 465},~2540--2558,
  \href{http://xxx.lanl.gov/abs/1609.01719}{{\normalfont [1609.01719]}}.
\newblock
  doi:{\changeurlcolor{black}\href{https://doi.org/10.1093/mnras/stw2847}{\detokenize{10.1093/mnras/stw2847}}}.

\bibitem[{Xiao} \em{et~al.}(2018){Xiao}, {Stanway}, and {Eldridge}]{Xiao2018}
{Xiao}, L.; {Stanway}, E.R.; {Eldridge}, J.J.
\newblock {Emission-line diagnostics of nearby H II regions including
  interacting binary populations}.
\newblock {\em \mnras} {\bf 2018}, {\em 477},~904--934,
  \href{http://xxx.lanl.gov/abs/1801.07068}{{\normalfont [1801.07068]}}.
\newblock
  doi:{\changeurlcolor{black}\href{https://doi.org/10.1093/mnras/sty646}{\detokenize{10.1093/mnras/sty646}}}.

\bibitem[{Safranek-Shrader} \em{et~al.}(2017){Safranek-Shrader}, {Krumholz},
  {Kim}, {Ostriker}, {Klein}, {Li}, {McKee}, and {Stone}]{Safranek-Shrader2017}
{Safranek-Shrader}, C.; {Krumholz}, M.R.; {Kim}, C.G.; {Ostriker}, E.C.;
  {Klein}, R.I.; {Li}, S.; {McKee}, C.F.; {Stone}, J.M.
\newblock {Chemistry and radiative shielding in star-forming galactic discs}.
\newblock {\em \mnras} {\bf 2017}, {\em 465},~885--905,
  \href{http://xxx.lanl.gov/abs/1605.07618}{{\normalfont [1605.07618]}}.
\newblock
  doi:{\changeurlcolor{black}\href{https://doi.org/10.1093/mnras/stw2647}{\detokenize{10.1093/mnras/stw2647}}}.

\bibitem[{Ginsburg} \em{et~al.}(2016){Ginsburg}, {Henkel}, {Ao}, {Riquelme},
  {Kauffmann}, {Pillai}, {Mills}, {Requena-Torres}, {Immer}, {Testi}, {Ott},
  {Bally}, {Battersby}, {Darling}, {Aalto}, {Stanke}, {Kendrew}, {Kruijssen},
  {Longmore}, {Dale}, {Guesten}, and {Menten}]{Ginsburg2016}
{Ginsburg}, A.; {Henkel}, C.; {Ao}, Y.; {Riquelme}, D.; {Kauffmann}, J.;
  {Pillai}, T.; {Mills}, E.A.C.; {Requena-Torres}, M.A.; {Immer}, K.; {Testi},
  L.; {Ott}, J.; {Bally}, J.; {Battersby}, C.; {Darling}, J.; {Aalto}, S.;
  {Stanke}, T.; {Kendrew}, S.; {Kruijssen}, J.M.D.; {Longmore}, S.; {Dale}, J.;
  {Guesten}, R.; {Menten}, K.M.
\newblock {Dense gas in the Galactic central molecular zone is warm and heated
  by turbulence}.
\newblock {\em \aap} {\bf 2016}, {\em 586},~A50,
  \href{http://xxx.lanl.gov/abs/1509.01583}{{\normalfont [1509.01583]}}.
\newblock
  doi:{\changeurlcolor{black}\href{https://doi.org/10.1051/0004-6361/201526100}{\detokenize{10.1051/0004-6361/201526100}}}.

\bibitem[{Pe{\~n}aloza} \em{et~al.}(2017){Pe{\~n}aloza}, {Clark}, {Glover},
  {Shetty}, and {Klessen}]{Penaloza2017}
{Pe{\~n}aloza}, C.H.; {Clark}, P.C.; {Glover}, S.C.O.; {Shetty}, R.; {Klessen},
  R.S.
\newblock {Using CO line ratios to trace the physical properties of molecular
  clouds}.
\newblock {\em \mnras} {\bf 2017}, {\em 465},~2277--2285,
  \href{http://xxx.lanl.gov/abs/1611.02127}{{\normalfont [1611.02127]}}.
\newblock
  doi:{\changeurlcolor{black}\href{https://doi.org/10.1093/mnras/stw2892}{\detokenize{10.1093/mnras/stw2892}}}.

\bibitem[{Pe{\~n}aloza} \em{et~al.}(2018){Pe{\~n}aloza}, {Clark}, {Glover}, and
  {Klessen}]{Penaloza2018}
{Pe{\~n}aloza}, C.H.; {Clark}, P.C.; {Glover}, S.C.O.; {Klessen}, R.S.
\newblock {CO line ratios in molecular clouds: the impact of environment}.
\newblock {\em \mnras} {\bf 2018}, {\em 475},~1508--1520,
  \href{http://xxx.lanl.gov/abs/1711.01221}{{\normalfont [1711.01221]}}.
\newblock
  doi:{\changeurlcolor{black}\href{https://doi.org/10.1093/mnras/stx3263}{\detokenize{10.1093/mnras/stx3263}}}.

\bibitem[{Li} \em{et~al.}(2008){Li}, {Hopkins}, {Hernquist}, {Finkbeiner},
  {Cox}, {Springel}, {Jiang}, {Fan}, and {Yoshida}]{Li2008}
{Li}, Y.; {Hopkins}, P.F.; {Hernquist}, L.; {Finkbeiner}, D.P.; {Cox}, T.J.;
  {Springel}, V.; {Jiang}, L.; {Fan}, X.; {Yoshida}, N.
\newblock {Modeling the Dust Properties of $z\sim6$ Quasars with ART$^{2}$ -
  All-Wavelength Radiative Transfer with Adaptive Refinement Tree}.
\newblock {\em \apj} {\bf 2008}, {\em 678},~41--63,
  \href{http://xxx.lanl.gov/abs/0706.3706}{{\normalfont [0706.3706]}}.
\newblock
  doi:{\changeurlcolor{black}\href{https://doi.org/10.1086/529364}{\detokenize{10.1086/529364}}}.

\bibitem[{Yajima} \em{et~al.}(2014){Yajima}, {Li}, {Zhu}, {Abel}, {Gronwall},
  and {Ciardullo}]{Yajima2014}
{Yajima}, H.; {Li}, Y.; {Zhu}, Q.; {Abel}, T.; {Gronwall}, C.; {Ciardullo}, R.
\newblock {Escape of Ly{$\alpha$} and continuum photons from star-forming
  galaxies}.
\newblock {\em \mnras} {\bf 2014}, {\em 440},~776--786,
  \href{http://xxx.lanl.gov/abs/1209.5842}{{\normalfont [1209.5842]}}.
\newblock
  doi:{\changeurlcolor{black}\href{https://doi.org/10.1093/mnras/stu299}{\detokenize{10.1093/mnras/stu299}}}.

\bibitem[{Gray} and {Scannapieco}(2016)]{Gray2016}
{Gray}, W.J.; {Scannapieco}, E.
\newblock {Atomic Chemistry in Turbulent Astrophysical Media. II. Effect of the
  Redshift Zero Metagalactic Background}.
\newblock {\em \apj} {\bf 2016}, {\em 818},~198,
  \href{http://xxx.lanl.gov/abs/1511.05158}{{\normalfont [1511.05158]}}.
\newblock
  doi:{\changeurlcolor{black}\href{https://doi.org/10.3847/0004-637X/818/2/198}{\detokenize{10.3847/0004-637X/818/2/198}}}.

\bibitem[{Gray} and {Scannapieco}(2017)]{Gray2017}
{Gray}, W.J.; {Scannapieco}, E.
\newblock {The Effect of Turbulence on Nebular Emission Line Ratios}.
\newblock {\em \apj} {\bf 2017}, {\em 849},~132,
  \href{http://xxx.lanl.gov/abs/1710.01312}{{\normalfont [1710.01312]}}.
\newblock
  doi:{\changeurlcolor{black}\href{https://doi.org/10.3847/1538-4357/aa9121}{\detokenize{10.3847/1538-4357/aa9121}}}.

\bibitem[{Dere} \em{et~al.}(1997){Dere}, {Landi}, {Mason}, {Monsignori Fossi},
  and {Young}]{dere97}
{Dere}, K.P.; {Landi}, E.; {Mason}, H.E.; {Monsignori Fossi}, B.C.; {Young},
  P.R.
\newblock {CHIANTI - an atomic database for emission lines}.
\newblock {\em \aaps} {\bf 1997}, {\em 125},~149--173.
\newblock
  doi:{\changeurlcolor{black}\href{https://doi.org/10.1051/aas:1997368}{\detokenize{10.1051/aas:1997368}}}.

\bibitem[{Del Zanna} \em{et~al.}(2015){Del Zanna}, {Dere}, {Young}, {Landi},
  and {Mason}]{delzanna15}
{Del Zanna}, G.; {Dere}, K.P.; {Young}, P.R.; {Landi}, E.; {Mason}, H.E.
\newblock {CHIANTI - An atomic database for emission lines. Version 8}.
\newblock {\em \aap} {\bf 2015}, {\em 582},~A56,
  \href{http://xxx.lanl.gov/abs/1508.07631}{{\normalfont
  [arXiv:astro-ph.SR/1508.07631]}}.
\newblock
  doi:{\changeurlcolor{black}\href{https://doi.org/10.1051/0004-6361/201526827}{\detokenize{10.1051/0004-6361/201526827}}}.

\bibitem[{Brickhouse} \em{et~al.}(2000){Brickhouse}, {Dupree}, {Edgar},
  {Liedahl}, {Drake}, {White}, and {Singh}]{brickhouse2000}
{Brickhouse}, N.S.; {Dupree}, A.K.; {Edgar}, R.J.; {Liedahl}, D.A.; {Drake},
  S.A.; {White}, N.E.; {Singh}, K.P.
\newblock {Coronal Structure and Abundances of Capella from Simultaneous EUVE
  and ASCA Spectroscopy}.
\newblock {\em \apj} {\bf 2000}, {\em 530},~387--402.
\newblock
  doi:{\changeurlcolor{black}\href{https://doi.org/10.1086/308350}{\detokenize{10.1086/308350}}}.

\bibitem[{Kaastra} \em{et~al.}(1996){Kaastra}, {Bleeker}, and
  {Mewe}]{kaastra96}
{Kaastra}, J.S.; {Bleeker}, J.A.M.; {Mewe}, R.
\newblock {X-ray diagnostics of supernova remnants.}
\newblock  UV and X-ray Spectroscopy of Astrophysical and Laboratory Plasmas;
  {Yamashita}, K.; {Watanabe}, T., Eds.,  1996, pp. 15--20.

\bibitem[{Bar-Shalom} \em{et~al.}(2000){Bar-Shalom}, {Oreg}, and
  {Klapisch}]{barshalom2000}
{Bar-Shalom}, A.; {Oreg}, J.; {Klapisch}, M.
\newblock {Recent developments in the SCROLL model.}
\newblock {\em \jqsrt} {\bf 2000}, {\em 65},~43--55.
\newblock
  doi:{\changeurlcolor{black}\href{https://doi.org/10.1016/S0022-4073(99)00054-0}{\detokenize{10.1016/S0022-4073(99)00054-0}}}.

\bibitem[{Del Zanna} \em{et~al.}(2004){Del Zanna}, {Berrington}, and
  {Mason}]{delzanna04}
{Del Zanna}, G.; {Berrington}, K.A.; {Mason}, H.E.
\newblock {Benchmarking atomic data for astrophysics: Fe X}.
\newblock {\em \aap} {\bf 2004}, {\em 422},~731--749.
\newblock
  doi:{\changeurlcolor{black}\href{https://doi.org/10.1051/0004-6361:20034432}{\detokenize{10.1051/0004-6361:20034432}}}.

\bibitem[{Keenan} \em{et~al.}(2007){Keenan}, {Drake}, and {Aggarwal}]{keenan07}
{Keenan}, F.P.; {Drake}, J.J.; {Aggarwal}, K.M.
\newblock {An investigation of FeXVI emission lines in solar and stellar
  extreme-ultraviolet and soft X-ray spectra}.
\newblock {\em \mnras} {\bf 2007}, {\em 381},~1727--1732,
  \href{http://xxx.lanl.gov/abs/0708.2640}{{\normalfont [0708.2640]}}.
\newblock
  doi:{\changeurlcolor{black}\href{https://doi.org/10.1111/j.1365-2966.2007.12365.x}{\detokenize{10.1111/j.1365-2966.2007.12365.x}}}.

\bibitem[{Kato} \em{et~al.}(2009){Kato}, {Sakaue}, {Murakami}, {Kato},
  {Nakamura}, {Ohtani}, {Yamamoto}, and {Watanabe}]{kato09}
{Kato}, D.; {Sakaue}, H.A.; {Murakami}, I.; {Kato}, T.; {Nakamura}, N.;
  {Ohtani}, S.; {Yamamoto}, N.; {Watanabe}, T.
\newblock {Electron-impact excitation of 2p$^{5}$3l ${\to}$ 2p$^{5}$3l' line
  emission of Fe XVII}.
\newblock  Journal of Physics Conference Series,  2009, Vol. 163, {\em Journal
  of Physics Conference Series}, p. 012076.
\newblock
  doi:{\changeurlcolor{black}\href{https://doi.org/10.1088/1742-6596/163/1/012076}{\detokenize{10.1088/1742-6596/163/1/012076}}}.

\bibitem[{Beiersdorfer} \em{et~al.}(2014){Beiersdorfer}, {Tr{\"a}bert},
  {Lepson}, {Brickhouse}, and {Golub}]{beiersdorfer14}
{Beiersdorfer}, P.; {Tr{\"a}bert}, E.; {Lepson}, J.K.; {Brickhouse}, N.S.;
  {Golub}, L.
\newblock {High-resolution Laboratory Measurements of Coronal Lines in the
  198-218 {\AA} Region}.
\newblock {\em \apj} {\bf 2014}, {\em 788},~25.
\newblock
  doi:{\changeurlcolor{black}\href{https://doi.org/10.1088/0004-637X/788/1/25}{\detokenize{10.1088/0004-637X/788/1/25}}}.

\bibitem[{Liang} \em{et~al.}(2009){Liang}, {Whiteford}, and {Badnell}]{liang09}
{Liang}, G.Y.; {Whiteford}, A.D.; {Badnell}, N.R.
\newblock {R-matrix inner-shell electron-impact excitation of the Na-like
  iso-electronic sequence}.
\newblock {\em Journal of Physics B Atomic Molecular Physics} {\bf 2009}, {\em
  42},~225002.
\newblock
  doi:{\changeurlcolor{black}\href{https://doi.org/10.1088/0953-4075/42/22/225002}{\detokenize{10.1088/0953-4075/42/22/225002}}}.

\bibitem[{Liang} and {Badnell}(2010)]{liang10}
{Liang}, G.Y.; {Badnell}, N.R.
\newblock {R-matrix electron-impact excitation data for the Ne-like
  iso-electronic sequence}.
\newblock {\em \aap} {\bf 2010}, {\em 518},~A64.
\newblock
  doi:{\changeurlcolor{black}\href{https://doi.org/10.1051/0004-6361/201014170}{\detokenize{10.1051/0004-6361/201014170}}}.

\bibitem[{Wang} \em{et~al.}(2014){Wang}, {Li}, {Liu}, {Han}, {Duan}, {Li},
  {Li}, {Guo}, {Chen}, and {Yan}]{wang14}
{Wang}, K.; {Li}, D.F.; {Liu}, H.T.; {Han}, X.Y.; {Duan}, B.; {Li}, C.Y.; {Li},
  J.G.; {Guo}, X.L.; {Chen}, C.Y.; {Yan}, J.
\newblock {Systematic Calculations of Energy Levels and Transition Rates of
  C-like Ions with Z = 13-36}.
\newblock {\em \apjs} {\bf 2014}, {\em 215},~26.
\newblock
  doi:{\changeurlcolor{black}\href{https://doi.org/10.1088/0067-0049/215/2/26}{\detokenize{10.1088/0067-0049/215/2/26}}}.

\bibitem[{Del Zanna} and {Badnell}(2016)]{delzanna16}
{Del Zanna}, G.; {Badnell}, N.R.
\newblock {Atomic data for astrophysics: Ni XII}.
\newblock {\em \aap} {\bf 2016}, {\em 585},~A118.
\newblock
  doi:{\changeurlcolor{black}\href{https://doi.org/10.1051/0004-6361/201527378}{\detokenize{10.1051/0004-6361/201527378}}}.

\bibitem[{Liedahl} \em{et~al.}(1995){Liedahl}, {Osterheld}, and
  {Goldstein}]{liedahl95}
{Liedahl}, D.A.; {Osterheld}, A.L.; {Goldstein}, W.H.
\newblock {New calculations of Fe L-shell X-ray spectra in high-temperature
  plasmas}.
\newblock {\em \apjl} {\bf 1995}, {\em 438},~L115--L118.
\newblock
  doi:{\changeurlcolor{black}\href{https://doi.org/10.1086/187729}{\detokenize{10.1086/187729}}}.

\bibitem[{Gu}(2008)]{gu08}
{Gu}, M.F.
\newblock {The flexible atomic code}.
\newblock {\em Canadian Journal of Physics} {\bf 2008}, {\em 86},~675--689.
\newblock
  doi:{\changeurlcolor{black}\href{https://doi.org/10.1139/P07-197}{\detokenize{10.1139/P07-197}}}.

\bibitem[{Dyall} \em{et~al.}(1989){Dyall}, {Grant}, {Johnson}, {Parpia}, and
  {Plummer}]{dyall89}
{Dyall}, K.G.; {Grant}, I.P.; {Johnson}, C.T.; {Parpia}, F.A.; {Plummer}, E.P.
\newblock {GRASP: A general-purpose relativistic atomic structure program}.
\newblock {\em Computer Physics Communications} {\bf 1989}, {\em 55},~425--456.
\newblock
  doi:{\changeurlcolor{black}\href{https://doi.org/10.1016/0010-4655(89)90136-7}{\detokenize{10.1016/0010-4655(89)90136-7}}}.

\bibitem[{Badnell}(1986)]{badnell86}
{Badnell}, N.R.
\newblock {Dielectric recombination of Fe(22+) and Fe(21+)}.
\newblock {\em Journal of Physics B Atomic Molecular Physics} {\bf 1986}, {\em
  19},~3827--3835.
\newblock
  doi:{\changeurlcolor{black}\href{https://doi.org/10.1088/0022-3700/19/22/023}{\detokenize{10.1088/0022-3700/19/22/023}}}.

\bibitem[{Aggarwal} and {Keenan}(2014)]{aggarwal14}
{Aggarwal}, K.M.; {Keenan}, F.P.
\newblock {Energy levels, radiative rates and electron impact excitation rates
  for transitions in Fe XIV}.
\newblock {\em \mnras} {\bf 2014}, {\em 445},~2015--2027,
  \href{http://xxx.lanl.gov/abs/1409.4214}{{\normalfont
  [arXiv:astro-ph.SR/1409.4214]}}.
\newblock
  doi:{\changeurlcolor{black}\href{https://doi.org/10.1093/mnras/stu1908}{\detokenize{10.1093/mnras/stu1908}}}.

\bibitem[{Del Zanna} and {Badnell}(2016)]{delzannabadnell16}
{Del Zanna}, G.; {Badnell}, N.R.
\newblock {Atomic data and density diagnostics for S IV}.
\newblock {\em \mnras} {\bf 2016}, {\em 456},~3720--3728.
\newblock
  doi:{\changeurlcolor{black}\href{https://doi.org/10.1093/mnras/stv2845}{\detokenize{10.1093/mnras/stv2845}}}.

\bibitem[{Landi} \em{et~al.}(2012){Landi}, {Del Zanna}, {Young}, {Dere}, and
  {Mason}]{landi12}
{Landi}, E.; {Del Zanna}, G.; {Young}, P.R.; {Dere}, K.P.; {Mason}, H.E.
\newblock {CHIANTI - An Atomic Database for Emission Lines. XII. Version 7 of
  the Database}.
\newblock {\em \apj} {\bf 2012}, {\em 744},~99.
\newblock
  doi:{\changeurlcolor{black}\href{https://doi.org/10.1088/0004-637X/744/2/99}{\detokenize{10.1088/0004-637X/744/2/99}}}.

\bibitem[{Del Zanna} and {Ishikawa}(2009)]{delzanna09}
{Del Zanna}, G.; {Ishikawa}, Y.
\newblock {Benchmarking atomic data for astrophysics: Fe XVII EUV lines}.
\newblock {\em \aap} {\bf 2009}, {\em 508},~1517--1526.
\newblock
  doi:{\changeurlcolor{black}\href{https://doi.org/10.1051/0004-6361/200911729}{\detokenize{10.1051/0004-6361/200911729}}}.

\bibitem[{Brown} \em{et~al.}(1998){Brown}, {Beiersdorfer}, {Liedahl},
  {Widmann}, and {Kahn}]{brown98}
{Brown}, G.V.; {Beiersdorfer}, P.; {Liedahl}, D.A.; {Widmann}, K.; {Kahn}, S.M.
\newblock {Laboratory Measurements and Modeling of the Fe XVII X-Ray Spectrum}.
\newblock {\em \apj} {\bf 1998}, {\em 502},~1015--1026.
\newblock
  doi:{\changeurlcolor{black}\href{https://doi.org/10.1086/305941}{\detokenize{10.1086/305941}}}.

\bibitem[Kramida \em{et~al.}(2014)Kramida, {Yu.~Ralchenko}, Reader, and {and
  NIST ASD Team}]{NIST_ASD}
Kramida, A.; {Yu.~Ralchenko}.; Reader, J.; {and NIST ASD Team}.
\newblock {NIST Atomic Spectra Database (ver. 5.2), [Online]. Available:
  {\tt{http://physics.nist.gov/asd}} [2014, November 28]. National Institute of
  Standards and Technology, Gaithersburg, MD.},  2014.

\bibitem[{Steinacker} \em{et~al.}(2013){Steinacker}, {Baes}, and
  {Gordon}]{dust3DRT}
{Steinacker}, J.; {Baes}, M.; {Gordon}, K.D.
\newblock {Three-Dimensional Dust Radiative Transfer*}.
\newblock {\em \araa} {\bf 2013}, {\em 51},~63--104,
  \href{http://xxx.lanl.gov/abs/1303.4998}{{\normalfont
  [arXiv:astro-ph.IM/1303.4998]}}.
\newblock
  doi:{\changeurlcolor{black}\href{https://doi.org/10.1146/annurev-astro-082812-141042}{\detokenize{10.1146/annurev-astro-082812-141042}}}.

\bibitem[{Gnedin} and {Kaurov}(2014)]{Gnedin2014}
{Gnedin}, N.Y.; {Kaurov}, A.A.
\newblock {Cosmic Reionization on Computers. II. Reionization History and Its
  Back-reaction on Early Galaxies}.
\newblock {\em \apj} {\bf 2014}, {\em 793},~30,
  \href{http://xxx.lanl.gov/abs/1403.4251}{{\normalfont [1403.4251]}}.
\newblock
  doi:{\changeurlcolor{black}\href{https://doi.org/10.1088/0004-637X/793/1/30}{\detokenize{10.1088/0004-637X/793/1/30}}}.

\bibitem[{So} \em{et~al.}(2014){So}, {Norman}, {Reynolds}, and {Wise}]{So2014}
{So}, G.C.; {Norman}, M.L.; {Reynolds}, D.R.; {Wise}, J.H.
\newblock {Fully Coupled Simulation of Cosmic Reionization. II. Recombinations,
  Clumping Factors, and the Photon Budget for Reionization}.
\newblock {\em \apj} {\bf 2014}, {\em 789},~149,
  \href{http://xxx.lanl.gov/abs/1311.2152}{{\normalfont [1311.2152]}}.
\newblock
  doi:{\changeurlcolor{black}\href{https://doi.org/10.1088/0004-637X/789/2/149}{\detokenize{10.1088/0004-637X/789/2/149}}}.

\bibitem[{Wise} \em{et~al.}(2014){Wise}, {Demchenko}, {Halicek}, {Norman},
  {Turk}, {Abel}, and {Smith}]{Wise2014}
{Wise}, J.H.; {Demchenko}, V.G.; {Halicek}, M.T.; {Norman}, M.L.; {Turk}, M.J.;
  {Abel}, T.; {Smith}, B.D.
\newblock {The birth of a galaxy - III. Propelling reionization with the
  faintest galaxies}.
\newblock {\em \mnras} {\bf 2014}, {\em 442},~2560--2579,
  \href{http://xxx.lanl.gov/abs/1403.6123}{{\normalfont [1403.6123]}}.
\newblock
  doi:{\changeurlcolor{black}\href{https://doi.org/10.1093/mnras/stu979}{\detokenize{10.1093/mnras/stu979}}}.

\bibitem[{Norman} \em{et~al.}(2015){Norman}, {Reynolds}, {So}, {Harkness}, and
  {Wise}]{Norman2015}
{Norman}, M.L.; {Reynolds}, D.R.; {So}, G.C.; {Harkness}, R.P.; {Wise}, J.H.
\newblock {Fully Coupled Simulation of Cosmic Reionization. I. Numerical
  Methods and Tests}.
\newblock {\em \apjs} {\bf 2015}, {\em 216},~16,
  \href{http://xxx.lanl.gov/abs/1306.0645}{{\normalfont
  [arXiv:astro-ph.IM/1306.0645]}}.
\newblock
  doi:{\changeurlcolor{black}\href{https://doi.org/10.1088/0067-0049/216/1/16}{\detokenize{10.1088/0067-0049/216/1/16}}}.

\bibitem[{Pawlik} \em{et~al.}(2015){Pawlik}, {Schaye}, and {Dalla
  Vecchia}]{Pawlik2015}
{Pawlik}, A.H.; {Schaye}, J.; {Dalla Vecchia}, C.
\newblock {Spatially adaptive radiation-hydrodynamical simulations of galaxy
  formation during cosmological reionization}.
\newblock {\em \mnras} {\bf 2015}, {\em 451},~1586--1605,
  \href{http://xxx.lanl.gov/abs/1501.01980}{{\normalfont [1501.01980]}}.
\newblock
  doi:{\changeurlcolor{black}\href{https://doi.org/10.1093/mnras/stv976}{\detokenize{10.1093/mnras/stv976}}}.

\bibitem[{Ocvirk} \em{et~al.}(2016){Ocvirk}, {Gillet}, {Shapiro}, {Aubert},
  {Iliev}, {Teyssier}, {Yepes}, {Choi}, {Sullivan}, {Knebe}, {Gottl{\"o}ber},
  {D'Aloisio}, {Park}, {Hoffman}, and {Stranex}]{Ocvirk2016}
{Ocvirk}, P.; {Gillet}, N.; {Shapiro}, P.R.; {Aubert}, D.; {Iliev}, I.T.;
  {Teyssier}, R.; {Yepes}, G.; {Choi}, J.H.; {Sullivan}, D.; {Knebe}, A.;
  {Gottl{\"o}ber}, S.; {D'Aloisio}, A.; {Park}, H.; {Hoffman}, Y.; {Stranex},
  T.
\newblock {Cosmic Dawn (CoDa): the First Radiation-Hydrodynamics Simulation of
  Reionization and Galaxy Formation in the Local Universe}.
\newblock {\em \mnras} {\bf 2016}, {\em 463},~1462--1485,
  \href{http://xxx.lanl.gov/abs/1511.00011}{{\normalfont [1511.00011]}}.
\newblock
  doi:{\changeurlcolor{black}\href{https://doi.org/10.1093/mnras/stw2036}{\detokenize{10.1093/mnras/stw2036}}}.

\bibitem[{Rosdahl} \em{et~al.}(2013){Rosdahl}, {Blaizot}, {Aubert}, {Stranex},
  and {Teyssier}]{Rosdahl2013}
{Rosdahl}, J.; {Blaizot}, J.; {Aubert}, D.; {Stranex}, T.; {Teyssier}, R.
\newblock {RAMSES-RT: radiation hydrodynamics in the cosmological context}.
\newblock {\em \mnras} {\bf 2013}, {\em 436},~2188--2231,
  \href{http://xxx.lanl.gov/abs/1304.7126}{{\normalfont [1304.7126]}}.
\newblock
  doi:{\changeurlcolor{black}\href{https://doi.org/10.1093/mnras/stt1722}{\detokenize{10.1093/mnras/stt1722}}}.

\bibitem[{Rosdahl} and {Teyssier}(2015)]{RAMSES-RT-dust}
{Rosdahl}, J.; {Teyssier}, R.
\newblock {A scheme for radiation pressure and photon diffusion with the M1
  closure in RAMSES-RT}.
\newblock {\em \mnras} {\bf 2015}, {\em 449},~4380--4403,
  \href{http://xxx.lanl.gov/abs/1411.6440}{{\normalfont
  [arXiv:astro-ph.IM/1411.6440]}}.
\newblock
  doi:{\changeurlcolor{black}\href{https://doi.org/10.1093/mnras/stv567}{\detokenize{10.1093/mnras/stv567}}}.

\bibitem[{Aubert} \em{et~al.}(2015){Aubert}, {Deparis}, and {Ocvirk}]{EMMA}
{Aubert}, D.; {Deparis}, N.; {Ocvirk}, P.
\newblock {EMMA: an adaptive mesh refinement cosmological simulation code with
  radiative transfer}.
\newblock {\em \mnras} {\bf 2015}, {\em 454},~1012--1037,
  \href{http://xxx.lanl.gov/abs/1508.07888}{{\normalfont
  [arXiv:astro-ph.IM/1508.07888]}}.
\newblock
  doi:{\changeurlcolor{black}\href{https://doi.org/10.1093/mnras/stv1896}{\detokenize{10.1093/mnras/stv1896}}}.

\bibitem[{Krumholz} \em{et~al.}(2007){Krumholz}, {Klein}, {McKee}, and
  {Bolstad}]{ORION}
{Krumholz}, M.R.; {Klein}, R.I.; {McKee}, C.F.; {Bolstad}, J.
\newblock {Equations and Algorithms for Mixed-frame Flux-limited Diffusion
  Radiation Hydrodynamics}.
\newblock {\em \apj} {\bf 2007}, {\em 667},~626--643,
  \href{http://xxx.lanl.gov/abs/astro-ph/0611003}{{\normalfont
  [astro-ph/0611003]}}.
\newblock
  doi:{\changeurlcolor{black}\href{https://doi.org/10.1086/520791}{\detokenize{10.1086/520791}}}.

\bibitem[{Davis} \em{et~al.}(2012){Davis}, {Stone}, and {Jiang}]{ATHENA-RT}
{Davis}, S.W.; {Stone}, J.M.; {Jiang}, Y.F.
\newblock {A Radiation Transfer Solver for Athena Using Short Characteristics}.
\newblock {\em \apjs} {\bf 2012}, {\em 199},~9,
  \href{http://xxx.lanl.gov/abs/1201.2222}{{\normalfont
  [arXiv:astro-ph.IM/1201.2222]}}.
\newblock
  doi:{\changeurlcolor{black}\href{https://doi.org/10.1088/0067-0049/199/1/9}{\detokenize{10.1088/0067-0049/199/1/9}}}.

\bibitem[{Elitzur} and {Asensio Ramos}(2006)]{Elizur2006-CEP}
{Elitzur}, M.; {Asensio Ramos}, A.
\newblock {A new exact method for line radiative transfer}.
\newblock {\em \mnras} {\bf 2006}, {\em 365},~779--791,
  \href{http://xxx.lanl.gov/abs/astro-ph/0510616}{{\normalfont
  [astro-ph/0510616]}}.
\newblock
  doi:{\changeurlcolor{black}\href{https://doi.org/10.1111/j.1365-2966.2005.09770.x}{\detokenize{10.1111/j.1365-2966.2005.09770.x}}}.

\bibitem[{Hubeny}(2001)]{EPvsERT}
{Hubeny}, I.
\newblock {From Escape Probabilities to Exact Radiative Transfer}.
\newblock  Spectroscopic Challenges of Photoionized Plasmas; {Ferland}, G.;
  {Savin}, D.W., Eds.,  2001, Vol. 247, {\em Astronomical Society of the
  Pacific Conference Series}, p. 197.

\bibitem[{Evans}(1999)]{Evans99}
{Evans}, II, N.J.
\newblock {Physical Conditions in Regions of Star Formation}.
\newblock {\em \araa} {\bf 1999}, {\em 37},~311--362,
  \href{http://xxx.lanl.gov/abs/astro-ph/9905050}{{\normalfont
  [astro-ph/9905050]}}.
\newblock
  doi:{\changeurlcolor{black}\href{https://doi.org/10.1146/annurev.astro.37.1.311}{\detokenize{10.1146/annurev.astro.37.1.311}}}.

\bibitem[{Shu}(1977)]{Shu77}
{Shu}, F.H.
\newblock {Self-similar collapse of isothermal spheres and star formation}.
\newblock {\em \apj} {\bf 1977}, {\em 214},~488--497.
\newblock
  doi:{\changeurlcolor{black}\href{https://doi.org/10.1086/155274}{\detokenize{10.1086/155274}}}.

\bibitem[{Keto} \em{et~al.}(2015){Keto}, {Caselli}, and {Rawlings}]{Keto+15}
{Keto}, E.; {Caselli}, P.; {Rawlings}, J.
\newblock {The dynamics of collapsing cores and star formation}.
\newblock {\em \mnras} {\bf 2015}, {\em 446},~3731--3740,
  \href{http://xxx.lanl.gov/abs/1410.5889}{{\normalfont
  [arXiv:astro-ph.SR/1410.5889]}}.
\newblock
  doi:{\changeurlcolor{black}\href{https://doi.org/10.1093/mnras/stu2247}{\detokenize{10.1093/mnras/stu2247}}}.

\bibitem[{Eldridge} \em{et~al.}(2017){Eldridge}, {Stanway}, {Xiao},
  {McClelland}, {Taylor}, {Ng}, {Greis}, and {Bray}]{Eldridge2017}
{Eldridge}, J.J.; {Stanway}, E.R.; {Xiao}, L.; {McClelland}, L.A.S.; {Taylor},
  G.; {Ng}, M.; {Greis}, S.M.L.; {Bray}, J.C.
\newblock {Binary Population and Spectral Synthesis Version 2.1: Construction,
  Observational Verification, and New Results}.
\newblock {\em \pasa} {\bf 2017}, {\em 34},~e058,
  \href{http://xxx.lanl.gov/abs/1710.02154}{{\normalfont
  [arXiv:astro-ph.SR/1710.02154]}}.
\newblock
  doi:{\changeurlcolor{black}\href{https://doi.org/10.1017/pasa.2017.51}{\detokenize{10.1017/pasa.2017.51}}}.

\bibitem[{Wofford} \em{et~al.}(2016){Wofford}, {Charlot}, {Bruzual},
  {Eldridge}, {Calzetti}, {Adamo}, {Cignoni}, {de Mink}, {Gouliermis},
  {Grasha}, {Grebel}, {Lee}, {{\"O}stlin}, {Smith}, {Ubeda}, and
  {Zackrisson}]{Wofford2016}
{Wofford}, A.; {Charlot}, S.; {Bruzual}, G.; {Eldridge}, J.J.; {Calzetti}, D.;
  {Adamo}, A.; {Cignoni}, M.; {de Mink}, S.E.; {Gouliermis}, D.A.; {Grasha},
  K.; {Grebel}, E.K.; {Lee}, J.C.; {{\"O}stlin}, G.; {Smith}, L.J.; {Ubeda},
  L.; {Zackrisson}, E.
\newblock {A comprehensive comparative test of seven widely used spectral
  synthesis models against multi-band photometry of young massive-star
  clusters}.
\newblock {\em \mnras} {\bf 2016}, {\em 457},~4296--4322,
  \href{http://xxx.lanl.gov/abs/1601.03850}{{\normalfont [1601.03850]}}.
\newblock
  doi:{\changeurlcolor{black}\href{https://doi.org/10.1093/mnras/stw150}{\detokenize{10.1093/mnras/stw150}}}.

\bibitem[{Izotov} \em{et~al.}(2016){Izotov}, {Schaerer}, {Thuan}, {Worseck},
  {Guseva}, {Orlitov{\'a}}, and {Verhamme}]{Izotov2016}
{Izotov}, Y.I.; {Schaerer}, D.; {Thuan}, T.X.; {Worseck}, G.; {Guseva}, N.G.;
  {Orlitov{\'a}}, I.; {Verhamme}, A.
\newblock {Detection of high Lyman continuum leakage from four low-redshift
  compact star-forming galaxies}.
\newblock {\em \mnras} {\bf 2016}, {\em 461},~3683--3701,
  \href{http://xxx.lanl.gov/abs/1605.05160}{{\normalfont [1605.05160]}}.
\newblock
  doi:{\changeurlcolor{black}\href{https://doi.org/10.1093/mnras/stw1205}{\detokenize{10.1093/mnras/stw1205}}}.

\bibitem[{Kroupa}(2001)]{Kroupa2001}
{Kroupa}, P.
\newblock {On the variation of the initial mass function}.
\newblock {\em \mnras} {\bf 2001}, {\em 322},~231--246,
  \href{http://xxx.lanl.gov/abs/astro-ph/0009005}{{\normalfont
  [astro-ph/0009005]}}.
\newblock
  doi:{\changeurlcolor{black}\href{https://doi.org/10.1046/j.1365-8711.2001.04022.x}{\detokenize{10.1046/j.1365-8711.2001.04022.x}}}.

\bibitem[{Meynet} \em{et~al.}(1994){Meynet}, {Maeder}, {Schaller}, {Schaerer},
  and {Charbonnel}]{Meynet1994}
{Meynet}, G.; {Maeder}, A.; {Schaller}, G.; {Schaerer}, D.; {Charbonnel}, C.
\newblock {Grids of massive stars with high mass loss rates. V. From 12 to 120
  M$_{sun}$\_ at Z=0.001, 0.004, 0.008, 0.020 and 0.040}.
\newblock {\em \aaps} {\bf 1994}, {\em 103},~97--105.

\bibitem[{Habing}(1968)]{Habing1968}
{Habing}, H.J.
\newblock {The interstellar radiation density between 912 A and 2400 A}.
\newblock {\em \bain} {\bf 1968}, {\em 19},~421.

\bibitem[{Veilleux} and {Osterbrock}(1987)]{Veilleux1987}
{Veilleux}, S.; {Osterbrock}, D.E.
\newblock {Spectral classification of emission-line galaxies}.
\newblock {\em \apjs} {\bf 1987}, {\em 63},~295--310.
\newblock
  doi:{\changeurlcolor{black}\href{https://doi.org/10.1086/191166}{\detokenize{10.1086/191166}}}.

\bibitem[{Revalski} \em{et~al.}(2018){Revalski}, {Crenshaw}, {Kraemer},
  {Fischer}, {Schmitt}, and {Machuca}]{Revalski2018}
{Revalski}, M.; {Crenshaw}, D.M.; {Kraemer}, S.B.; {Fischer}, T.C.; {Schmitt},
  H.R.; {Machuca}, C.
\newblock {Quantifying Feedback from Narrow Line Region Outflows in Nearby
  Active Galaxies. I. Spatially Resolved Mass Outflow Rates for the Seyfert 2
  Galaxy Markarian 573}.
\newblock {\em \apj} {\bf 2018}, {\em 856},~46,
  \href{http://xxx.lanl.gov/abs/1802.07734}{{\normalfont [1802.07734]}}.
\newblock
  doi:{\changeurlcolor{black}\href{https://doi.org/10.3847/1538-4357/aab107}{\detokenize{10.3847/1538-4357/aab107}}}.

\bibitem[{Sutherland} and {Dopita}(2017)]{Sutherland2017}
{Sutherland}, R.S.; {Dopita}, M.A.
\newblock {Effects of Preionization in Radiative Shocks. I. Self-consistent
  Models}.
\newblock {\em \apjs} {\bf 2017}, {\em 229},~34,
  \href{http://xxx.lanl.gov/abs/1702.07453}{{\normalfont
  [arXiv:astro-ph.IM/1702.07453]}}.
\newblock
  doi:{\changeurlcolor{black}\href{https://doi.org/10.3847/1538-4365/aa6541}{\detokenize{10.3847/1538-4365/aa6541}}}.

\bibitem[{Ballesteros-Paredes} \em{et~al.}(1999){Ballesteros-Paredes},
  {V{\'a}zquez-Semadeni}, and {Scalo}]{BP+99}
{Ballesteros-Paredes}, J.; {V{\'a}zquez-Semadeni}, E.; {Scalo}, J.
\newblock {Clouds as Turbulent Density Fluctuations: Implications for Pressure
  Confinement and Spectral Line Data Interpretation}.
\newblock {\em \apj} {\bf 1999}, {\em 515},~286--303,
  \href{http://xxx.lanl.gov/abs/astro-ph/9806059}{{\normalfont
  [astro-ph/9806059]}}.
\newblock
  doi:{\changeurlcolor{black}\href{https://doi.org/10.1086/307007}{\detokenize{10.1086/307007}}}.

\bibitem[{Hennebelle} and {P{\'e}rault}(1999)]{HP99}
{Hennebelle}, P.; {P{\'e}rault}, M.
\newblock {Dynamical condensation in a thermally bistable flow. Application to
  interstellar cirrus}.
\newblock {\em \aap} {\bf 1999}, {\em 351},~309--322.

\bibitem[{Koyama} and {Inutsuka}(2002)]{KI02}
{Koyama}, H.; {Inutsuka}, S.i.
\newblock {An Origin of Supersonic Motions in Interstellar Clouds}.
\newblock {\em \apjl} {\bf 2002}, {\em 564},~L97--L100,
  \href{http://xxx.lanl.gov/abs/astro-ph/0112420}{{\normalfont
  [astro-ph/0112420]}}.
\newblock
  doi:{\changeurlcolor{black}\href{https://doi.org/10.1086/338978}{\detokenize{10.1086/338978}}}.

\bibitem[{Audit} and {Hennebelle}(2005)]{AH05}
{Audit}, E.; {Hennebelle}, P.
\newblock {Thermal condensation in a turbulent atomic hydrogen flow}.
\newblock {\em \aap} {\bf 2005}, {\em 433},~1--13,
  \href{http://xxx.lanl.gov/abs/astro-ph/0410062}{{\normalfont
  [astro-ph/0410062]}}.
\newblock
  doi:{\changeurlcolor{black}\href{https://doi.org/10.1051/0004-6361:20041474}{\detokenize{10.1051/0004-6361:20041474}}}.

\bibitem[{Heitsch} \em{et~al.}(2005){Heitsch}, {Burkert}, {Hartmann}, {Slyz},
  and {Devriendt}]{Heitsch+05}
{Heitsch}, F.; {Burkert}, A.; {Hartmann}, L.W.; {Slyz}, A.D.; {Devriendt},
  J.E.G.
\newblock {Formation of Structure in Molecular Clouds: A Case Study}.
\newblock {\em \apjl} {\bf 2005}, {\em 633},~L113--L116,
  \href{http://xxx.lanl.gov/abs/astro-ph/0507567}{{\normalfont
  [astro-ph/0507567]}}.
\newblock
  doi:{\changeurlcolor{black}\href{https://doi.org/10.1086/498413}{\detokenize{10.1086/498413}}}.

\bibitem[{V{\'a}zquez-Semadeni} \em{et~al.}(2007){V{\'a}zquez-Semadeni},
  {G{\'o}mez}, {Jappsen}, {Ballesteros-Paredes}, {Gonz{\'a}lez}, and
  {Klessen}]{VS+07}
{V{\'a}zquez-Semadeni}, E.; {G{\'o}mez}, G.C.; {Jappsen}, A.K.;
  {Ballesteros-Paredes}, J.; {Gonz{\'a}lez}, R.F.; {Klessen}, R.S.
\newblock {Molecular Cloud Evolution. II. From Cloud Formation to the Early
  Stages of Star Formation in Decaying Conditions}.
\newblock {\em \apj} {\bf 2007}, {\em 657},~870--883,
  \href{http://xxx.lanl.gov/abs/astro-ph/0608375}{{\normalfont
  [astro-ph/0608375]}}.
\newblock
  doi:{\changeurlcolor{black}\href{https://doi.org/10.1086/510771}{\detokenize{10.1086/510771}}}.

\bibitem[{Banerjee} \em{et~al.}(2009){Banerjee}, {V{\'a}zquez-Semadeni},
  {Hennebelle}, and {Klessen}]{Banerjee+09}
{Banerjee}, R.; {V{\'a}zquez-Semadeni}, E.; {Hennebelle}, P.; {Klessen}, R.S.
\newblock {Clump morphology and evolution in MHD simulations of molecular cloud
  formation}.
\newblock {\em \mnras} {\bf 2009}, {\em 398},~1082--1092,
  \href{http://xxx.lanl.gov/abs/0808.0986}{{\normalfont [0808.0986]}}.
\newblock
  doi:{\changeurlcolor{black}\href{https://doi.org/10.1111/j.1365-2966.2009.15115.x}{\detokenize{10.1111/j.1365-2966.2009.15115.x}}}.

\bibitem[{Dobbs} \em{et~al.}(2012){Dobbs}, {Pringle}, and {Burkert}]{Dobbs+12}
{Dobbs}, C.L.; {Pringle}, J.E.; {Burkert}, A.
\newblock {Giant molecular clouds: what are they made from, and how do they get
  there?}
\newblock {\em \mnras} {\bf 2012}, {\em 425},~2157--2168,
  \href{http://xxx.lanl.gov/abs/1206.4904}{{\normalfont [1206.4904]}}.
\newblock
  doi:{\changeurlcolor{black}\href{https://doi.org/10.1111/j.1365-2966.2012.21558.x}{\detokenize{10.1111/j.1365-2966.2012.21558.x}}}.

\bibitem[{Olsen} \em{et~al.}(2017){Olsen}, {Greve}, {Narayanan}, {Thompson},
  {Dav{\'e}}, {Niebla Rios}, and {Stawinski}]{olsen17}
{Olsen}, K.; {Greve}, T.R.; {Narayanan}, D.; {Thompson}, R.; {Dav{\'e}}, R.;
  {Niebla Rios}, L.; {Stawinski}, S.
\newblock {S{\'{I}}GAME Simulations of the [CII], [OI], and [OIII] Line
  Emission from Star-forming Galaxies at $z\simeq 6$}.
\newblock {\em \apj} {\bf 2017}, {\em 846},~105,
  \href{http://xxx.lanl.gov/abs/1708.04936}{{\normalfont [1708.04936]}}.
\newblock
  doi:{\changeurlcolor{black}\href{https://doi.org/10.3847/1538-4357/aa86b4}{\detokenize{10.3847/1538-4357/aa86b4}}}.

\bibitem[{Pallottini} \em{et~al.}(2017){Pallottini}, {Ferrara}, {Bovino},
  {Vallini}, {Gallerani}, {Maiolino}, and {Salvadori}]{pallottini:2017chem}
{Pallottini}, A.; {Ferrara}, A.; {Bovino}, S.; {Vallini}, L.; {Gallerani}, S.;
  {Maiolino}, R.; {Salvadori}, S.
\newblock {The impact of chemistry on the structure of high-z galaxies}.
\newblock {\em \mnras} {\bf 2017}, {\em 471},~4128--4143,
  \href{http://xxx.lanl.gov/abs/1707.04259}{{\normalfont [1707.04259]}}.
\newblock
  doi:{\changeurlcolor{black}\href{https://doi.org/10.1093/mnras/stx1792}{\detokenize{10.1093/mnras/stx1792}}}.

\bibitem[{Capelo} \em{et~al.}(2018){Capelo}, {Bovino}, {Lupi}, {Schleicher},
  and {Grassi}]{capelo:2018}
{Capelo}, P.R.; {Bovino}, S.; {Lupi}, A.; {Schleicher}, D.R.G.; {Grassi}, T.
\newblock {The effect of non-equilibrium metal cooling on the interstellar
  medium}.
\newblock {\em \mnras} {\bf 2018}, {\em 475},~3283--3304,
  \href{http://xxx.lanl.gov/abs/1710.01302}{{\normalfont [1710.01302]}}.
\newblock
  doi:{\changeurlcolor{black}\href{https://doi.org/10.1093/mnras/stx3355}{\detokenize{10.1093/mnras/stx3355}}}.

\bibitem[{Rosdahl} \em{et~al.}(2015){Rosdahl}, {Schaye}, {Teyssier}, and
  {Agertz}]{rosdahl:2015MNRAS}
{Rosdahl}, J.; {Schaye}, J.; {Teyssier}, R.; {Agertz}, O.
\newblock {Galaxies that shine: radiation-hydrodynamical simulations of disc
  galaxies}.
\newblock {\em \mnras} {\bf 2015}, {\em 451},~34--58,
  \href{http://xxx.lanl.gov/abs/1501.04632}{{\normalfont [1501.04632]}}.
\newblock
  doi:{\changeurlcolor{black}\href{https://doi.org/10.1093/mnras/stv937}{\detokenize{10.1093/mnras/stv937}}}.

\bibitem[{Katz} \em{et~al.}(2017){Katz}, {Kimm}, {Sijacki}, and
  {Haehnelt}]{katz2017}
{Katz}, H.; {Kimm}, T.; {Sijacki}, D.; {Haehnelt}, M.G.
\newblock {Interpreting ALMA observations of the ISM during the epoch of
  reionization}.
\newblock {\em \mnras} {\bf 2017}, {\em 468},~4831--4861,
  \href{http://xxx.lanl.gov/abs/1612.01786}{{\normalfont [1612.01786]}}.
\newblock
  doi:{\changeurlcolor{black}\href{https://doi.org/10.1093/mnras/stx608}{\detokenize{10.1093/mnras/stx608}}}.

\bibitem[{Grassi} \em{et~al.}(2017){Grassi}, {Bovino}, {Haugb{\o}lle}, and
  {Schleicher}]{Grassi2017}
{Grassi}, T.; {Bovino}, S.; {Haugb{\o}lle}, T.; {Schleicher}, D.R.G.
\newblock {A detailed framework to incorporate dust in hydrodynamical
  simulations}.
\newblock {\em \mnras} {\bf 2017}, {\em 466},~1259--1274,
  \href{http://xxx.lanl.gov/abs/1606.01229}{{\normalfont [1606.01229]}}.
\newblock
  doi:{\changeurlcolor{black}\href{https://doi.org/10.1093/mnras/stw2871}{\detokenize{10.1093/mnras/stw2871}}}.

\bibitem[{Popping} \em{et~al.}(2017){Popping}, {Somerville}, and
  {Galametz}]{Popping2017dust}
{Popping}, G.; {Somerville}, R.S.; {Galametz}, M.
\newblock {The dust content of galaxies from z = 0 to z = 9}.
\newblock {\em \mnras} {\bf 2017}, {\em 471},~3152--3185,
  \href{http://xxx.lanl.gov/abs/1609.08622}{{\normalfont [1609.08622]}}.
\newblock
  doi:{\changeurlcolor{black}\href{https://doi.org/10.1093/mnras/stx1545}{\detokenize{10.1093/mnras/stx1545}}}.

\bibitem[{McKinnon} \em{et~al.}(2017){McKinnon}, {Torrey}, {Vogelsberger},
  {Hayward}, and {Marinacci}]{McKinnon2017}
{McKinnon}, R.; {Torrey}, P.; {Vogelsberger}, M.; {Hayward}, C.C.; {Marinacci},
  F.
\newblock {Simulating the dust content of galaxies: successes and failures}.
\newblock {\em \mnras} {\bf 2017}, {\em 468},~1505--1521,
  \href{http://xxx.lanl.gov/abs/1606.02714}{{\normalfont [1606.02714]}}.
\newblock
  doi:{\changeurlcolor{black}\href{https://doi.org/10.1093/mnras/stx467}{\detokenize{10.1093/mnras/stx467}}}.

\bibitem[{Aoyama} \em{et~al.}(2018){Aoyama}, {Hou}, {Hirashita}, {Nagamine},
  and {Shimizu}]{Aoyama2018}
{Aoyama}, S.; {Hou}, K.C.; {Hirashita}, H.; {Nagamine}, K.; {Shimizu}, I.
\newblock {Cosmological simulation with dust formation and destruction}.
\newblock {\em ArXiv e-prints} {\bf 2018},
  \href{http://xxx.lanl.gov/abs/1802.04027}{{\normalfont [1802.04027]}}.

\bibitem[{Behrens} \em{et~al.}(2018){Behrens}, {Pallottini}, {Ferrara},
  {Gallerani}, and {Vallini}]{Behrens2018}
{Behrens}, C.; {Pallottini}, A.; {Ferrara}, A.; {Gallerani}, S.; {Vallini}, L.
\newblock {Dusty galaxies in the Epoch of Reionization: simulations}.
\newblock {\em \mnras} {\bf 2018}, {\em 477},~552--565,
  \href{http://xxx.lanl.gov/abs/1802.07772}{{\normalfont [1802.07772]}}.
\newblock
  doi:{\changeurlcolor{black}\href{https://doi.org/10.1093/mnras/sty552}{\detokenize{10.1093/mnras/sty552}}}.

\bibitem[{Narayanan} \em{et~al.}(2018){Narayanan}, {Conroy}, {Dave}, {Johnson},
  and {Popping}]{Narayanan2018}
{Narayanan}, D.; {Conroy}, C.; {Dave}, R.; {Johnson}, B.; {Popping}, G.
\newblock {A Theory for the Variation of Dust Attenuation Laws in Galaxies}.
\newblock {\em ArXiv e-prints} {\bf 2018},
  \href{http://xxx.lanl.gov/abs/1805.06905}{{\normalfont [1805.06905]}}.

\bibitem[{Shirazi} and {Brinchmann}(2012)]{Shirazi2012}
{Shirazi}, M.; {Brinchmann}, J.
\newblock {Strongly star forming galaxies in the local Universe with nebular He
  II{$\lambda$}4686 emission}.
\newblock {\em \mnras} {\bf 2012}, {\em 421},~1043--1063,
  \href{http://xxx.lanl.gov/abs/1201.1290}{{\normalfont [1201.1290]}}.
\newblock
  doi:{\changeurlcolor{black}\href{https://doi.org/10.1111/j.1365-2966.2012.20439.x}{\detokenize{10.1111/j.1365-2966.2012.20439.x}}}.

\bibitem[{Gutkin} \em{et~al.}(2016){Gutkin}, {Charlot}, and
  {Bruzual}]{Gutkin2016}
{Gutkin}, J.; {Charlot}, S.; {Bruzual}, G.
\newblock {Modelling the nebular emission from primeval to present-day
  star-forming galaxies}.
\newblock {\em \mnras} {\bf 2016}, {\em 462},~1757--1774,
  \href{http://xxx.lanl.gov/abs/1607.06086}{{\normalfont [1607.06086]}}.
\newblock
  doi:{\changeurlcolor{black}\href{https://doi.org/10.1093/mnras/stw1716}{\detokenize{10.1093/mnras/stw1716}}}.

\bibitem[{Senchyna} \em{et~al.}(2017){Senchyna}, {Stark},
  {Vidal-Garc{\'{\i}}a}, {Chevallard}, {Charlot}, {Mainali}, {Jones},
  {Wofford}, {Feltre}, and {Gutkin}]{Senchyna2017}
{Senchyna}, P.; {Stark}, D.P.; {Vidal-Garc{\'{\i}}a}, A.; {Chevallard}, J.;
  {Charlot}, S.; {Mainali}, R.; {Jones}, T.; {Wofford}, A.; {Feltre}, A.;
  {Gutkin}, J.
\newblock {Ultraviolet spectra of extreme nearby star-forming regions -
  approaching a local reference sample for JWST}.
\newblock {\em \mnras} {\bf 2017}, {\em 472},~2608--2632,
  \href{http://xxx.lanl.gov/abs/1706.00881}{{\normalfont [1706.00881]}}.
\newblock
  doi:{\changeurlcolor{black}\href{https://doi.org/10.1093/mnras/stx2059}{\detokenize{10.1093/mnras/stx2059}}}.

\bibitem[{Gallerani} \em{et~al.}(2014){Gallerani}, {Ferrara}, {Neri}, and
  {Maiolino}]{Gallerani2014}
{Gallerani}, S.; {Ferrara}, A.; {Neri}, R.; {Maiolino}, R.
\newblock {First CO(17-16) emission line detected in a z > 6 quasar}.
\newblock {\em \mnras} {\bf 2014}, {\em 445},~2848--2853,
  \href{http://xxx.lanl.gov/abs/1409.4413}{{\normalfont [1409.4413]}}.
\newblock
  doi:{\changeurlcolor{black}\href{https://doi.org/10.1093/mnras/stu2031}{\detokenize{10.1093/mnras/stu2031}}}.

\bibitem[{Meijerink} and {Spaans}(2005)]{Meijerink2005}
{Meijerink}, R.; {Spaans}, M.
\newblock {Diagnostics of irradiated gas in galaxy nuclei. I. A far-ultraviolet
  and X-ray dominated region code}.
\newblock {\em \aap} {\bf 2005}, {\em 436},~397--409,
  \href{http://xxx.lanl.gov/abs/astro-ph/0502454}{{\normalfont
  [astro-ph/0502454]}}.
\newblock
  doi:{\changeurlcolor{black}\href{https://doi.org/10.1051/0004-6361:20042398}{\detokenize{10.1051/0004-6361:20042398}}}.

\bibitem[{Mingozzi} \em{et~al.}(2018){Mingozzi}, {Vallini}, {Pozzi}, {Vignali},
  {Mignano}, {Gruppioni}, {Talia}, {Cimatti}, {Cresci}, and
  {Massardi}]{Mingozzi2018}
{Mingozzi}, M.; {Vallini}, L.; {Pozzi}, F.; {Vignali}, C.; {Mignano}, A.;
  {Gruppioni}, C.; {Talia}, M.; {Cimatti}, A.; {Cresci}, G.; {Massardi}, M.
\newblock {CO excitation in the Seyfert galaxy NGC 34: stars, shock or AGN
  driven?}
\newblock {\em \mnras} {\bf 2018}, {\em 474},~3640--3648,
  \href{http://xxx.lanl.gov/abs/1711.07995}{{\normalfont [1711.07995]}}.
\newblock
  doi:{\changeurlcolor{black}\href{https://doi.org/10.1093/mnras/stx3011}{\detokenize{10.1093/mnras/stx3011}}}.

\bibitem[{Crenshaw} \em{et~al.}(2003){Crenshaw}, {Kraemer}, and
  {George}]{crenshaw2003}
{Crenshaw}, D.M.; {Kraemer}, S.B.; {George}, I.M.
\newblock {Mass Loss from the Nuclei of Active Galaxies}.
\newblock {\em \araa} {\bf 2003}, {\em 41},~117--167.
\newblock
  doi:{\changeurlcolor{black}\href{https://doi.org/10.1146/annurev.astro.41.082801.100328}{\detokenize{10.1146/annurev.astro.41.082801.100328}}}.

\bibitem[{Hopkins} \em{et~al.}(2005){Hopkins}, {Hernquist}, {Cox}, {Di Matteo},
  {Martini}, {Robertson}, and {Springel}]{hopkins2005}
{Hopkins}, P.F.; {Hernquist}, L.; {Cox}, T.J.; {Di Matteo}, T.; {Martini}, P.;
  {Robertson}, B.; {Springel}, V.
\newblock {Black Holes in Galaxy Mergers: Evolution of Quasars}.
\newblock {\em \apj} {\bf 2005}, {\em 630},~705--715,
  \href{http://xxx.lanl.gov/abs/astro-ph/0504190}{{\normalfont
  [astro-ph/0504190]}}.
\newblock
  doi:{\changeurlcolor{black}\href{https://doi.org/10.1086/432438}{\detokenize{10.1086/432438}}}.

\bibitem[{Kormendy} and {Ho}(2013)]{kormendy2013}
{Kormendy}, J.; {Ho}, L.C.
\newblock {Coevolution (Or Not) of Supermassive Black Holes and Host Galaxies}.
\newblock {\em \araa} {\bf 2013}, {\em 51},~511--653,
  \href{http://xxx.lanl.gov/abs/1304.7762}{{\normalfont [1304.7762]}}.
\newblock
  doi:{\changeurlcolor{black}\href{https://doi.org/10.1146/annurev-astro-082708-101811}{\detokenize{10.1146/annurev-astro-082708-101811}}}.

\bibitem[{Batiste} \em{et~al.}(2017){Batiste}, {Bentz}, {Raimundo},
  {Vestergaard}, and {Onken}]{batiste2017}
{Batiste}, M.; {Bentz}, M.C.; {Raimundo}, S.I.; {Vestergaard}, M.; {Onken},
  C.A.
\newblock {Recalibration of the M$_{BH}$-$\sigma_\star$ Relation for AGN}.
\newblock {\em \apjl} {\bf 2017}, {\em 838},~L10,
  \href{http://xxx.lanl.gov/abs/1612.02815}{{\normalfont [1612.02815]}}.
\newblock
  doi:{\changeurlcolor{black}\href{https://doi.org/10.3847/2041-8213/aa6571}{\detokenize{10.3847/2041-8213/aa6571}}}.

\bibitem[{Fiore} \em{et~al.}(2017){Fiore}, {Feruglio}, {Shankar}, {Bischetti},
  {Bongiorno}, {Brusa}, {Carniani}, {Cicone}, {Duras}, {Lamastra}, {Mainieri},
  {Marconi}, {Menci}, {Maiolino}, {Piconcelli}, {Vietri}, and
  {Zappacosta}]{fiore2017}
{Fiore}, F.; {Feruglio}, C.; {Shankar}, F.; {Bischetti}, M.; {Bongiorno}, A.;
  {Brusa}, M.; {Carniani}, S.; {Cicone}, C.; {Duras}, F.; {Lamastra}, A.;
  {Mainieri}, V.; {Marconi}, A.; {Menci}, N.; {Maiolino}, R.; {Piconcelli}, E.;
  {Vietri}, G.; {Zappacosta}, L.
\newblock {AGN wind scaling relations and the co-evolution of black holes and
  galaxies}.
\newblock {\em \aap} {\bf 2017}, {\em 601},~A143,
  \href{http://xxx.lanl.gov/abs/1702.04507}{{\normalfont [1702.04507]}}.
\newblock
  doi:{\changeurlcolor{black}\href{https://doi.org/10.1051/0004-6361/201629478}{\detokenize{10.1051/0004-6361/201629478}}}.

\bibitem[{Choi} \em{et~al.}(2016){Choi}, {Ostriker}, {Naab}, {Somerville},
  {Hirschmann}, {N{\'u}{\~n}ez}, {Hu}, and {Oser}]{Choi2016}
{Choi}, E.; {Ostriker}, J.P.; {Naab}, T.; {Somerville}, R.S.; {Hirschmann}, M.;
  {N{\'u}{\~n}ez}, A.; {Hu}, C.Y.; {Oser}, L.
\newblock {Physics of Galactic Metals: Evolutionary Effects due to Production,
  Distribution, Feedback, and Interaction with Black Holes}.
\newblock {\em ArXiv:1610.09389} {\bf 2016},
  \href{http://xxx.lanl.gov/abs/1610.09389}{{\normalfont [1610.09389]}}.

\bibitem[{Richings} and {Faucher-Gigu{\`e}re}(2018)]{richings2018}
{Richings}, A.J.; {Faucher-Gigu{\`e}re}, C.A.
\newblock {The origin of fast molecular outflows in quasars: molecule formation
  in AGN-driven galactic winds}.
\newblock {\em \mnras} {\bf 2018}, {\em 474},~3673--3699,
  \href{http://xxx.lanl.gov/abs/1706.03784}{{\normalfont [1706.03784]}}.
\newblock
  doi:{\changeurlcolor{black}\href{https://doi.org/10.1093/mnras/stx3014}{\detokenize{10.1093/mnras/stx3014}}}.

\bibitem[{Tumlinson} \em{et~al.}(2017){Tumlinson}, {Peeples}, and
  {Werk}]{tumlinson2017}
{Tumlinson}, J.; {Peeples}, M.S.; {Werk}, J.K.
\newblock {The Circumgalactic Medium}.
\newblock {\em \araa} {\bf 2017}, {\em 55},~389--432,
  \href{http://xxx.lanl.gov/abs/1709.09180}{{\normalfont [1709.09180]}}.
\newblock
  doi:{\changeurlcolor{black}\href{https://doi.org/10.1146/annurev-astro-091916-055240}{\detokenize{10.1146/annurev-astro-091916-055240}}}.

\bibitem[{Suresh} \em{et~al.}(2017){Suresh}, {Rubin}, {Kannan}, {Werk},
  {Hernquist}, and {Vogelsberger}]{Suresh2017}
{Suresh}, J.; {Rubin}, K.H.R.; {Kannan}, R.; {Werk}, J.K.; {Hernquist}, L.;
  {Vogelsberger}, M.
\newblock {On the OVI abundance in the circumgalactic medium of low-redshift
  galaxies}.
\newblock {\em \mnras} {\bf 2017}, {\em 465},~2966--2982,
  \href{http://xxx.lanl.gov/abs/1511.00687}{{\normalfont [1511.00687]}}.
\newblock
  doi:{\changeurlcolor{black}\href{https://doi.org/10.1093/mnras/stw2499}{\detokenize{10.1093/mnras/stw2499}}}.

\bibitem[{Hummels} \em{et~al.}(2017){Hummels}, {Smith}, and
  {Silvia}]{hummels2017}
{Hummels}, C.B.; {Smith}, B.D.; {Silvia}, D.W.
\newblock {Trident: A Universal Tool for Generating Synthetic Absorption
  Spectra from Astrophysical Simulations}.
\newblock {\em \apj} {\bf 2017}, {\em 847},~59,
  \href{http://xxx.lanl.gov/abs/1612.03935}{{\normalfont
  [arXiv:astro-ph.IM/1612.03935]}}.
\newblock
  doi:{\changeurlcolor{black}\href{https://doi.org/10.3847/1538-4357/aa7e2d}{\detokenize{10.3847/1538-4357/aa7e2d}}}.

\bibitem[{Dubois} and {Teyssier}(2008)]{Dubois2008}
{Dubois}, Y.; {Teyssier}, R.
\newblock {On the onset of galactic winds in quiescent star forming galaxies}.
\newblock {\em \aap} {\bf 2008}, {\em 477},~79--94,
  \href{http://xxx.lanl.gov/abs/0707.3376}{{\normalfont [0707.3376]}}.
\newblock
  doi:{\changeurlcolor{black}\href{https://doi.org/10.1051/0004-6361:20078326}{\detokenize{10.1051/0004-6361:20078326}}}.

\bibitem[{Padoan} and {Nordlund}(2011)]{Padoan2011}
{Padoan}, P.; {Nordlund}, {\AA}.
\newblock {The Star Formation Rate of Supersonic Magnetohydrodynamic
  Turbulence}.
\newblock {\em \apj} {\bf 2011}, {\em 730},~40,
  \href{http://xxx.lanl.gov/abs/0907.0248}{{\normalfont
  [arXiv:astro-ph.GA/0907.0248]}}.
\newblock
  doi:{\changeurlcolor{black}\href{https://doi.org/10.1088/0004-637X/730/1/40}{\detokenize{10.1088/0004-637X/730/1/40}}}.

\bibitem[{Federrath} and {Klessen}(2012)]{Federrath2012}
{Federrath}, C.; {Klessen}, R.S.
\newblock {The Star Formation Rate of Turbulent Magnetized Clouds: Comparing
  Theory, Simulations, and Observations}.
\newblock {\em \apj} {\bf 2012}, {\em 761},~156,
  \href{http://xxx.lanl.gov/abs/1209.2856}{{\normalfont
  [arXiv:astro-ph.SR/1209.2856]}}.
\newblock
  doi:{\changeurlcolor{black}\href{https://doi.org/10.1088/0004-637X/761/2/156}{\detokenize{10.1088/0004-637X/761/2/156}}}.

\bibitem[{Kimm} and {Cen}(2014)]{Kimm2014}
{Kimm}, T.; {Cen}, R.
\newblock {Escape Fraction of Ionizing Photons during Reionization: Effects due
  to Supernova Feedback and Runaway OB Stars}.
\newblock {\em \apj} {\bf 2014}, {\em 788},~121,
  \href{http://xxx.lanl.gov/abs/1405.0552}{{\normalfont [1405.0552]}}.
\newblock
  doi:{\changeurlcolor{black}\href{https://doi.org/10.1088/0004-637X/788/2/121}{\detokenize{10.1088/0004-637X/788/2/121}}}.

\bibitem[{Werk} \em{et~al.}(2016){Werk}, {Prochaska}, {Cantalupo}, {Fox},
  {Oppenheimer}, {Tumlinson}, {Tripp}, {Lehner}, and {McQuinn}]{Werk2016}
{Werk}, J.K.; {Prochaska}, J.X.; {Cantalupo}, S.; {Fox}, A.J.; {Oppenheimer},
  B.; {Tumlinson}, J.; {Tripp}, T.M.; {Lehner}, N.; {McQuinn}, M.
\newblock {The COS-Halos Survey: Origins of the Highly Ionized Circumgalactic
  Medium of Star-Forming Galaxies}.
\newblock {\em \apj} {\bf 2016}, {\em 833},~54,
  \href{http://xxx.lanl.gov/abs/1609.00012}{{\normalfont [1609.00012]}}.
\newblock
  doi:{\changeurlcolor{black}\href{https://doi.org/10.3847/1538-4357/833/1/54}{\detokenize{10.3847/1538-4357/833/1/54}}}.

\bibitem[{McCourt} \em{et~al.}(2018){McCourt}, {Oh}, {O'Leary}, and
  {Madigan}]{McCourt2018}
{McCourt}, M.; {Oh}, S.P.; {O'Leary}, R.; {Madigan}, A.M.
\newblock {A characteristic scale for cold gas}.
\newblock {\em \mnras} {\bf 2018}, {\em 473},~5407--5431,
  \href{http://xxx.lanl.gov/abs/1610.01164}{{\normalfont [1610.01164]}}.
\newblock
  doi:{\changeurlcolor{black}\href{https://doi.org/10.1093/mnras/stx2687}{\detokenize{10.1093/mnras/stx2687}}}.

\bibitem[{Oppenheimer} \em{et~al.}(2018){Oppenheimer}, {Segers}, {Schaye},
  {Richings}, and {Crain}]{Oppenheimer2018}
{Oppenheimer}, B.D.; {Segers}, M.; {Schaye}, J.; {Richings}, A.J.; {Crain},
  R.A.
\newblock {Flickering AGN can explain the strong circumgalactic O VI observed
  by COS-Halos}.
\newblock {\em \mnras} {\bf 2018}, {\em 474},~4740--4755,
  \href{http://xxx.lanl.gov/abs/1705.07897}{{\normalfont [1705.07897]}}.
\newblock
  doi:{\changeurlcolor{black}\href{https://doi.org/10.1093/mnras/stx2967}{\detokenize{10.1093/mnras/stx2967}}}.

\bibitem[{Del Zanna} and {DeLuca}(2018)]{delzanna18}
{Del Zanna}, G.; {DeLuca}, E.E.
\newblock {Solar Coronal Lines in the Visible and Infrared: A Rough Guide}.
\newblock {\em \apj} {\bf 2018}, {\em 852},~52,
  \href{http://xxx.lanl.gov/abs/1708.03626}{{\normalfont
  [arXiv:astro-ph.SR/1708.03626]}}.
\newblock
  doi:{\changeurlcolor{black}\href{https://doi.org/10.3847/1538-4357/aa9edf}{\detokenize{10.3847/1538-4357/aa9edf}}}.

\end{thebibliography}

\end{document}